\documentclass[prc,superscriptaddress,showpacs,twocolumn,amssymb,amsmath,amsfonts,aps]{revtex4}

\usepackage{graphicx}
\usepackage{dcolumn}
\usepackage{bm}

\begin{document}

\preprint{CMU/JLab/8-2005}

%
%
%
%

\newcommand*{\CMU}{Carnegie Mellon University, Pittsburgh, Pennsylvania 15213}
\affiliation{\CMU}
\newcommand*{\ASU}{Arizona State University, Tempe, Arizona 85287-1504}
\affiliation{\ASU}
\newcommand*{\UCLA}{University of California at Los Angeles, Los Angeles, California  90095-1547}
\affiliation{\UCLA}
\newcommand*{\CUA}{Catholic University of America, Washington, D.C. 20064}
\affiliation{\CUA}
\newcommand*{\SACLAY}{CEA-Saclay, Service de Physique Nucl\'eaire, F91191 Gif-sur-Yvette,Cedex, France}
\affiliation{\SACLAY}
\newcommand*{\CNU}{Christopher Newport University, Newport News, Virginia 23606}
\affiliation{\CNU}
\newcommand*{\UCONN}{University of Connecticut, Storrs, Connecticut 06269}
\affiliation{\UCONN}
\newcommand*{\DUKE}{Duke University, Durham, North Carolina 27708-0305}
\affiliation{\DUKE}
\newcommand*{\ECOSSEE}{Edinburgh University, Edinburgh EH9 3JZ, United Kingdom}
\affiliation{\ECOSSEE}
\newcommand*{\FIU}{Florida International University, Miami, Florida 33199}
\affiliation{\FIU}
\newcommand*{\FSU}{Florida State University, Tallahassee, Florida 32306}
\affiliation{\FSU}
\newcommand*{\GWU}{The George Washington University, Washington, DC 20052}
\affiliation{\GWU}
\newcommand*{\ECOSSEG}{University of Glasgow, Glasgow G12 8QQ, United Kingdom}
\affiliation{\ECOSSEG}
\newcommand*{\ISU}{Idaho State University, Pocatello, Idaho 83209}
\affiliation{\ISU}
\newcommand*{\INFNFR}{INFN, Laboratori Nazionali di Frascati, Frascati, Italy}
\affiliation{\INFNFR}
\newcommand*{\INFNGE}{INFN, Sezione di Genova, 16146 Genova, Italy}
\affiliation{\INFNGE}
\newcommand*{\ORSAY}{Institut de Physique Nucleaire ORSAY, Orsay, France}
\affiliation{\ORSAY}
\newcommand*{\ITEP}{Institute of Theoretical and Experimental Physics, Moscow, 117259, Russia}
\affiliation{\ITEP}
\newcommand*{\JMU}{James Madison University, Harrisonburg, Virginia 22807}
\affiliation{\JMU}
\newcommand*{\KYUNGPOOK}{Kyungpook National University, Daegu 702-701, South Korea}
\affiliation{\KYUNGPOOK}
\newcommand*{\MIT}{Massachusetts Institute of Technology, Cambridge, Massachusetts  02139-4307}
\affiliation{\MIT}
\newcommand*{\UMASS}{University of Massachusetts, Amherst, Massachusetts  01003}
\affiliation{\UMASS}
\newcommand*{\MOSCOW}{Moscow State University, General Nuclear Physics Institute, 119899 Moscow, Russia}
\affiliation{\MOSCOW}
\newcommand*{\UNH}{University of New Hampshire, Durham, New Hampshire 03824-3568}
\affiliation{\UNH}
\newcommand*{\NSU}{Norfolk State University, Norfolk, Virginia 23504}
\affiliation{\NSU}
\newcommand*{\OHIOU}{Ohio University, Athens, Ohio  45701}
\affiliation{\OHIOU}
\newcommand*{\ODU}{Old Dominion University, Norfolk, Virginia 23529}
\affiliation{\ODU}
\newcommand*{\PITT}{University of Pittsburgh, Pittsburgh, Pennsylvania 15260}
\affiliation{\PITT}
\newcommand*{\RPI}{Rensselaer Polytechnic Institute, Troy, New York 12180-3590}
\affiliation{\RPI}
\newcommand*{\RICE}{Rice University, Houston, Texas 77005-1892}
\affiliation{\RICE}
\newcommand*{\URICH}{University of Richmond, Richmond, Virginia 23173}
\affiliation{\URICH}
\newcommand*{\SCAROLINA}{University of South Carolina, Columbia, South Carolina 29208}
\affiliation{\SCAROLINA}
\newcommand*{\JLAB}{Thomas Jefferson National Accelerator Facility, Newport News, Virginia 23606}
\affiliation{\JLAB}
\newcommand*{\UNIONC}{Union College, Schenectady, NY 12308}
\affiliation{\UNIONC}
\newcommand*{\VT}{Virginia Polytechnic Institute and State University, Blacksburg, Virginia   24061-0435}
\affiliation{\VT}
\newcommand*{\VIRGINIA}{University of Virginia, Charlottesville, Virginia 22901}
\affiliation{\VIRGINIA}
\newcommand*{\WM}{College of William and Mary, Williamsburg, Virginia 23187-8795}
\affiliation{\WM}
\newcommand*{\YEREVAN}{Yerevan Physics Institute, 375036 Yerevan, Armenia}
\affiliation{\YEREVAN}
\newcommand*{\NOWOHIOU}{Ohio University, Athens, Ohio  45701}
\newcommand*{\NOWINDSTRA}{Systems Planning and Analysis, Alexandria, Virginia 22311}
\newcommand*{\NOWUNH}{University of New Hampshire, Durham, New Hampshire 03824-3568}
\newcommand*{\NOWMOSCOW}{Moscow State University, General Nuclear Physics Institute, 119899 Moscow, Russia}
\newcommand*{\NOWCUA}{Catholic University of America, Washington, D.C. 20064}
\newcommand*{\NOWUMASS}{University of Massachusetts, Amherst, Massachusetts  01003}
\newcommand*{\NOWMIT}{Massachusetts Institute of Technology, Cambridge, Massachusetts  02139-4307}
\newcommand*{\NOWODU}{Old Dominion University, Norfolk, Virginia 23529}
\newcommand*{\NOWGEISSEN}{Physikalisches Institut der Universitaet Giessen, 35392 Giessen, Germany}
\newcommand*{\NOWROCH}{Univ. of Rochester, New York 14627}
\newcommand*{\NOWTURK}{Univ. of Sakarya, Turkey}

\author{R.~Bradford}
\altaffiliation[Current address:]{\NOWROCH}
\affiliation{\CMU}
\author{R.A.~Schumacher}
\affiliation{\CMU}
\author{J.W.C.~McNabb}
\affiliation{\CMU}
\author{L.~Todor} 
\affiliation{\CMU}
\author {G.~Adams} 
\affiliation{\RPI}
\author {P.~Ambrozewicz} 
\affiliation{\FIU}
\author {E.~Anciant} 
\affiliation{\SACLAY}
\author {M.~Anghinolfi} 
\affiliation{\INFNGE}
\author {B.~Asavapibhop} 
\affiliation{\UMASS}
\author {G.~Asryan} 
\affiliation{\YEREVAN}
% ...
\collaboration{The CLAS Collaboration}
     \noaffiliation

\title{Differential Cross Sections for $\gamma + p \rightarrow K^+ + Y$ for 
$\Lambda$ and $\Sigma^0$ Hyperons.} 

\date{9-26-05}

\begin{abstract} 

High-statistics cross sections for the reactions $\gamma + p
\rightarrow K^+ + \Lambda$ and $\gamma + p \rightarrow K^+ + \Sigma^0$
have been measured using CLAS at Jefferson Lab for center-of-mass
energies $W$ between 1.6 and 2.53 GeV, and for
$-0.85<\cos\theta_{K^+}^{c.m.}< +0.95$.  In the $K^+\Lambda$ channel
we confirm a resonance-like structure near $W=1.9$ GeV at backward
kaon angles.  The position and width of this structure change with
angle, indicating that more than one resonance is likely playing a
role.  The $K^+\Lambda$ channel at forward angles and all energies is
well described by a $t$-channel scaling characteristic of Regge
exchange, while the same scaling applied to the $K^+\Sigma^0$ channel
is less successful.  Several existing theoretical models are compared
to the data, but none provide a good representation of the results.

\end{abstract}

\maketitle

\section{\label{intro}INTRODUCTION}

\begin{figure}
\centering
\resizebox{0.45\textwidth}{!}{\includegraphics{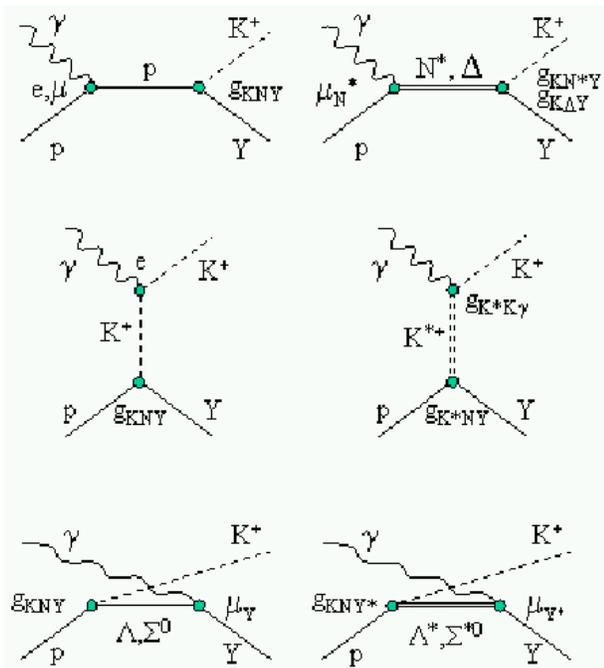}}
\caption{
Representative tree-level diagrams illustrating $s-$ (top), $t-$
(middle), and $u-$ channel (bottom) exchanges.  Born terms (left
column), baryon resonance excitations (top right), and other exchanges
(right middle, bottom) lead to production of $K^+Y$. Models differ in their 
electromagnetic transition moments ($\mu$'s),the strong couplings
(g$_{MBB}$'s), and form factors, as well as the effects of channel
couplings.
}
\label{fig:born}       
\end{figure}

We report on measurements of the photoproduction from the proton of
two ground state hyperons, namely the reactions $\gamma + p
\rightarrow K^+ + \Lambda$ and $\gamma + p \rightarrow K^+ +
\Sigma^0$.  Intermediate baryonic states in these reactions can be the
$N^*$ resonances in the case of $\Lambda$ production, and $N^*$ or $\Delta$
resonances in the case of $\Sigma^0$ production.  In either case one
expects strange meson exchange in the $t$ channel and hyperon exchange
in the $u$ channel. This is illustrated in Fig.~\ref{fig:born}.  To
unravel the production mechanism in these reactions, highly detailed
measurements of as many observables as possible are needed.

In this paper we present results for the differential cross sections,
$d\sigma/d\cos(\theta_{K^+}^{c.m.})$, obtained with the CLAS system in
Hall B at Jefferson Lab.  Following our previous publication,
Ref.~\cite{mcnabb}, these results are based on additional data
accumulated by CLAS and use a different analysis technique.  In
another forthcoming paper we will present results for the beam-recoil
double polarization observables, $C_x$ and $C_z$, for the same
reactions obtained from the same data set.

The main motivation for this work was to provide data to investigate
the spectrum of non-strange ($N^*$ and $\Delta$) baryon resonances above
the strangeness-production threshold at $W = \sqrt{s} = 1.6$~GeV.
Between this threshold and the upper limit of our data set, at
$W=2.53$~GeV, many baryon resonances are predicted by quark
models~\cite{cap}, but relatively few are clearly
established~\cite{pdg}. These resonances are broad and overlapping,
making partial wave analysis challenging, but it is also possible that
some dynamical aspect of hadronic structure may act to restrict the
quark models' spectrum of states to something closer to what has
already been established~\cite{klempt}.  This is the so-called
``missing resonance'' problem.  While the branching fractions of most
high-mass resonances to $KY$ final states are expected to be small
(cross sections $\sim 1~\mu b$) compared to three-body modes such as
$\pi\pi\ N$ ($\sim 100~\mu b$), the study of these decays do have
advantages.  First, two-body final states are often easier to analyze
than three-body final states.  Second, couplings of nucleon resonances
to $KY$ final states will differ from coupling to $\pi N$, $\eta N$,
or $\pi\pi N$ final states~\cite{cap}.  Thus, one can hope that this
alternate light cast on the baryon resonance spectrum may emphasize
resonances not otherwise revealed.  Some ``missing'' resonances may
only be ``hidden'' when sought in more well-studied reaction channels.

The $\Lambda$ and $\Sigma^0$ hyperons have isospin 0 and 1,
respectively, and so intermediate baryonic states leading to the
production of $\Lambda$'s can only have isospin 1/2 ($N^*$ only),
whereas for the $\Sigma^0$'s, intermediate states with both isospin 1/2
and 3/2 ($N^*$ or $\Delta$) can contribute. Thus, simultaneous study
of these reactions provides a kind of isospin selectivity of the sort
used in comparing $\eta$ and $\pi$ photoproduction reactions.  To
date, however, the PDG compilation~\cite{pdg} gives poorly-known
$K\Lambda$ couplings for only five well-established resonances, and no
$K\Sigma$ couplings for any resonances. The most widely-available
model calculation of the $K\Lambda$ photoproduction, the Kaon-MAID
code~\cite{maid}, includes a mere three well-established $N^*$ states:
the $S_{11}(1650)$, the $P_{11}(1710)$, and the $P_{13}(1720)$.  Thus,
it is timely and interesting to have additional good-quality
photoproduction data of these channels to see what additional
resonance formation and decay information can be obtained.

Section II of this paper discusses briefly the reaction models that
will be compared with the present data.  Section III discusses the
experimental setup of the CLAS system for this experiment.  The steps
taken to obtain the cross sections from the raw data are discussed in
Section IV.  Section V presents the results for the measured angular
distributions and $W$-dependence of the cross sections.  In Section VI
we discuss the results in light of previous measurements, and in
relation to several previously-published reaction models.  We also
show how the data can be parameterized in terms of $t$-channel scaling
using a simple Regge-based picture, and in terms of simple Legendre
polynomials.  In Section VII we recapitulate the main results.

\section{\label{sec:theory}THEORETICAL MODELS}

The results in this experiment will be compared to model calculations
that fall into two classes: tree-level Effective Lagrangian models and
Reggeized meson exchange models.  Effective Lagrangian models evaluate
tree-level Feynman diagrams as in Fig.~\ref{fig:born}, including
resonant and non-resonant exchanges of baryons and mesons. A complete
description of the physics processes will require taking into account
all possible channels which could couple to the one being measured,
but the advantages of the tree-level approach are to limit complexity
and to identify the dominant trends. In the one-channel tree-level
approach, some tens of parameters (in particular, the couplings of the
non-strange baryon resonances to the hyperon-kaon systems) must be
fixed by fitting to data, since they are poorly known from other
sources.  An alternative approach is to use no baryon resonance terms
and instead model the cross sections in a Reggeized meson exchange
picture.  While this is not expected to reproduce the results in
detail, it will show where the high-energy phenomenology of
$t$-channel dominance blends into the nucleon resonance region
picture.

For $K^+\Lambda$ production, the model of Mart and
Bennhold~\cite{mart} has four baryon resonance contributions.
Near threshold, the steep rise of the cross section is accounted for
with the $N^*$ states $S_{11}(1650)$, the $P_{11}(1710)$, and
$P_{13}(1720)$.  To explain the broad cross-section bump in the mass
range above these resonances, they introduced the $D_{13}(1895)$
resonance that was predicted in the relativized quark models of
Capstick and Roberts~\cite{cap} and L\"oring, Metsch, and
Petry~\cite{loring} to have especially strong coupling to the
$K^+\Lambda$ channel.  In addition, the higher mass region has
contributions, in this model, from the exchange of vector $K^*(892)$
and pseudovector $K_1(1270)$ mesons.  The hadronic form factors,
cutoff masses, and the prescription for enforcing gauge invariance
were elements of the model for which specific choices were made.  The
content of this model is embedded in the Kaon-MAID code~\cite{maid}
which was used for the comparisons in this paper.  This model was
tuned to results from the experiment at Bonn/{\small
SAPHIR}~\cite{bonn1}, and offers a fair description of those results.

On the other hand, analysis by Saghai {\it et al.}~\cite{sag} using
the same data set showed that, by tuning the background processes
involved, the need for the extra resonance was removed. Janssen {\it
et al.}~\cite{jan,jan_a} showed that the same data set was not
complete enough to make firm statements since models with and without
the presence of a hypothesized $N^*(1895) D_{13}$ resulted in equally good
fits to data. A subsequent analysis ~\cite{ireland}, which also fitted
calculations to photon beam asymmetry measurements from
SPring-8~\cite{zegers} and electroproduction data measured at
Jefferson Lab~\cite{mohring}, indicated weak evidence for one or more
of $S_{11}$, $P_{11}$, $P_{13}$, or $D_{13} (1895)$, with the $P_{11}$
solution giving the best fit. The conclusion was that a more
comprehensive data set would be required to make further progress.

More elaborate model calculations have been undertaken in which
channel coupling is considered, in addition to the tree-level
approaches mentioned above.  Penner and Mosel~\cite{penner} found fair
agreement for the $K^+\Lambda$ data without invoking a new $D_{13}$
structure.  Chiang {\it et al.}~\cite{chiang} showed that coupled
channel effects are significant at the $20\%$ level in the total cross
sections when including pionic final states.  Shklyar, Lenske, and
Mosel~\cite{shklyar} used a unitary coupled-channel effective
Lagrangian model applied to $\pi$ and $\gamma$ -induced reactions to
find dominant resonant contributions from $S_{11}(1650)$,
$P_{13}(1720)$, and $P_{13}(1895)$ states, but not from $P_{11}(1710)$
or $D_{13}(1895)$. This conclusion was true despite the discrepancies
between previous data from CLAS~\cite{mcnabb} and
{\small SAPHIR}~\cite{bonn2}.  Recently, Sarantsev {\it et
al.}~\cite{sarantsev} did a phenomenological multi-channel fit for $K
\Lambda$, $K \Sigma$, as well as $\pi$ and $\eta$ photoproduction
data.  They found fairly strong evidence for a $P_{11}$ at 1840 MeV
and two $D_{13}$ states at 1870 and 2170 MeV.  Even better quality
$KY$ data such as we are presenting here are needed to solidify these
conclusions.  We will not compare the present results to those models in
this paper, however.

While it is to be expected that $s$-channel resonance structure is a
significant component of the $K^+\Lambda$ and $K^+\Sigma^0$ reaction
mechanisms, it is instructive to compare to a model that has no such
content at all.  The model of Guidal, Laget, and
Vanderhaeghen~\cite{lag1,lag2} is such a model, in which the exchanges
are restricted to two linear Regge trajectories corresponding to the
vector $K^*$ and the pseudovector $K_1$.  The model was fit to
higher-energy photoproduction data where there is little doubt of the
dominance of these exchanges.  In this paper, we extend that model
into the resonance region in order to make a critical comparison.

\section{\label{sec:setup}EXPERIMENTAL SETUP}

Differential cross section data were obtained with the CLAS system in
Hall B at the Thomas Jefferson National Accelerator Facility.
Electron beam energies of 2.4 and 3.1 GeV contributed to the data set,
each of typically 10 nA current.  Real photons were produced via
bremsstrahlung from a $1\times10^{-4}$ radiation length gold radiator
and ``tagged'' using the recoiling electrons analyzed in a dipole
magnet and scintillator hodoscopes~\cite{tagger}.  The energy tagging
range was from 20\% to 95\% of the beam endpoint energy, and the
integrated rate of tagged photons was typically $5\times10^6$~/sec.
Using the tagger and the accelerator RF signal, photon timing at the
physics target was defined with an {\it rms} precision of 180 psec.
The useful energy range for this experiment was from the
strangeness-production threshold at $E_\gamma=0.911$ GeV ($W$ = 1.61
GeV) up to 2.95 GeV ($W$ = 2.53 GeV). In this range, the tagger
resolution was typically 5 MeV, set by the size of the hodoscope
elements, but the data were analyzed in bins of 25 MeV photon energy
to be commensurate with any energy-dependent structure expected in the
hadronic cross sections.  The centroids of these bins were adjusted in
the analysis by between $-6$ and $+5$ MeV to compensate for mechanical
sag of the hodoscope array measured by kinematically fitting
$p(\gamma,p\pi^+\pi^-)$ data; hence our final results are given in
unequal energy steps.

The physics target consisted of a 17.9 cm long liquid hydrogen cell of
diameter 4.0 cm.  Temperature and pressure were monitored continuously
to determine the density to 0.3\% precision.  The target cell was
surrounded by a set of six 3 mm thick scintillators to help define the
starting time for particle tracks leaving the target, though actually
the timing given by the photon tagger was used to define the event
times.

The CLAS system, described in detail elsewhere~\cite{clas0}, consisted
of a toroidal magnetic field, with drift chamber tracking of charged
particles.  The overall geometry was six-fold symmetric viewed along
the beam line.  Particles could be tracked from $8^\circ$ to
$140^\circ$ in laboratory polar angle, and over about $80\%$ of $2\pi$ in
the azimuthal direction.  Outside the magnetic field region a set of
288 scintillators was used for triggering and for later particle
identification using the time-of-flight technique. The momentum
resolution of the system was $\approx 0.5\%$, with variations due to
multiple scattering and tracking resolution considerations.  The
low-momentum cut-off was set in the analysis at 200 MeV/c.  Other
components of the CLAS system, such as the electromagnetic calorimeter
and the Cerenkov counters, were not used for these measurements.

The event trigger required an electron signal from the photon tagger,
and at least one charged-track coincidence between the time-of-flight
`Start' counters near the target and the time-of-flight `Stop'
counters surrounding the drift chambers.  The photon tagger signal
consisted of the OR of coincidences among hits in a two-plane
hodoscope, which had 61 timing scintillators in coincidence with their
matching energy-defining scintillators.  The charged-track trigger in
CLAS was a coincidence of six OR'd start counter elements and the OR
of the outer time-of-flight scintillators.  Events were accumulated at
the rate of $\sim 2500$ hadronic events per second, though only a
sub-percent fraction of these events contained the kaons and hyperons
of interest for the present analysis.

\section{\label{sec:dataanal}DATA ANALYSIS}

\subsection{\label{sec:dselect}Data and event selection}

The data used in this experiment were obtained in late 1999 as part of
the CLAS ``g1c'' data taking period.  Since the electronic trigger was
loose, data for several photoproduction studies were contained in the
data set.  Off-line calibration was performed to align the timing
spectra of the elements of the photon tagger, the six elements of
the start counter, and the 288 elements of the time-of-flight (TOF) 
counters.  Drift-time calibrations were made for the 18 drift chamber
packages.  Pulse height calibrations and timing-walk corrections were
made for the time-of-flight counters.  The raw data were then processed
to reconstruct tracks in the drift chambers and to associate them with
hits in the time-of-flight counters.

\subsection{\label{sec:pid}Particle Identification}

Kaon, proton, and pion tracks were separated using momentum and
time-of-flight measurements.  The momentum, $\vec{p}$, of each track
was measured directly via track reconstruction through the CLAS
magnetic field; this measurement also gave the path length, $d$, from
the reaction vertex to the time-of-flight counter hit by the track.
The starting time of the track was determined by projecting the tagger
signal time, synchronized with the accelerator RF timing, to the
reaction vertex inside the hydrogen target.  The stopping time was
determined by the element hit in the array of TOF scintillators.  The
difference, $T$, between these two times was the measured time of
flight, which in CLAS could range between about 4 and 100 nsec.  From
$T$ the speed, $\beta$, could be obtained as $\beta=d/(c T)$.  The
mass, $m_x$, was then computed according to $m_x=
\sqrt{1-\beta^2}\times pT/d$.  In CLAS, the dominant mass uncertainty
in this situation came from the time-of-flight resolution, $\delta
T$. $\delta T$ was independent of particle momentum, so particle
selection based on time of flight was largely independent of momentum
as well.  For kaon identification we used the time-of-flight
difference technique, where the measured time, $T$, of the track was
compared to the expected time, $T_h$, for a hadron of mass $m_h$ and
momentum $p$.  For a hypothesized value of $m_h$ we can define $\Delta
tof = T - T_h$ and write

\begin{equation}
\Delta tof  = T \left(1- \sqrt{ \frac{(m_hc)^2 + p^2}{(m_xc)^2 + p^2}}\right).
\end{equation}

\begin{figure}
\resizebox{0.48\textwidth}{!}{\includegraphics{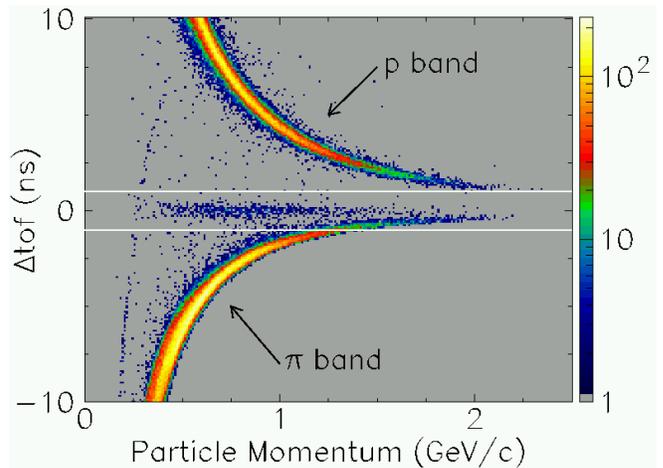}}
\caption{(Color online) 
Time-of-flight difference spectrum for a sample of tracks, assuming
the mass of the particle is that of a kaon.  White lines indicate the
cut limits for selecting kaons in a time window of $\pm 1.0$ nsec.  
Note the logarithmic scale on the intensity axis.
}
\label{fig:tof1}       
\end{figure}

Figure~\ref{fig:tof1} shows an example of such a time difference
spectrum when we took $m_h$ to be the kaon mass.  The candidate kaon
tracks were selected using a $\pm1$ nsec cut centered at zero.  Pion
and proton bands are well separated from the kaons up to 1 and 2 GeV,
respectively. A crossing band due to a badly-calibrated detector
element is shown for illustration; such tracks were later rejected by
removing the detector element and/or by the kinematic cuts and fits
applied later.  Above 1 GeV some pions leak into the set of candidate
kaon tracks.  These were rejected by subsequent event reconstruction
cuts and by background rejection fitting.  Protons were identified
using a similar $\Delta tof$ correlation but with looser cuts due to
the straggling effects which broadened the distribution.  

\begin{figure}
\resizebox{0.50\textwidth}{!}{\includegraphics{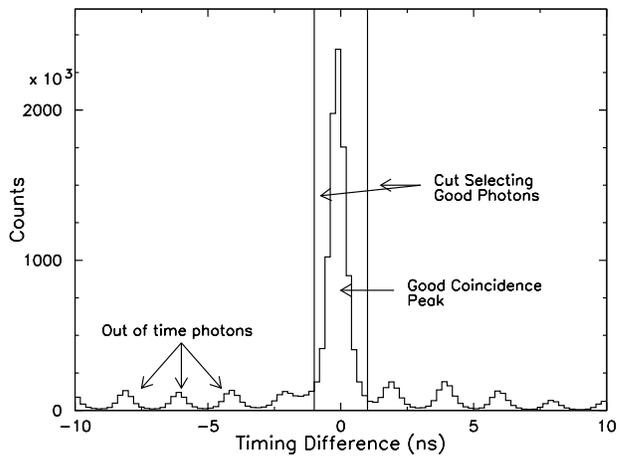}}
\caption{
Time difference between photon tagger time and target start counter
time showing the peak at zero of good matches between the photons and
the hadrons at the event vertex in CLAS.  Coincidences due to hadrons
mismatched to random photons in the tagger show the 2 nsec bunch
structure of CEBAF.
}
\label{fig:tof2}       
\end{figure}

Photons matching the hadronic tracks in CLAS were selected using the
time difference between the hadronic track projected back to the event
vertex and the photon tagger time projected forward to the event
vertex.  Figure~\ref{fig:tof2} shows such a spectrum, which
illustrates the presence of random coincidences between the photons
and the hadronic tracks.  The 2~nsec RF time structure of the
accelerator is clearly seen.  A $\pm 1.0$~nsec cut was used to reject
out-of-time combinations.  In-time accidentals under the central peak
were treated as potentially-correct photons, and such particle-photon
combinations were retained in the analysis.  Since ambiguous photons
were generally widely separated in energy, the $(\gamma,K^+)$ missing
mass for incorrect combinations fell into the broad background under
the hyperons, and were then rejected at the peak-fitting stage of the
analysis discussed below.

In this analysis we demanded detection of positive kaons and protons.
Negative pions from $\Lambda$ decay or photons from $\Sigma^0$ decay
were not required.  Fiducial cuts were applied in track angle and
momentum to restrict events to the well-described portions of the
detector.  This included removal of 9 out of 288 time-of-flight
elements due to poor timing properties. Corrections were applied for
the mean energy losses of kaons and protons as they passed through the
production target, target walls, beam pipe, and air.  The nominal CLAS
momentum reconstruction algorithms were found to provide sufficient
hyperon mass resolution (see below) that no higher-order momentum
corrections were applied.

A missing mass cut was applied to $p(\gamma,K^+p)\pi^-(\gamma)$ to select
events consistent with a missing pion and (for the $\Sigma^0$) a missing
photon.  The losses incurred by this cut due to multiple scattering
effects on the part of the kaons and protons were studied in the real data
and in Monte Carlo.  The estimated residual uncertainty due to the cut
and its compensation via the acceptance calculation was $1\% - 2\%$.  

\subsection{\label{sec:yields}Yield of hyperons}

\begin{figure}
\resizebox{0.50\textwidth}{!}{\includegraphics{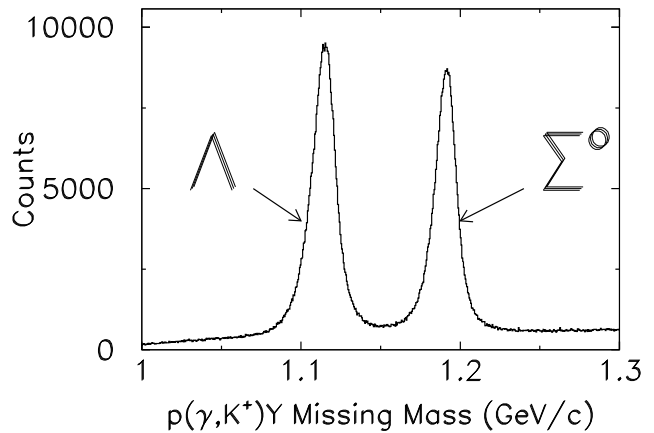}}
\caption{
Hyperon spectrum via missing mass using the photon and detected kaon,
integrated over all kaon angles and photon energies using a 3.1 GeV
endpoint energy. 
 }
\label{fig:mm}       
\end{figure}

The extraction of kaon yields in each bin of photon energy and kaon
angle depended on fits to the missing mass spectrum given by
$p(\gamma,K^+)Y$.  When integrating over all of our 3.1 GeV data,
for all energies and angles, the resulting missing mass spectrum is
shown in Fig.~\ref{fig:mm}.  This figure illustrates that the overall
missing mass resolution of the system was $\sigma=8.9$ MeV for the
$\Lambda$ and $\sigma=8.2$ MeV for the $\Sigma^0$.  The overall
resolution averaged 6.3 MeV in the 2.4 GeV data set, where all the
average momenta were lower.  However, the width of the peaks and the
extent of the background to be removed from under the peaks via
fitting varied substantially across the measured range of energy and
angle, so a careful fitting procedure was needed to obtain
well-controlled hyperon yields.

The main source of background in the hyperon mass spectra was due to
events where the kaons were actually mis-identified pions.  The yields
of $\Lambda$ and $\Sigma^0$ hyperons were obtained using lineshape
fits to missing-mass spectra in each of over 1,450 kinematic bins of
photon energy and kaon angle.  The data were binned in 25 MeV steps in
$E_\gamma$ and in 18 bins of kaon center-of-mass (c.m.) angle,
$\cos(\theta_{K^+}^{c.m.})$, centered in steps of 0.1 between $-0.8$
and $+0.9$.

\begin{figure*}
\resizebox{0.30\textwidth}{!}{\includegraphics{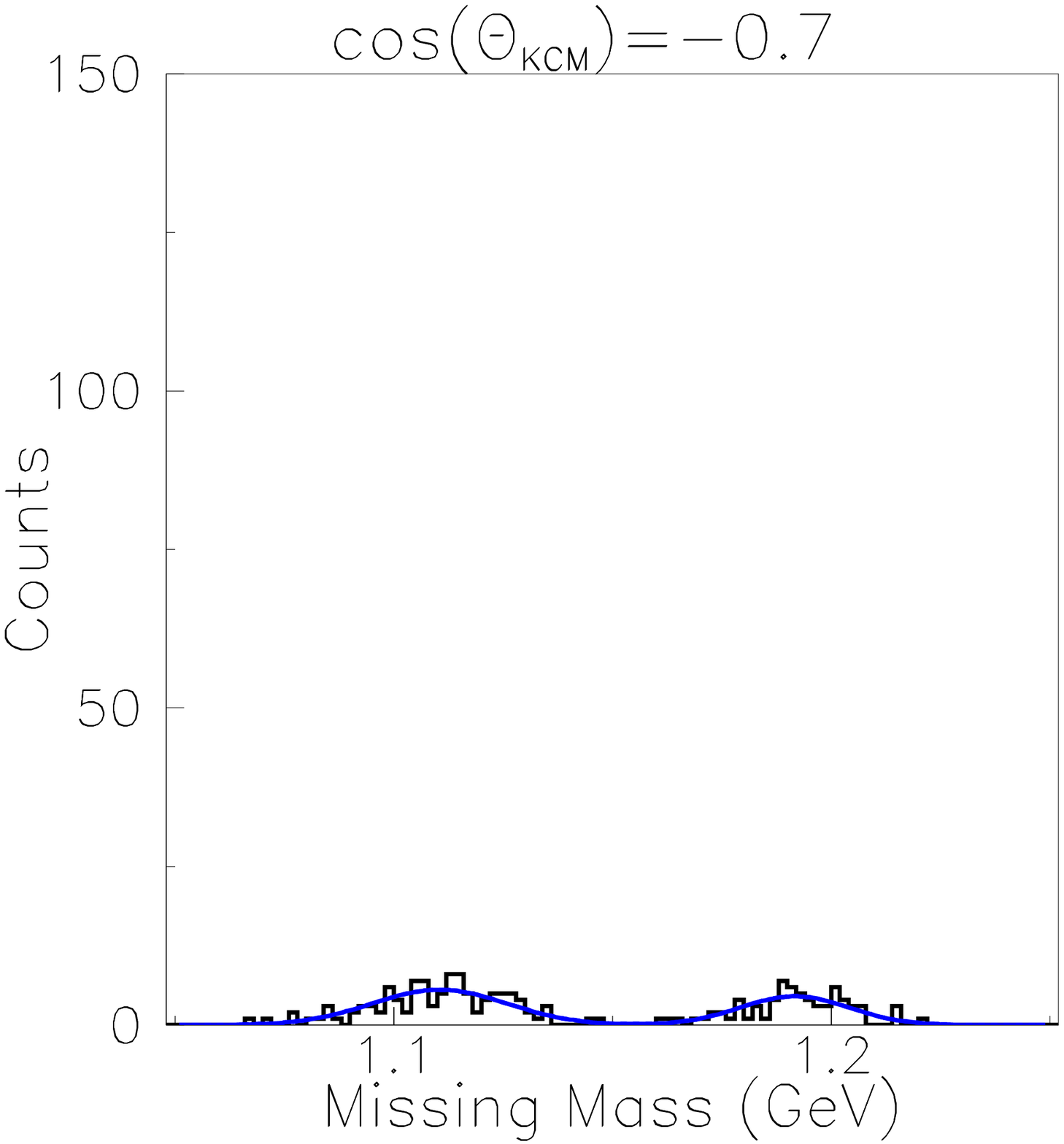}}
\resizebox{0.30\textwidth}{!}{\includegraphics{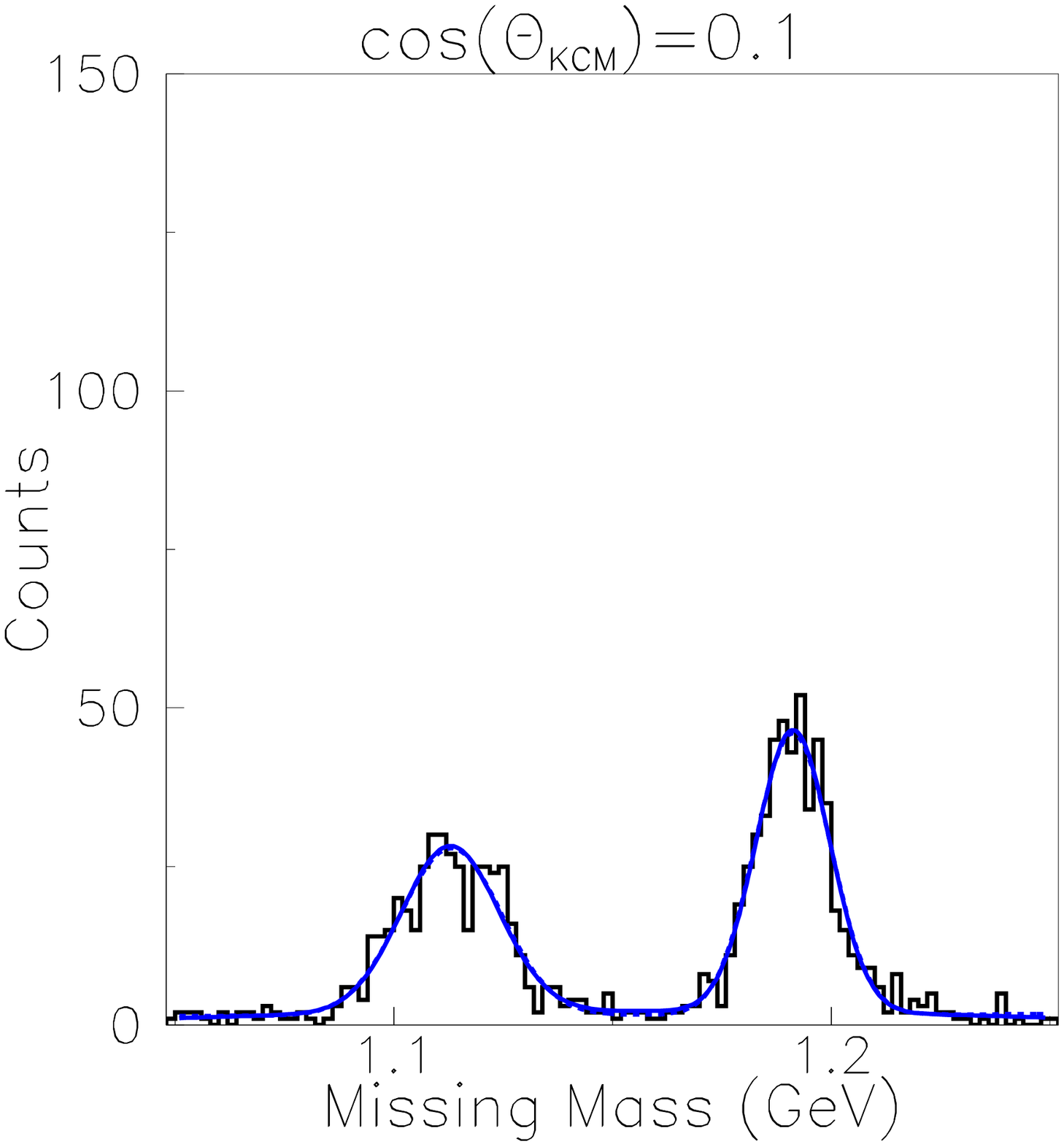}}
\resizebox{0.30\textwidth}{!}{\includegraphics{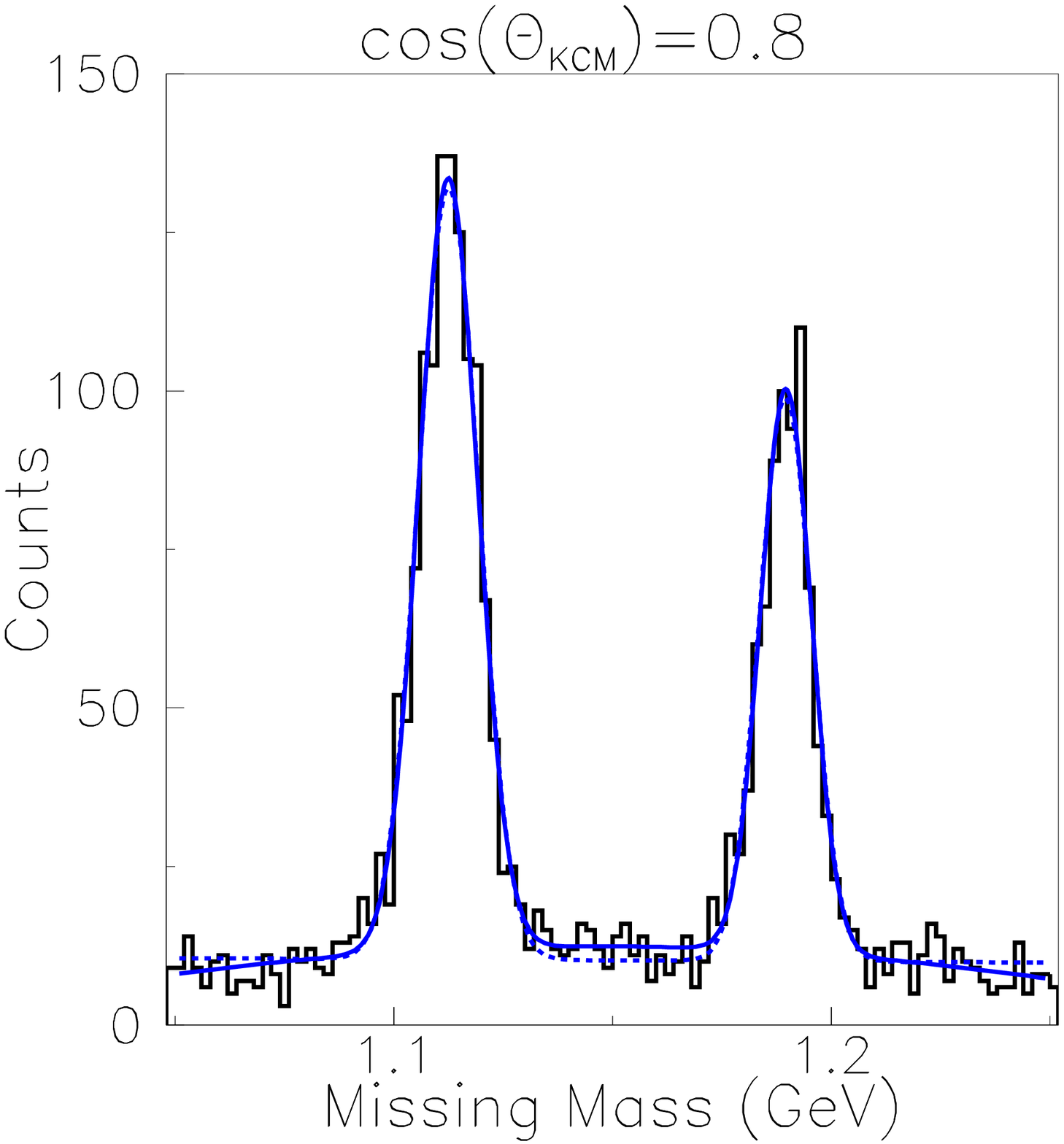}}
\caption{
Sample missing mass fits used for the determination of hyperon yields
at $E_\gamma = 1.825$ GeV and for three representative $K^+$ angles:
$\cos(\theta_{K^+}^{c.m.}) = -0.7, +0.1, +0.8$.  The fits to 
different orders of polynomial background are nearly indistinguishable: 
solid lines for quadratic and dashed lines for linear.  
}
\label{fig:mmfit}       
\end{figure*}

Typical hyperon yield fits of $p(\gamma,K^+)Y$ for the middle of the
photon energy range are shown in Fig.~\ref{fig:mmfit}.  The fits were
performed in two passes.  In the first pass, events for all kaon
center-of-mass angles were summed together.  These first fits served
to determine and fix the centroids and widths of the Gaussian peaks
for the two hyperons.  These were 7 to 9 parameter fits, depending on
the background model employed.  A log-likelihood fitting algorithm was
used.  The background was modeled as polynomials of order up to 2
(quadratic).  In the second pass, fits were made with 3 to 5
parameters for the yields in each kaon angle bin, allowing only the
integrated counts of the peaks to vary in addition to the background
parameters.  The two-pass method was used to stabilize fits of
low-yield bins at low photon energy and backward kaon
angle. Acceptable fits all had $\chi^2$ per degree of freedom of less
than 2.0.

Background parameterizations that were a simple constant or a sloped
line were sufficient to yield good fits over most of the kinematic
range.  At more forward kaon angles the effect of background due to
mis-identified pions increased and the quadratic fits generally gave the
best results.  Above $E_\gamma=2.3$ GeV the momentum resolution of
CLAS broadened to the degree that the forward-angle quadratic fits
became less stable, so the linear fits were preferred.  This led to an
extra estimated systematic uncertainty of 10\% on both the forward-angle
differential and total cross sections above this energy.  In some
low-yield back-angle bins, where no good fits were obtained, side-band
subtraction was used to determine the yield.

The final cross sections were based on the following numbers of fully
reconstructed events: from the 2.4 and the 3.1 GeV data sets we had
236,260 and 325,792 $K^+\Lambda$ events, respectively, and 169,796 and
269,216 $K^+\Sigma^0$ events, respectively.

\subsection{\label{sec:acceptance}Acceptance calculation}

The acceptance and efficiency were modeled using a CLAS-standard
GEANT-based simulation code (``GSIM'').  Events were initially modeled
using a phase space distribution for $\gamma+p\rightarrow K^+ + Y$.
The GSIM code simulated the events in the CLAS detector at the level
of ADC and TDC hits in the scintillators and drift time information in
the tracking chambers.  The events were further fine-tuned such that
the time distributions in the TDC's accurately matched the actual data
using another well-tested CLAS software package (``GPP'').  These
simulated events were then processed through the same analysis codes
as the real data, and thus the acceptance was computed in each
kinematic bin.  Dead regions of the drift chambers were removed both
from the real data and from the simulated data during track
reconstruction (``A1C'').  Detector efficiency was simultaneously
accounted for through the simulation: sources of inefficiency included
track reconstruction failures and time-of-flight paddle removals.  The
only particle background in this physics Monte Carlo was due to
particle decays, especially the kaons, and multiple scattering
effects.  Thus, we relied on the yield extractions discussed earlier
to remove background due to mis-identified pions or protons.

The effect of using a phase-space event generator to compute the
acceptance, $\eta_{P.S.}$, was studied by using the fits to the
angular distributions presented in Section~\ref{sec:angle} to
regenerate the acceptance, $\eta_{Data}$, with an improved
representation of the reactions.  Since these cross sections vary
quite slowly with angle, and since the kinematic bins were each small
on the scale of these variations, no large effects were to be
expected.  We found agreement between the two acceptance models at the
level of $0.25\%$ $rms$ over essentially the whole of the kinematic
space, consistent with the statistical variations of the simulations.
The exception was in the forward-most angle bin ($0.85 <
\cos(\theta_{K^+}^{c.m.}) <0.95$) for both hyperons.  There, because
of the extrapolation of the analysis into CLAS's forward acceptance
hole, the ratio of acceptances $\eta_{Data}/\eta_{P.S.}$ dropped from
1.0 to 0.85 over the range $E_\gamma \sim 1.75$ GeV to $E_\gamma \sim
2.90$ GeV.  Theoretical models of the behavior of the cross section in
the very forward direction differ strongly, as shown later, so it was
not known whether a ``flat'' or a ``forward-peaked'' or a
``forward-dipped'' acceptance model was more accurate.  Thus, the
forward-most angle results at $\cos(\theta_{K^+}^{c.m.}) = +.9$ have
an additional systematic uncertainty on the cross section which is, on
the average, $\pm8\%$.

\begin{figure}
\resizebox{0.45\textwidth}{!}{\includegraphics{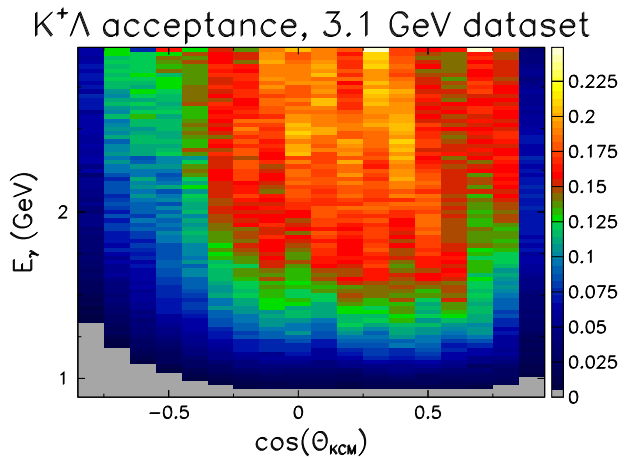}}
\caption{(Color online) 
Computed CLAS acceptance for the $E_\gamma=3.1$ GeV data
set for the $\gamma + p \rightarrow K^+ + \Lambda$ reaction. The scale
on the right gives the value of the acceptance for each kinematic bin.
}
\label{fig:accept}       
\end{figure}

A sample of the acceptances computed for CLAS for these reactions is
shown in Fig.~\ref{fig:accept}.  It was largest at mid to forward kaon
angles and at higher photon energies.  The maximum acceptance was
about $22\%$ for $K^+\Lambda$ and $23\%$ for $K^+\Sigma^0$.  A lower
cut-off was applied, such that the smallest allowed acceptance in the
experiment was $0.5\%$.  For each hyperon, 10 million events were
generated at each beam endpoint energy.  Non-uniformities in the
distribution arise from the effects of detector element removals and
track reconstruction efficiencies.  Since the kinematics of the two
hyperon reactions are very similar, the acceptance function for the
$\Sigma^0$ looked very similar, apart from the higher production
threshold.

\subsection{\label{sec:photons}Photon flux}

The number of photons striking the target was computed from the
measured rate of electrons detected at the tagger hodoscope.  TDC
spectra of the tagger elements recorded the hits of electrons in a 150
nsec time range around each event.  This flux was scaled and
integrated in ten-second intervals.  After statistical corrections for
multiple hits and electronic live time, the flux was summed over whole
runs.  The fine granularity of the tagging system was grouped into
bins of 25 MeV in photon energy.  

Photon losses in the beam line due to tagger acceptance, beam
collimation, and thin windows were determined using a separate
total-absorption counter downstream of CLAS.  This low-rate lead glass
detector was periodically put in the beam line to monitor the tagging
efficiency.  For the 2.4 GeV data set the average efficiency for
tagging photons was 78\%, and the stability of this efficiency, which
was measured periodically throughout the data taking period, was
$\pm0.5\%$.

By taking data at 2.4 and 3.1 GeV endpoint energies it was possible to
test the flux normalization of many elements of the tagging system, as
discussed in the next section.  At energies above $E_\gamma = 2.325$
GeV the two data sets no longer overlapped, however, and defective
electronics in a few channels of the tagger led to a gap in our final
spectra.  Bins at $E_\gamma = $ 2.375 and 2.400 GeV were removed
because of this.

\subsection{\label{sec:systematics}Systematic uncertainties}

The 2.4 and 3.1 GeV photon beam endpoint data sets were compared to
investigate variations in the photon tagger efficiency.  The
photon-normalized yield of particles at any given energy had to be
independent of bremsstrahlung endpoint energy, so consistency of this
quantity tested stability of the electronics.  Localized regions of
tagger inefficiency ``moved'' in photon energy when the endpoint
energy changed.  We took the higher normalized yield between the two
data sets as the correct one. Localized regions of high inefficiency
were found in the 3.1 GeV data set at 1.1, 1.4, and 1.8 GeV; in those
regions we made corrections of up to 50\% in one data set to
compensate for tagger efficiency losses in the other.  Much smaller
corrections ($\sim 3\%$) were made at other energies.  The absolute
uncertainty on these corrections was estimated to be $\pm3\%$.

\begin{figure}
\resizebox{0.50\textwidth}{!}{\includegraphics{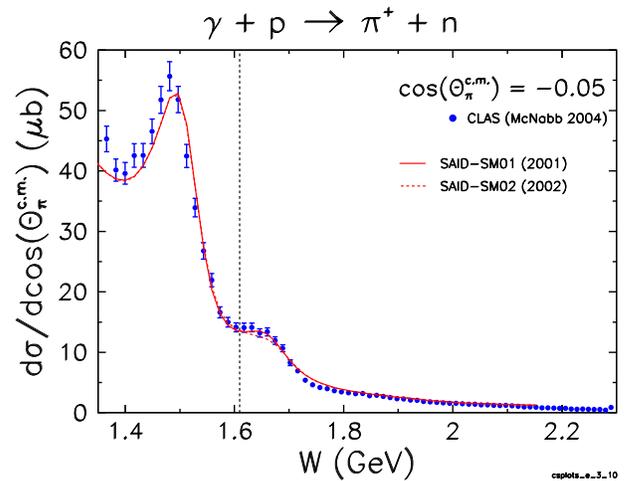}}
\caption{ 
CLAS differential cross section as a function of $W$ for
$\gamma + p \rightarrow \pi^+ + n$. Shown for comparison are two
versions of the SAID parameterization of world pion data.  They are
essentially indistinguishable.  The vertical dotted line is the
strangeness production threshold.  }
\label{fig:pion}       
\end{figure}

As a check on our results, the $p(\gamma,\pi^+)n$ cross section was
measured using the same analysis chain, as far as possible, as the
$p(\gamma,K^+)Y$ data.  The same procedure was also used to generate
the acceptance for the $p(\gamma,\pi^+)n$ cross sections used to check
the whole analysis process, except that the SAID code was used to
generate the initial events.  Figure~\ref{fig:pion} shows the pion
cross section measured in this analysis as a function of $W$ for a
mid-range c.m. pion angle.  The CLAS pion cross section was found to
be in fair agreement with the SAID~\cite{said} parameterization of the
world's data between $W=1.6$ and $2.1$ GeV, albeit lower by an average
scale factor of 0.95.  As a function of pion center of mass angle, the
CLAS to SAID ratio was $\sim 1.0$ at back angles and $\sim .92$ at
forward angles. Thus the pion results indicate a possible systematic
error in the acceptance calculation at the level of $\pm3\%$, apart
from the average scale factor.  The absolute accuracy of the pion
cross sections, as parameterized by SAID over the range of comparison
we used, is similar to this.  Therefore, we chose not to make a
renormalization of our results to the average pion cross sections.
The results presented in this paper are on an absolute scale.  The
kaon analysis was not identical to the pion analysis, since the kaon
decay corrections are much larger in the former case, and since the
final kaon analysis included detection of the proton from the hyperon
decays.  Hence, it was difficult to translate the systematic trends in
our pion results compared to SAID to the kaon results presented here.
However, based on the comparison to the pion data analysis, we
estimated the overall systematic uncertainty in our kaon cross section
to be less than $\pm7\%$.  This was the largest single contribution to
our cross section systematic uncertainty.

The analysis of this experiment was done twice, in largely independent ways.
The first analysis~\cite{john,mcnabb} computed cross sections based on
detection of the kaon alone, or $p(\gamma,K^+)Y$.  Starting from the
same data set, the second analysis~\cite{bradford} detected the kaon
and the proton for each event, or $p(\gamma,K^+ p)\pi^-(\gamma)$,
where the $\pi^-$ (and the possible $\gamma$ from $\Sigma^0$ decay)
were ignored.  Particle identification and acceptances were developed
independently.  For the results presented here, the first analysis was
revised to take into account more advanced modeling of the CLAS
detector in the acceptance; both analyses used the standard CLAS GEANT
package for computing acceptances.  The same flux normalization
procedures were used.  The first analysis used only data from the 2.4
GeV endpoint data set, while the second analysis also included data
from 3.1 GeV endpoint.  Consistency checks were then made between the
two analyses. Results for the final cross sections from the two
studies were in very good agreement across the full range of energies
and angles where they overlapped.  Isolated differences of $\sim5\%$
in small ranges of angle were attributed to details of the acceptance
modeling.  By comparing the acceptances developed over the course of
the $p(\gamma,K^+)Y$ studies, we estimated that average systematic
uncertainty across the kinematics of the experiment was $\pm2.7\%$,
arising from variations in the implementation of the detector model
and the track-reconstruction algorithms.

On the energy axis, our results are precise to $\pm2$~MeV.  This
systematic uncertainty arises from an energy bin centering correction
that was applied to each data point due to the calibration of the
photon tagger.  In an independent study, kinematic fitting to the
reaction $p(\gamma,p\pi^+\pi^-)$ showed that the CLAS tagger and the
photon beam were mismatched by up to $\pm10$ MeV due to mechanical
effects in the structure of the tagger.  The correction 
shifted the centroids of each energy bin by an amount
estimated to be precise as stated above.  The indirect effect
that this centroid shift had on the acceptance of CLAS was considered
negligible, since the cross sections vary slowly in energy and the energy bins 
for the results are 25 MeV wide.

The estimated systematic uncertainties discussed above were combined
with contributions due to particle yield extraction $(3.6\%)$, photon
attenuation in the beam line $(0.2\%)$, target density uncertainty
$(0.14\%)$, and target length uncertainty $(0.28\%)$.  This led to an
estimate of the global scale uncertainty of $\pm8\%$.  Due to
additional systematic uncertainty about extrapolation of the data to
zero degrees, the forward-most angle bin above $E_\gamma = 1.75$ GeV
has an overall uncertainty of $\pm 11\%$.

\section{\label{sec:results}RESULTS}

\subsection{\label{sec:angle}Angular Distributions}

Since the differential cross sections in this measurement are
symmetric in the azimuthal angle $\phi$, we present the results in the
partially integrated form
\begin{equation}
\frac{d\sigma}{d(\cos \theta_{K^+}^{c.m.})} = 2\pi \frac {d\sigma}{d\Omega}
\end{equation}
since this also puts the values on a convenient scale of order 
$1~\mu b$.  

\begin{figure*}
\resizebox{0.90\textwidth}{!}{\includegraphics{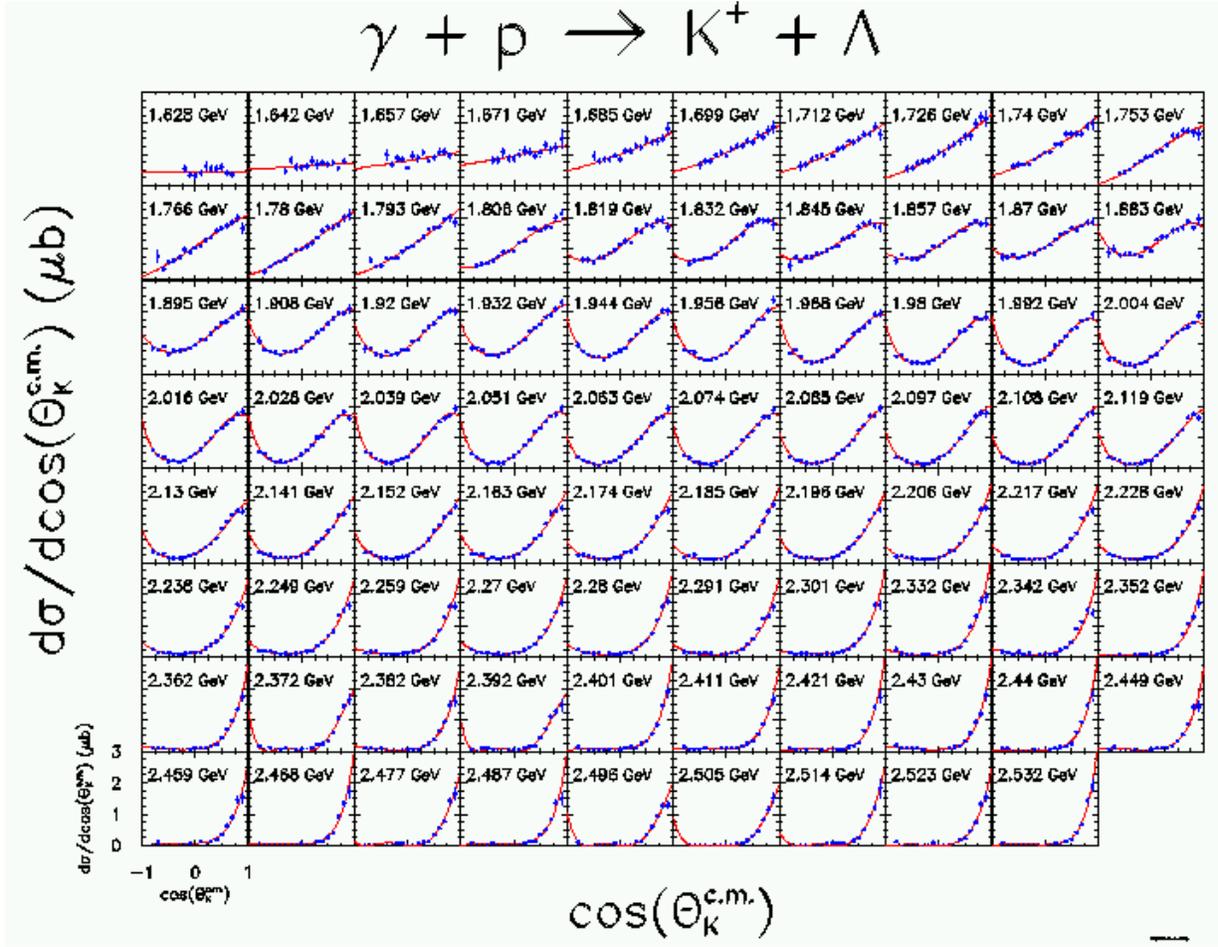}}
\vspace{0.15in}
\caption{
Differential cross sections for $\gamma + p \rightarrow K^+ +
\Lambda$. The number in each panel designates $W$ ($=\sqrt{s}$).  
The solid lines are results of the amplitude fits (Eq.~\ref{eq:amp}) 
discussed in the text.
}
\label{fig:lang}       
\end{figure*}

The angular distribution results for the reaction $\gamma + p
\rightarrow K^+ + \Lambda$ are shown in Fig.~\ref{fig:lang}.  The
results are shown as a function of $\cos(\theta_{K^+}^{c.m.})$ for 79
bins in $W$.  The step sizes in $W$ were determined by the 25-MeV step
size in the nominal photon energy, $E_\gamma$, at which the cross
sections were extracted, together with a few-MeV correction for tagger
re-calibration.  There are 18 bins in $\cos(\theta_{K^+}^{c.m.})$,
each of width 0.1, centered from $-0.80$ to $+0.90$.  The cross
sections are the averages within each angle bin, with no bin
centering.  The results are the weighted means of the 2.4 and 3.1 GeV
beam energy data sets.  The error bars are dominated by the
statistical uncertainties of the hyperon yield extraction fits, but
also include the statistics from the Monte Carlo acceptances.  The
overall systematic uncertainty, as discussed previously, is $\pm 8\%$,
except in the forward-most bin where above $E_\gamma = 1.75$ GeV it is
$\pm 11\%$.  There are 1,377 data points in the $K^+\Lambda$ set.

The curves in Fig.~\ref{fig:lang} arise from fits intended to capture
the main features of the decay amplitudes contributing to the angular
distributions.  The form is
\begin{equation}
\frac{d\sigma}{d\cos(\theta_{K^+}^{c.m.})} = \frac{q}{k} 
\left\{
\sum_{i=0}^{4}{a_i P_i (\cos\theta_{K^+}^{c.m.}) }
\right\}^2
\label{eq:amp}
\end{equation}
where the $P_i$ are the Legendre polynomials of order $i$, and the
$a_i$ are the fit coefficients which represent the $L = 0,1,2,3,4
\equiv S, P, D, F, G $ -wave amplitudes for the decay distributions.
The factor $q/k$ is the phase space ratio of the reaction, where $k$
and $q$ are the center-of-mass frame momenta of the initial and final
states, respectively. The value of this ratio ranges from zero at
threshold up to .86 at our highest energy.

Qualitatively, the cross section is flat as a function of
$\cos(\theta_{K^+}^{c.m.})$ near threshold, as would be expected for
$S$-wave behavior.  As the energy rises to about 1.8 GeV the cross
section develops a significant forward peaking consistent either with
$t$-channel contributions or with $s$-channel interference effects
between even and odd waves.  As the energy rises further the
cross section develops a tendency toward a slower rise in the extreme
forward direction and also a rise in the backward direction.  Above
about 2.3 GeV the cross section is dominantly forward peaked,
consistent with $t$-channel exchange dominance, though on a
logarithmic scale (see discussion in Sec.~\ref{sec:scaling}) the
fall-off is not exponential all the way to back angles.

\begin{figure*}
\vspace{-0.50in}
\resizebox{0.80\textwidth}{!}{\includegraphics{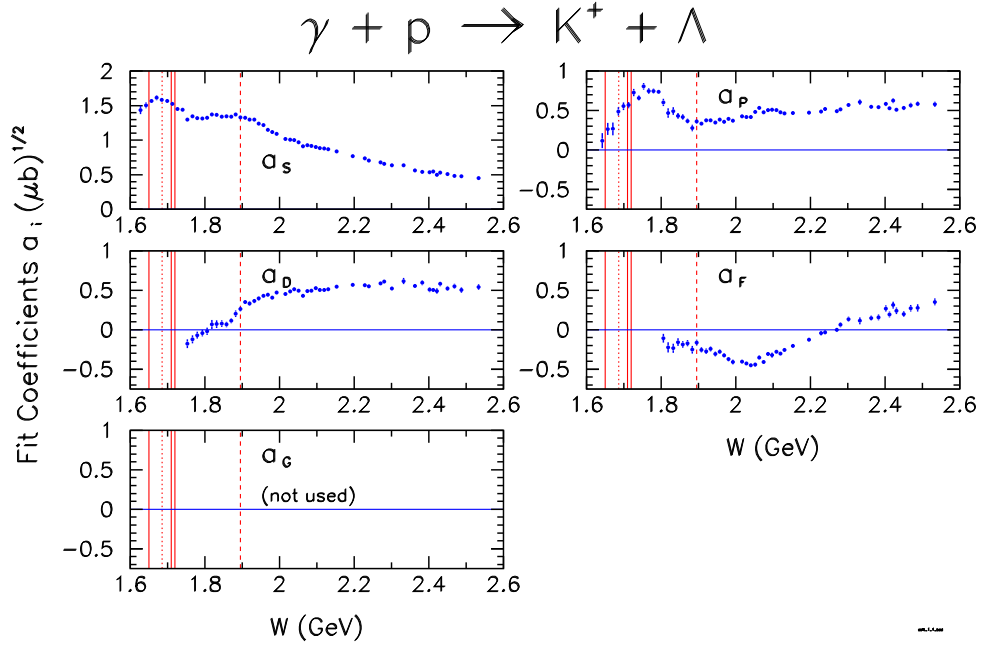}}
\vspace{-0.25in}
\caption{(Color online)
Amplitude fit to the differential cross sections for $\gamma + p
\rightarrow K^+ + \Lambda$.  The coefficients are defined in
Eq.~\ref{eq:amp}.  The solid vertical lines mark the well-known $N^*$
resonances $S_{11}(1650)$, $P_{11}(1710)$, and $P_{13}(1720)$.  The
dotted line marks the $\Sigma^0$ threshold, and the dashed line marks
the $D_{13}(1895)$ position.  }
\label{fig:lamp}       
\end{figure*}

The parameters of the fit may be used to gain some insight into the
reaction mechanism, unraveling effects due to interference among partial
waves.  Figure~\ref{fig:lamp} shows the coefficients from the fit using
Eq.~\ref{eq:amp}. The $a_i$ were taken to be purely real numbers.
The range over which each parameter is plotted depended upon its
significance, as estimated by the statistical F-test.  Mostly, the
higher partial waves are not significant near threshold, but our
angular coverage is also less complete near threshold, disallowing
higher-order fits.  One may note a prominent bump in the $P$-wave
amplitude between threshold and 1.9 GeV, centered near 1.7 GeV.  The
$D$-wave amplitude turns on quite strongly near 1.9 GeV, and the
$F$-wave amplitude has a broad dip centered at 2.05 GeV.  In the
$K^+\Lambda$ case, the $G$-wave was not significant at any energy.

\begin{figure*}
\vspace{-0.50in}
\resizebox{0.80\textwidth}{!}{\includegraphics{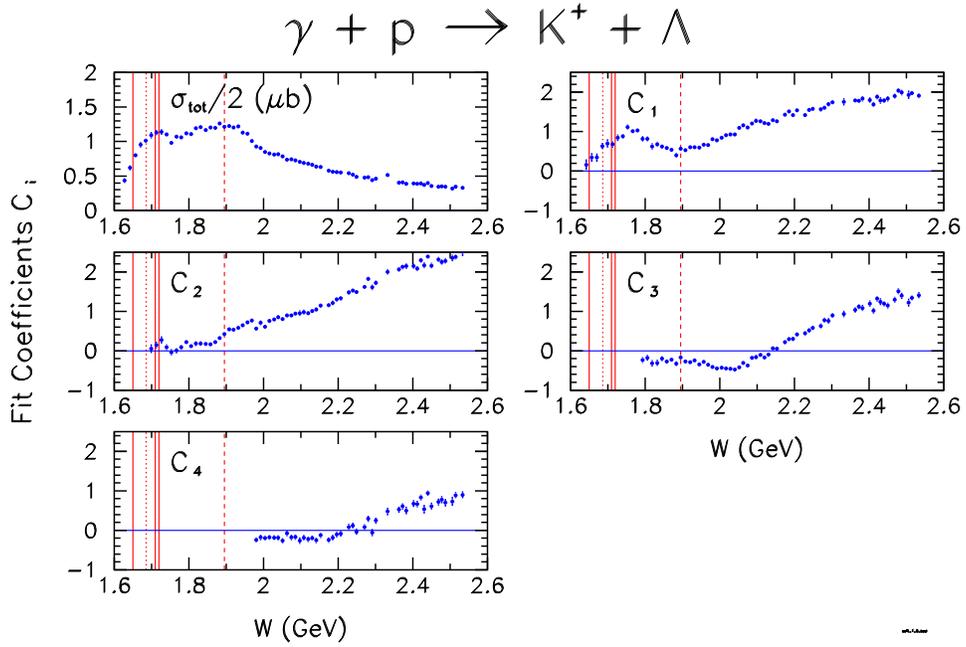}}
\vspace{-0.25in}
\caption{(Color online)
Fit to the magnitude of the differential cross sections for $\gamma +
p \rightarrow K^+ + \Lambda$.  The coefficients are defined in
Eq.~\ref{eq:mag}.  The solid vertical lines mark the well-known $N^*$
resonances $S_{11}(1650)$, $P_{11}(1710)$, and $P_{13}(1720)$.  The
dotted line marks the $\Sigma^0$ threshold, and the dashed line marks
the $D_{13}(1895)$ position.  }
\label{fig:lmag}       
\end{figure*}

An alternative fitting procedure was performed that decomposes the
angular distribution magnitudes directly into Legendre coefficients,
rather than amplitude-level partial wave Legendre coefficients.  The
fits were of the form
\begin{equation}
\frac{d\sigma}{d\cos(\theta_{K^+}^{c.m.})} = \frac{\sigma_{tot}}{2}
\left\{
1 + \sum_{i=1}^{4}{C_i P_i (\cos\theta_{K^+}^{c.m.}) }
\right\}
\label{eq:mag}
\end{equation}
and are shown in Fig.~\ref{fig:lmag}.  The total cross section,
$\sigma_{tot}$, was used as a parameter in order to obtain a proper
estimate of its uncertainty, which, due to parameter covariances, is
more difficult with the fits using Eq.~\ref{eq:amp}.  The coefficients
$C_i$ are dimensionless ratios of the $i^{th}$ moments of the angular
distribution to the total cross section.  This fit procedure, to
magnitudes rather than amplitudes of the distributions, is less useful
in revealing interference effects.  Nevertheless, some structure is
visible.  The $C_1$ parameter shows a bump below 1.9 GeV which arises
either from S-P or higher wave interference, and the $C_3$ parameter
has a change in slope near 2.05 GeV.  Overall, the increasingly
forward-peaked cross section with increasing energy forces all the
$C_i$'s to rise with $W$.

\begin{figure*}
\resizebox{0.88\textwidth}{!}{\includegraphics{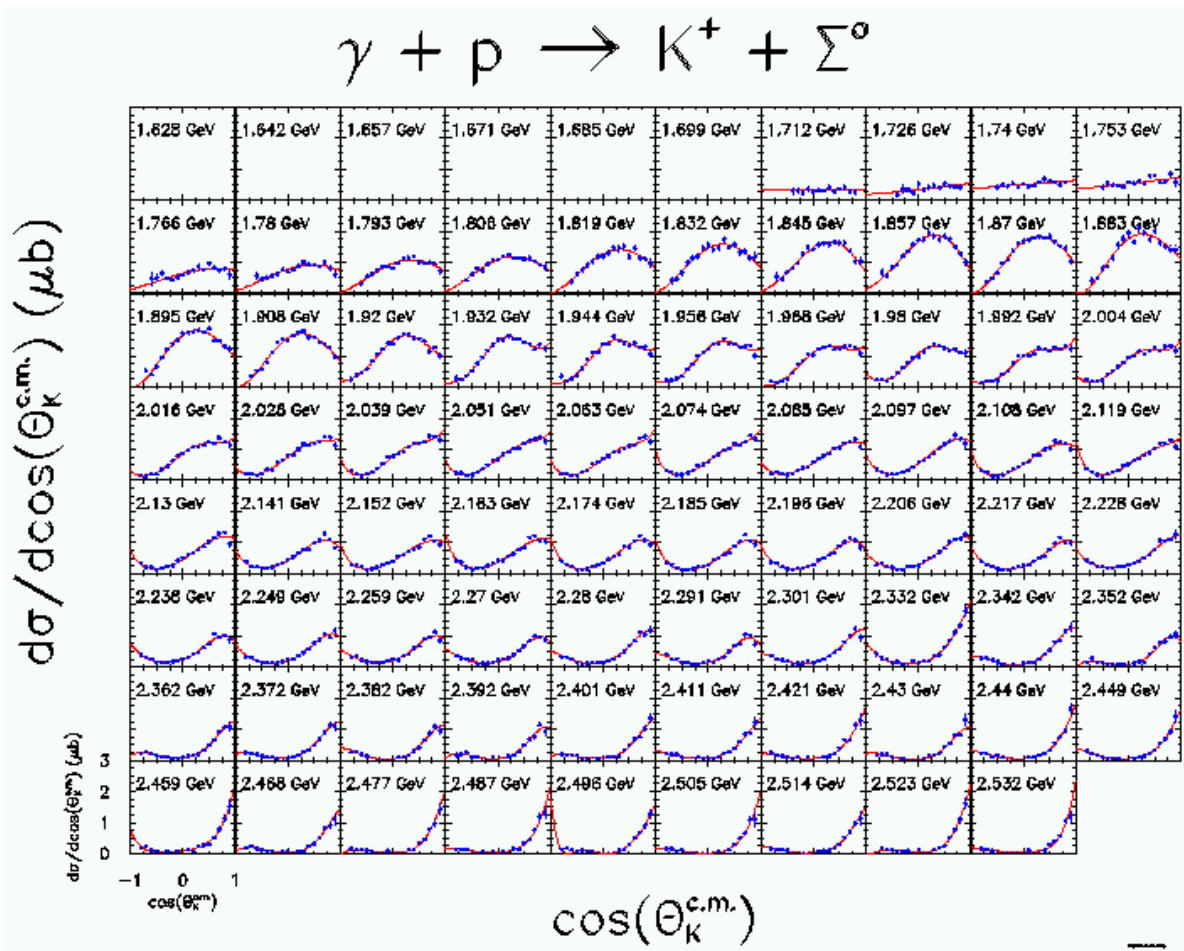}}
\vspace{-0.15in}
\caption{
Differential cross sections for $\gamma + p \rightarrow K^+ +
\Sigma^0$. The number in each panel designates $W$ ($=\sqrt{s}$).  
The solid lines are results of the amplitude fits (Eq.~\ref{eq:amp}) 
discussed in the text.
}
\label{fig:sang}       
\end{figure*}

The differential cross sections for the $\Lambda$ can be compared to
the angular distributions for $\Sigma^0$ production shown in
Fig.~\ref{fig:sang}.  The bins in $W$ are the same as before, allowing
direct comparison of the panels in Figs.~\ref{fig:lang} and
\ref{fig:sang}.  Results for both hyperons were extracted together, using
identical procedures discussed previously.  There are 1,280 data
points in the $K^+\Sigma^0$ angular distributions.

Besides the higher reaction threshold, the most significant
qualitative difference is that the $\Sigma^0$ cross section is not
forward peaked in the energy range below 2 GeV.  At $W=1.85$ GeV, for
example, the cross section peaks near $\cos(\theta_{K^+}^{c.m.}) =
0.35$, or $70^\circ$ in the center-of-mass frame.  This is consistent with a
reaction mechanism for $\Sigma^0$ production that is less influenced
by $t$-channel exchanges and is more $s$-channel resonance dominated than
$\Lambda$ production.  The back-angle cross section is less prominent
than for the $\Lambda$ case in this energy range as well.  Above the
nucleon resonance region (above about 2.4 GeV), however, the two
channels look quite similar, with characteristic $t$-channel forward
peaking.

\begin{figure*}
\vspace{-0.50in}
\resizebox{0.80\textwidth}{!}{\includegraphics{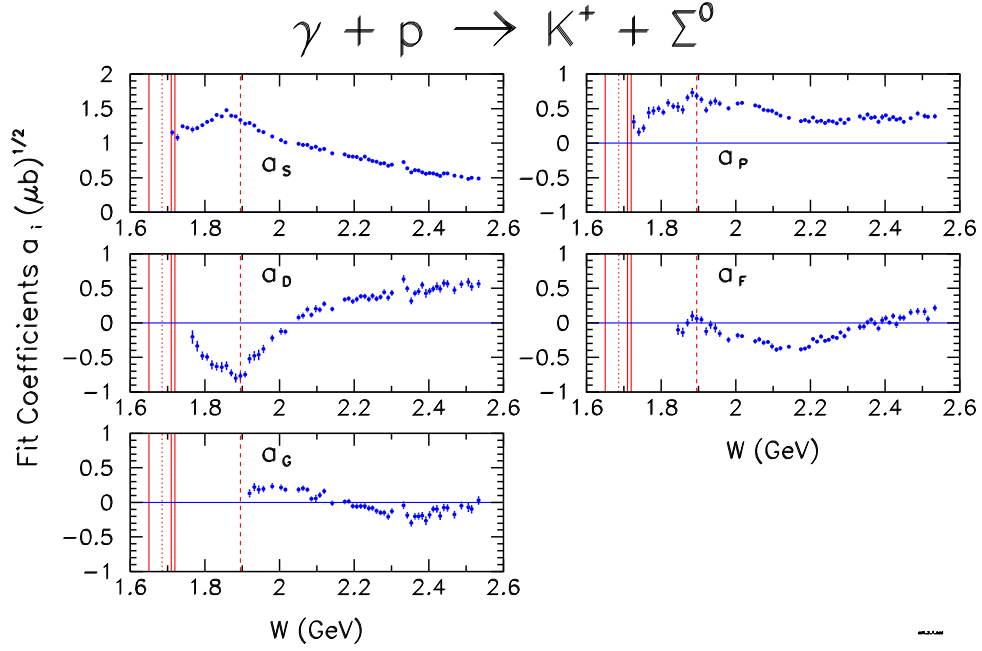}}
\vspace{-0.25in}
\caption{(Color online)
Amplitude fit to the differential cross sections for $\gamma + p
\rightarrow K^+ + \Sigma^0$.  The coefficients are defined in
Eq.~\ref{eq:amp}.  The solid vertical lines mark the well-known $N^*$
resonances $S_{11}(1650)$, $P_{11}(1710)$, and $P_{13}(1720)$.  The
dotted line marks the $\Sigma^0$ threshold, and the dashed line marks
the $D_{13}(1895)$ position.  
}
\label{fig:samp}       
\end{figure*}

\begin{figure*}
\vspace{-0.50in}
\resizebox{0.80\textwidth}{!}{\includegraphics{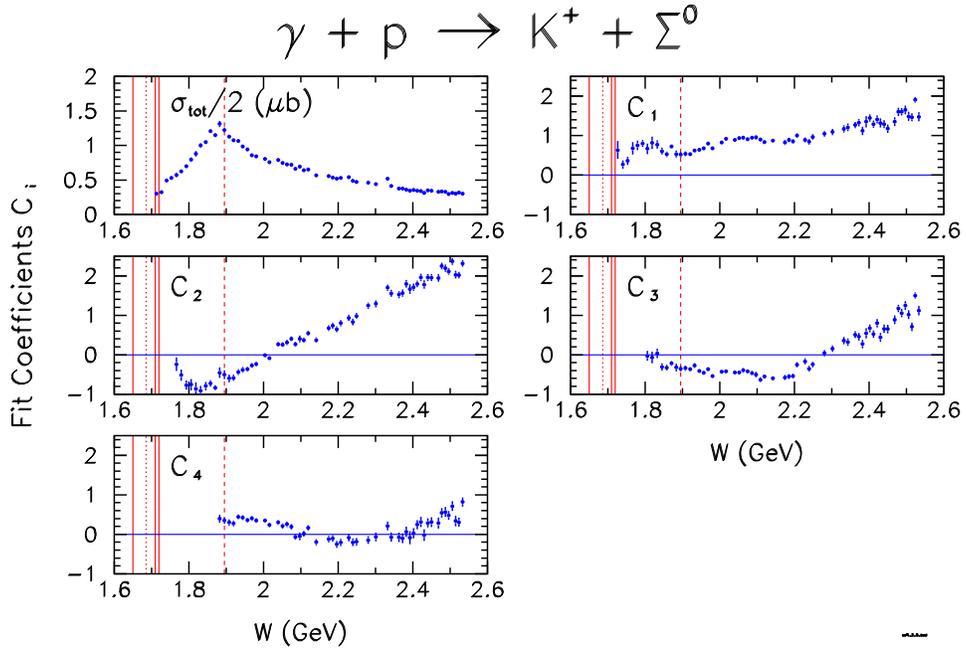}}
\vspace{-0.25in}
\caption{(Color online)
Fit to the magnitude of the differential cross sections for $\gamma +
p \rightarrow K^+ + \Sigma^0$.  The coefficients are defined in
Eq.~\ref{eq:mag}.  The solid vertical lines mark the well-known $N^*$
resonances $S_{11}(1650)$, $P_{11}(1710)$, and $P_{13}(1720)$.  The
dotted line marks the $\Sigma^0$ threshold, and the dashed line marks
the $D_{13}(1895)$ position.
}
\label{fig:smag}       
\end{figure*}

The coefficients of the amplitude-level fit in Eq.~\ref{eq:amp} for
the $\Sigma^0$ angular distributions are shown in Fig.~\ref{fig:samp}.
Comparing the $\Lambda$ to the $\Sigma^0$ shows that in the $\Sigma^0$
case the $D$ wave amplitude plays a more important role, falling and
rising with a centroid near 1.85 GeV.  The $P$ wave shows no strong
bump in the $\Sigma^0$, unlike the $\Lambda$.  In this case, the $G$
wave coefficient is statistically significant but shows little
structure.  For completeness, we also show the magnitude-level fit
according to Eq.~\ref{eq:mag} in Fig.~\ref{fig:smag}.  The coefficient
$C_1$ shows some structure, again due to $S-P$ or higher-wave
interference.  The coefficient $C_2$ clearly falls and rises, which
can be due to $P$ wave activity or interferences between $S$ and $D$
waves, for example.

\begin{figure*}
\resizebox{0.46\textwidth}{!}{\includegraphics{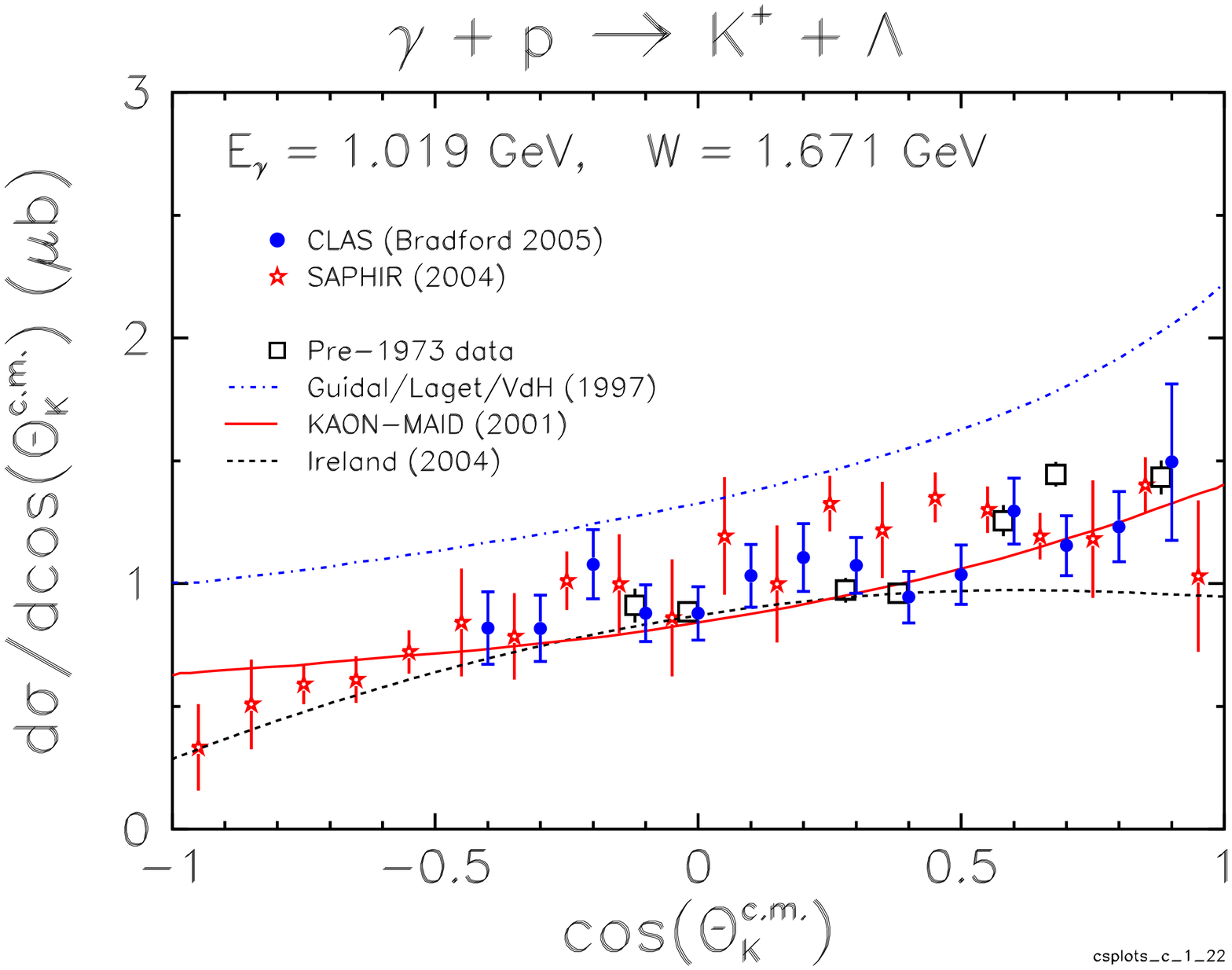}}
\resizebox{0.46\textwidth}{!}{\includegraphics{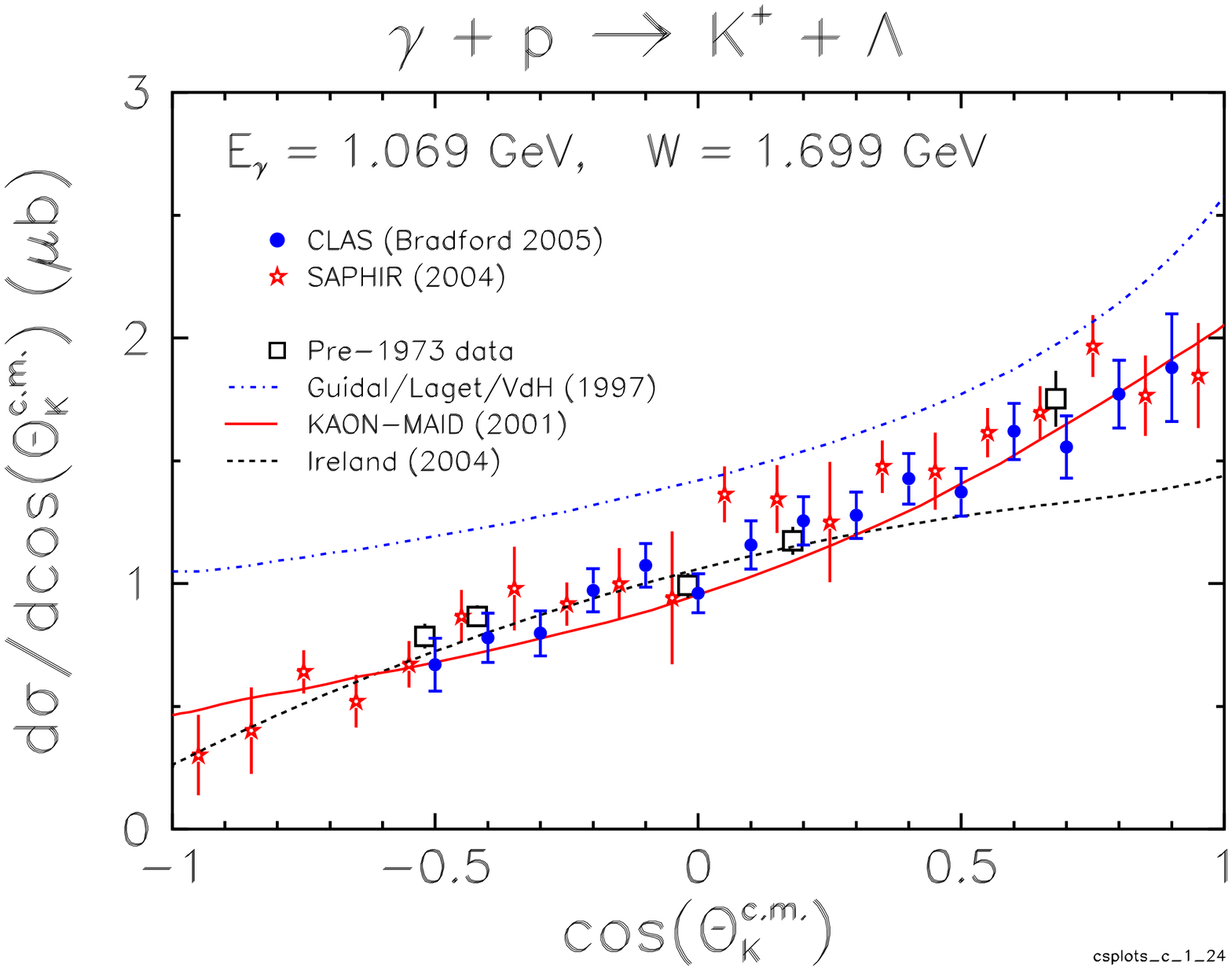}}
\resizebox{0.46\textwidth}{!}{\includegraphics{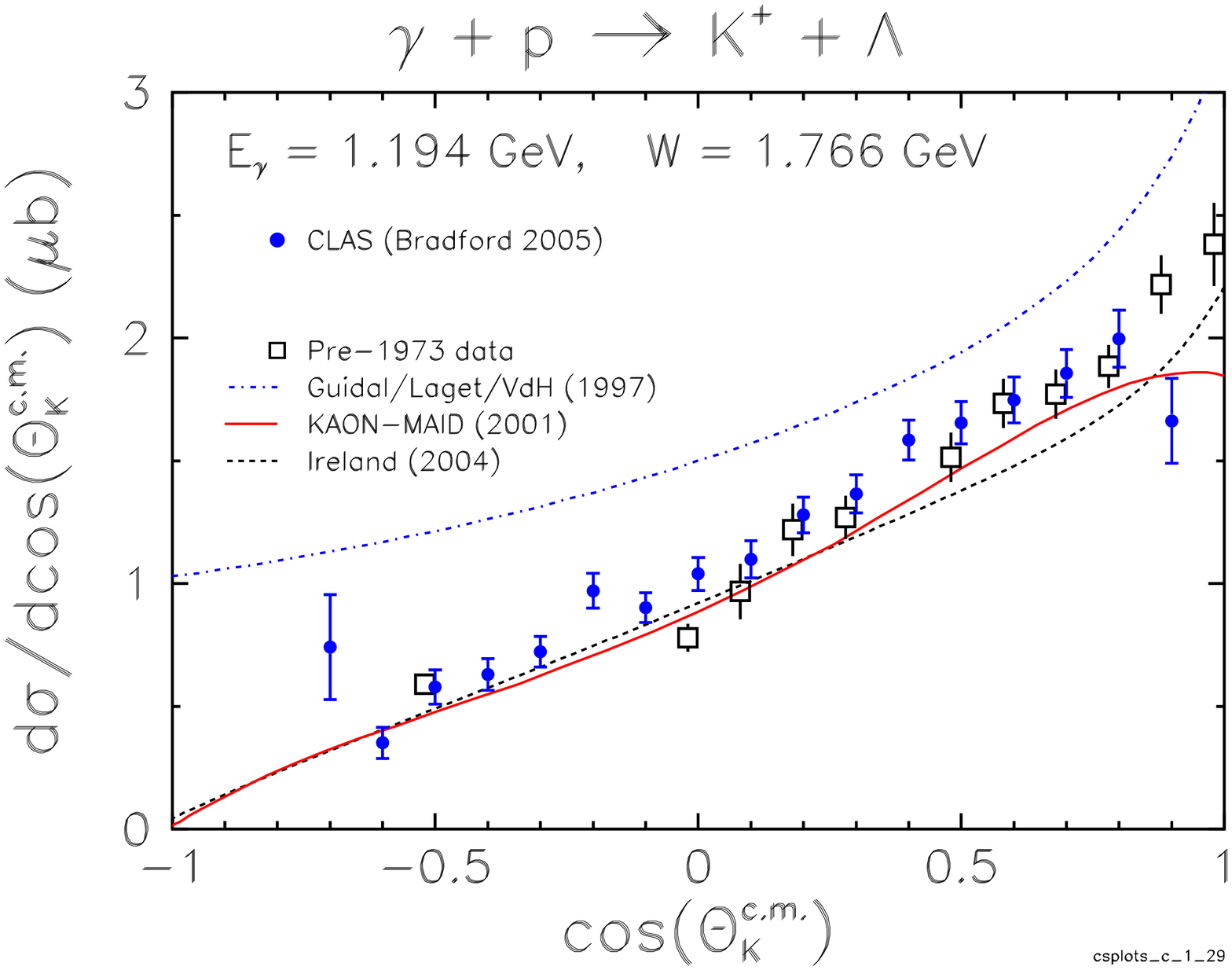}}
\resizebox{0.46\textwidth}{!}{\includegraphics{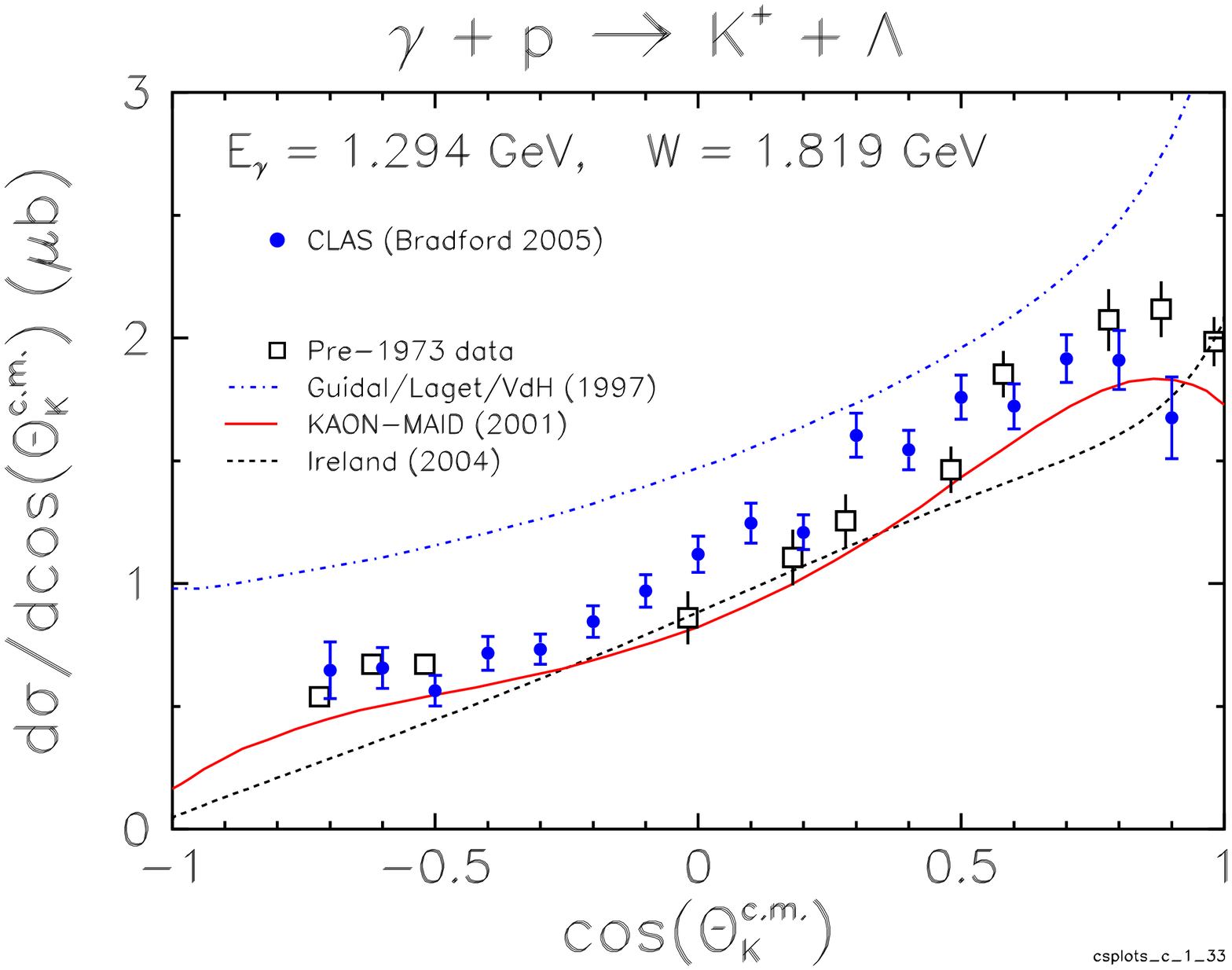}}
\resizebox{0.46\textwidth}{!}{\includegraphics{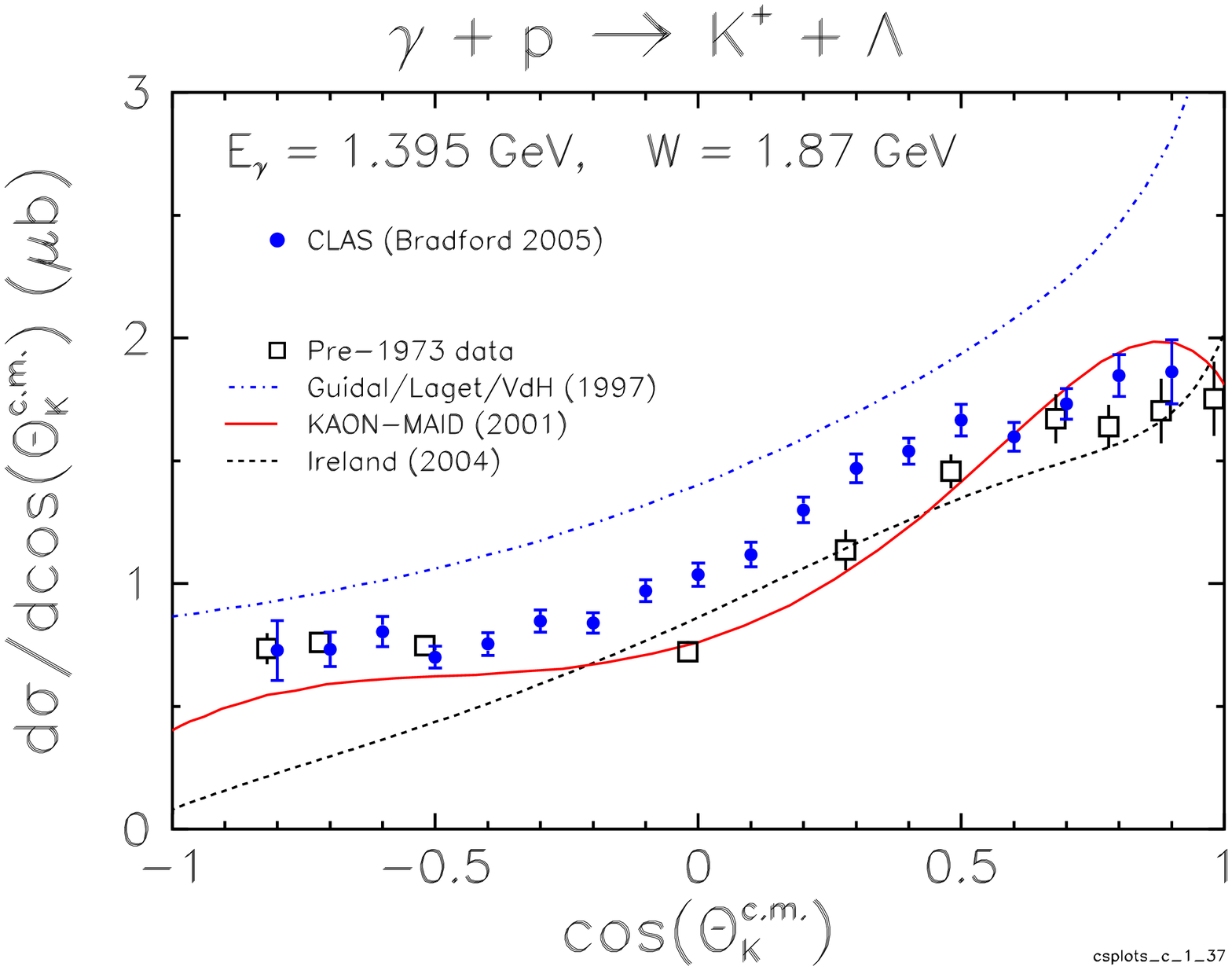}}
\resizebox{0.46\textwidth}{!}{\includegraphics{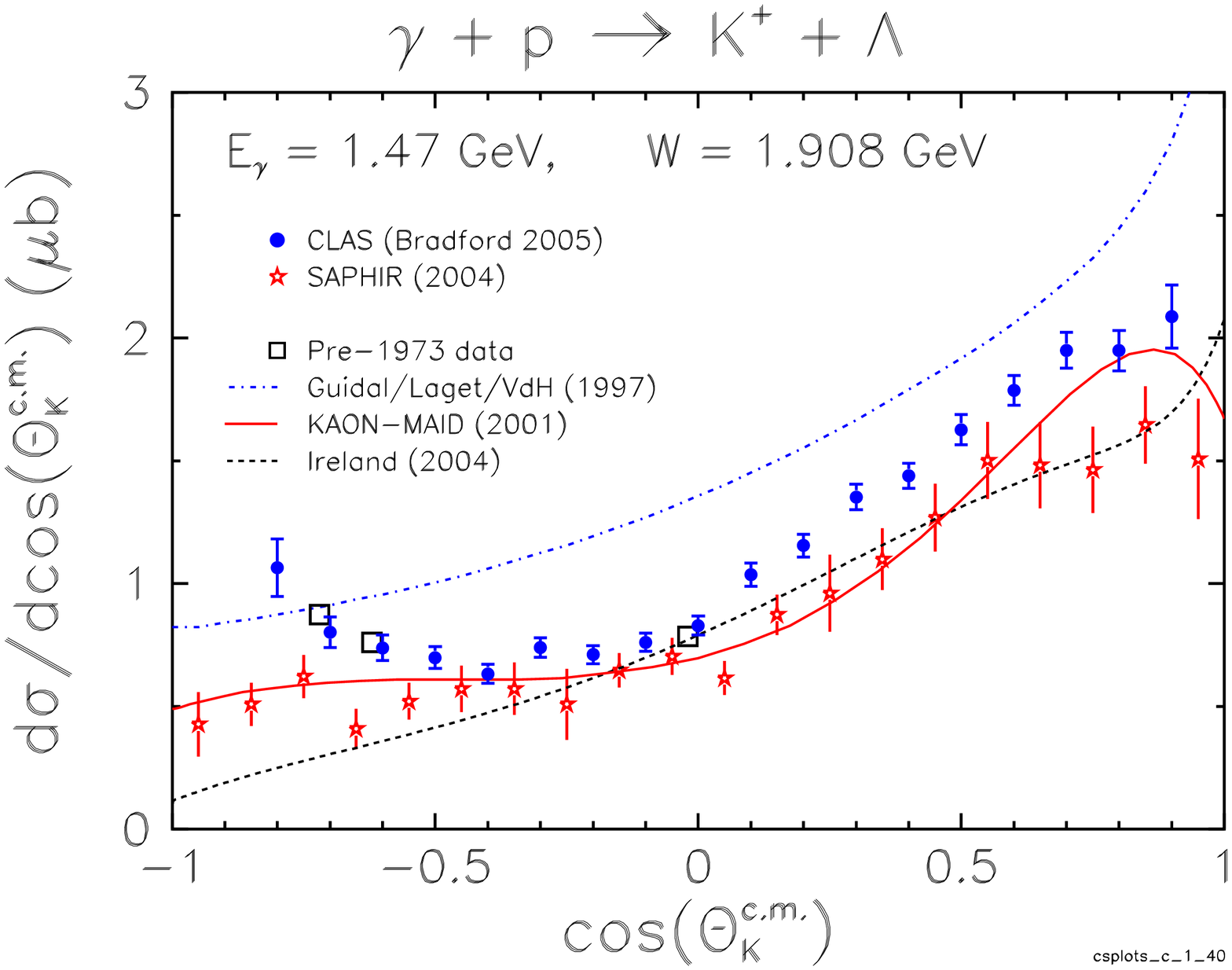}}
\caption{(Color online)
Angular distributions for $\gamma + p \rightarrow K^+ + \Lambda$ for
selected bins of total energy $W$.  The present CLAS results (blue circles)
are shown with statistical and yield-fit uncertainties.  Data from {\small
SAPHIR}~\cite{bonn2} (open red stars) and from older
experiments~\cite{land} (black squares) are also shown.  The curves are for
effective Lagrangian calculations computed by Kaon-MAID~\cite{maid}
(solid red) and Ireland {\it et al.}~\cite{ireland} (dashed black), and a
Regge-model calculation of Guidal {\it et al.}~\cite{lag1,lag2}
(dot-dashed blue).  
}
\label{fig:dsdo_l_c1}       
\end{figure*}

\begin{figure*}
\resizebox{0.46\textwidth}{!}{\includegraphics{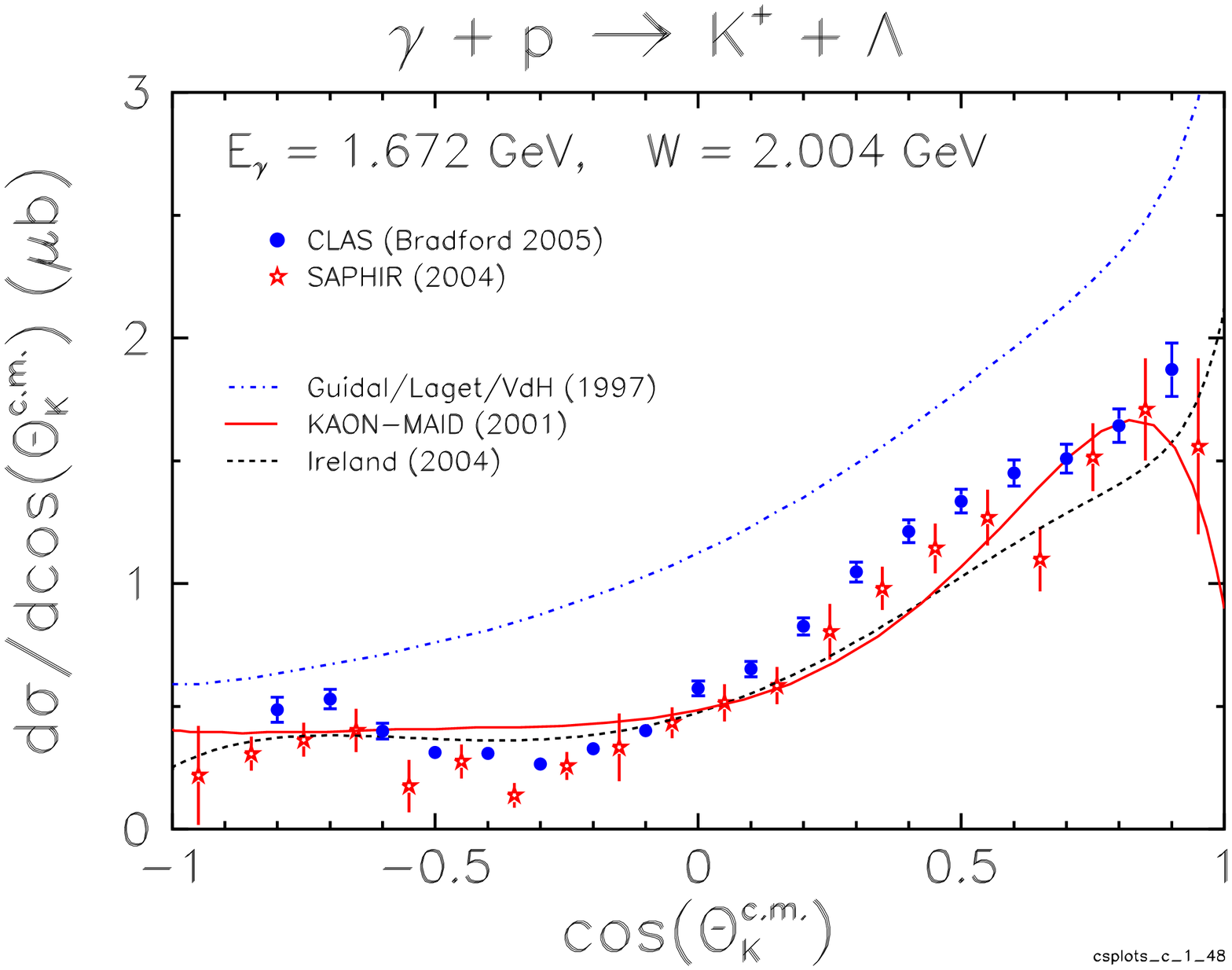}}
\resizebox{0.46\textwidth}{!}{\includegraphics{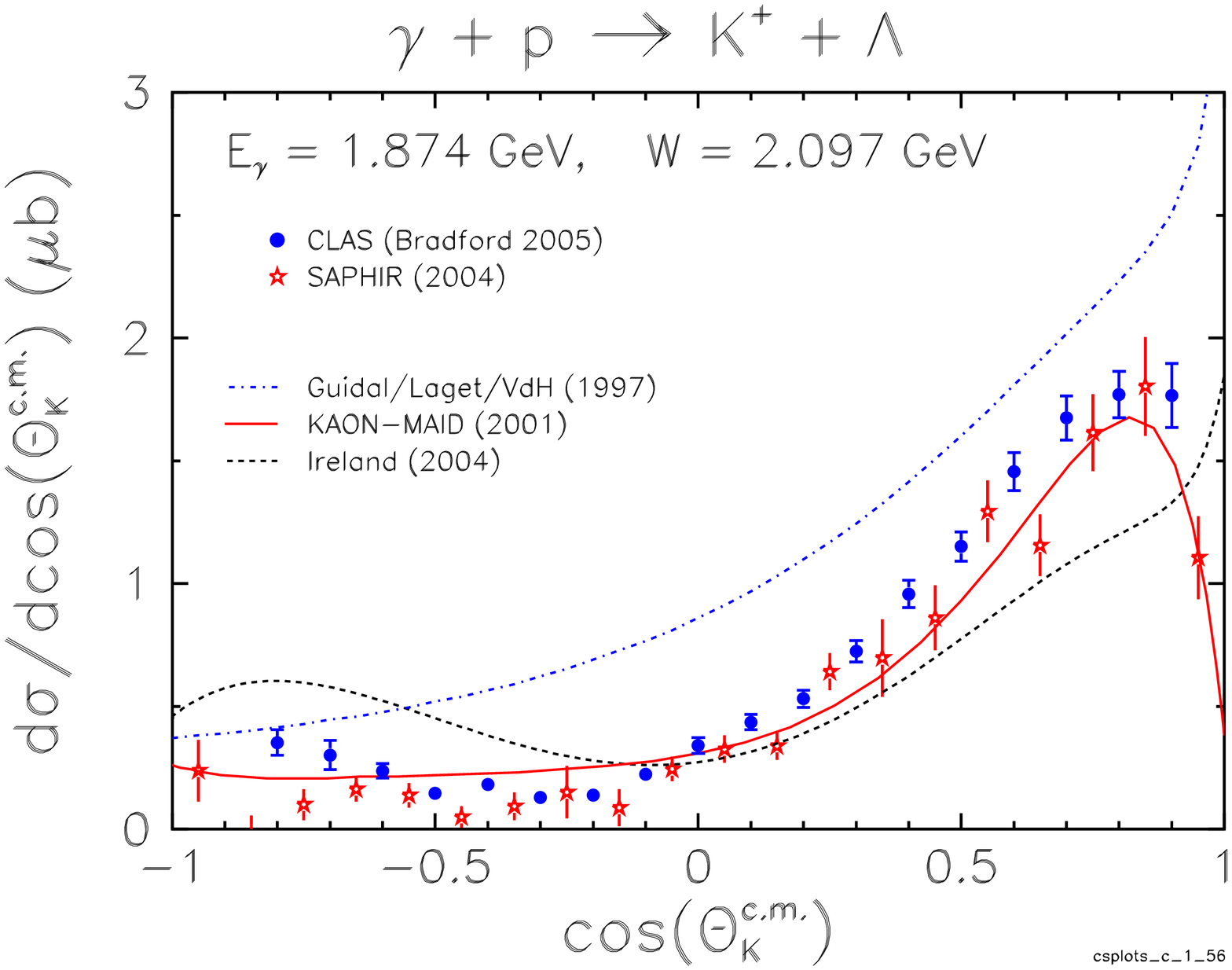}}
\resizebox{0.46\textwidth}{!}{\includegraphics{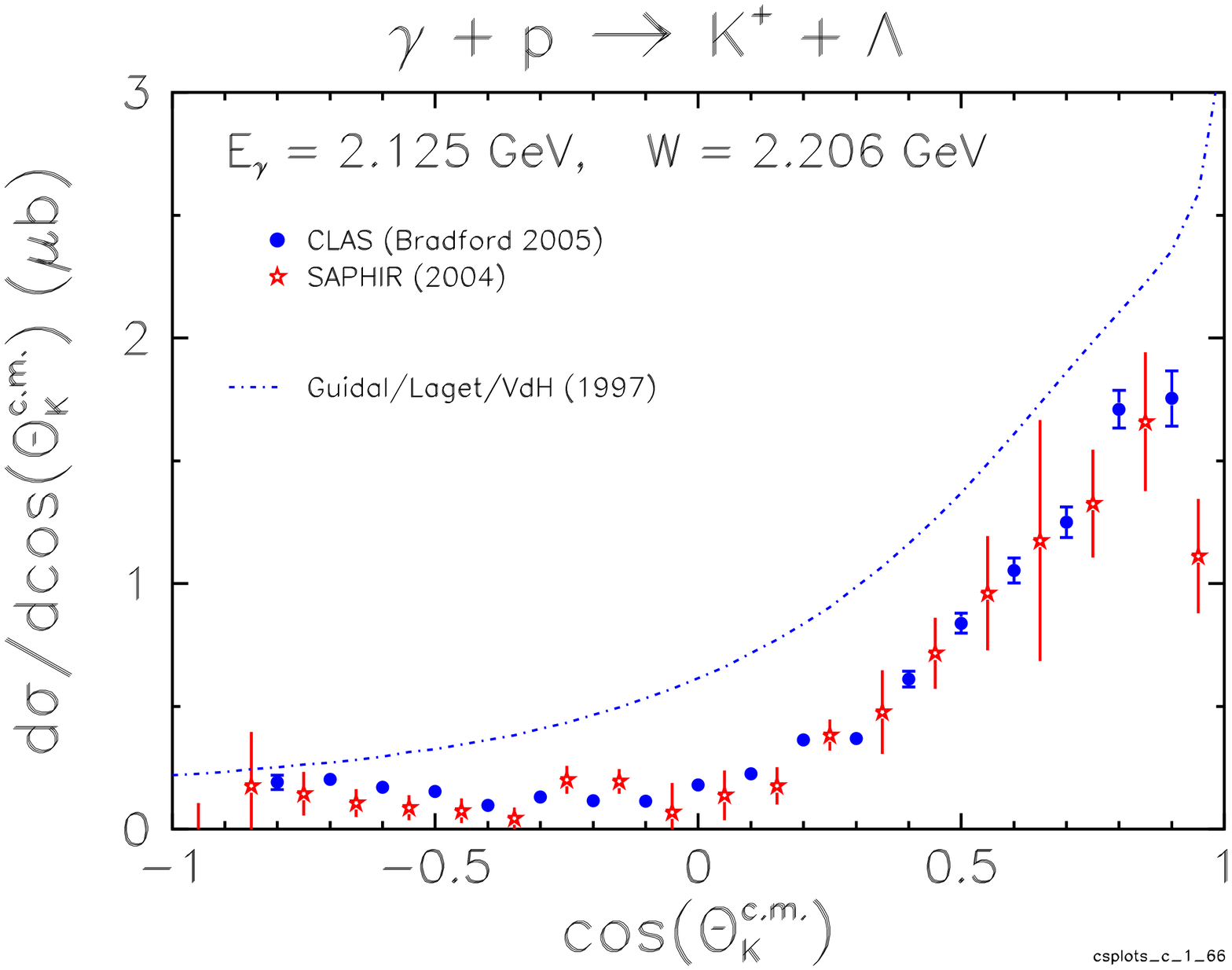}}
\resizebox{0.46\textwidth}{!}{\includegraphics{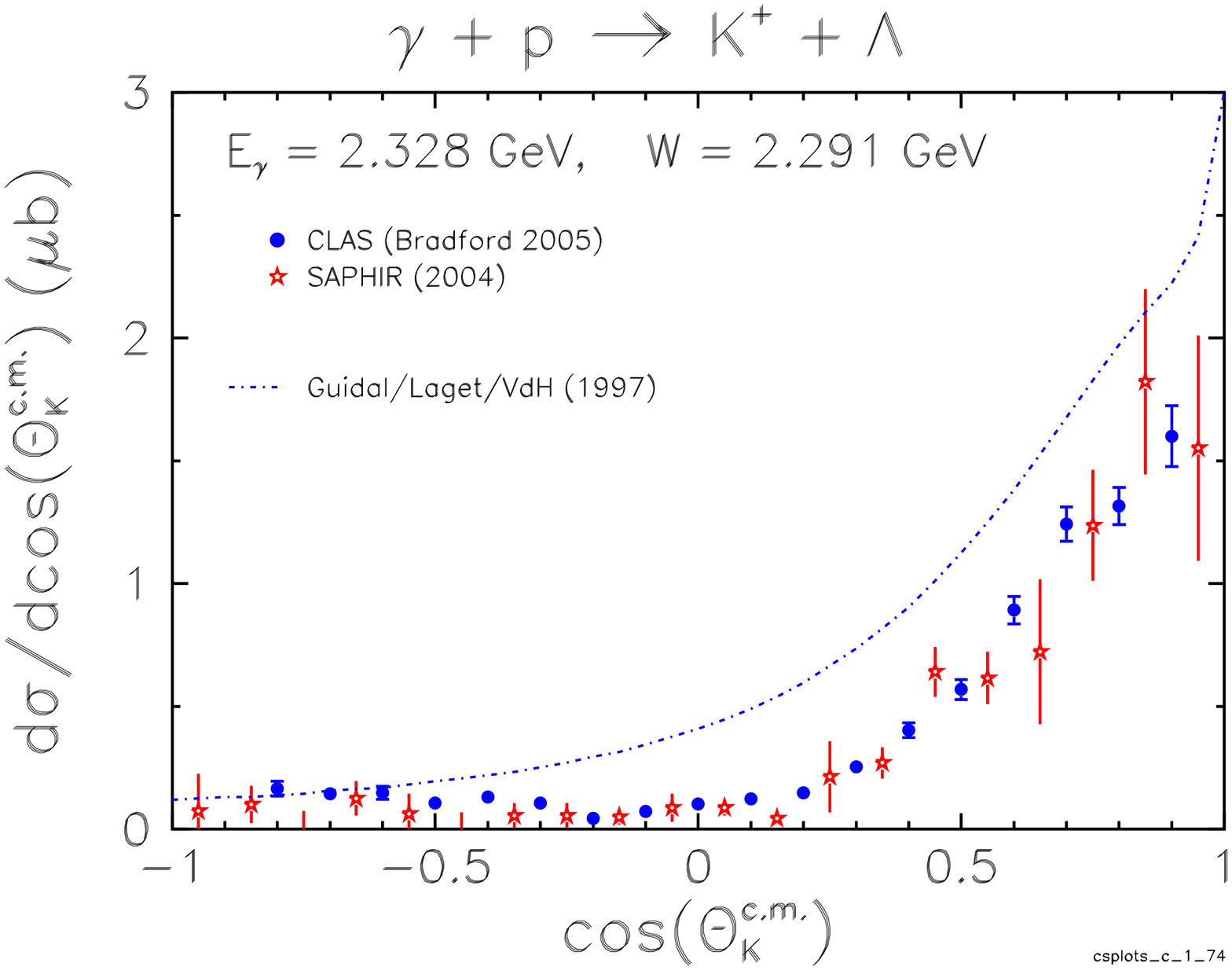}}
\resizebox{0.46\textwidth}{!}{\includegraphics{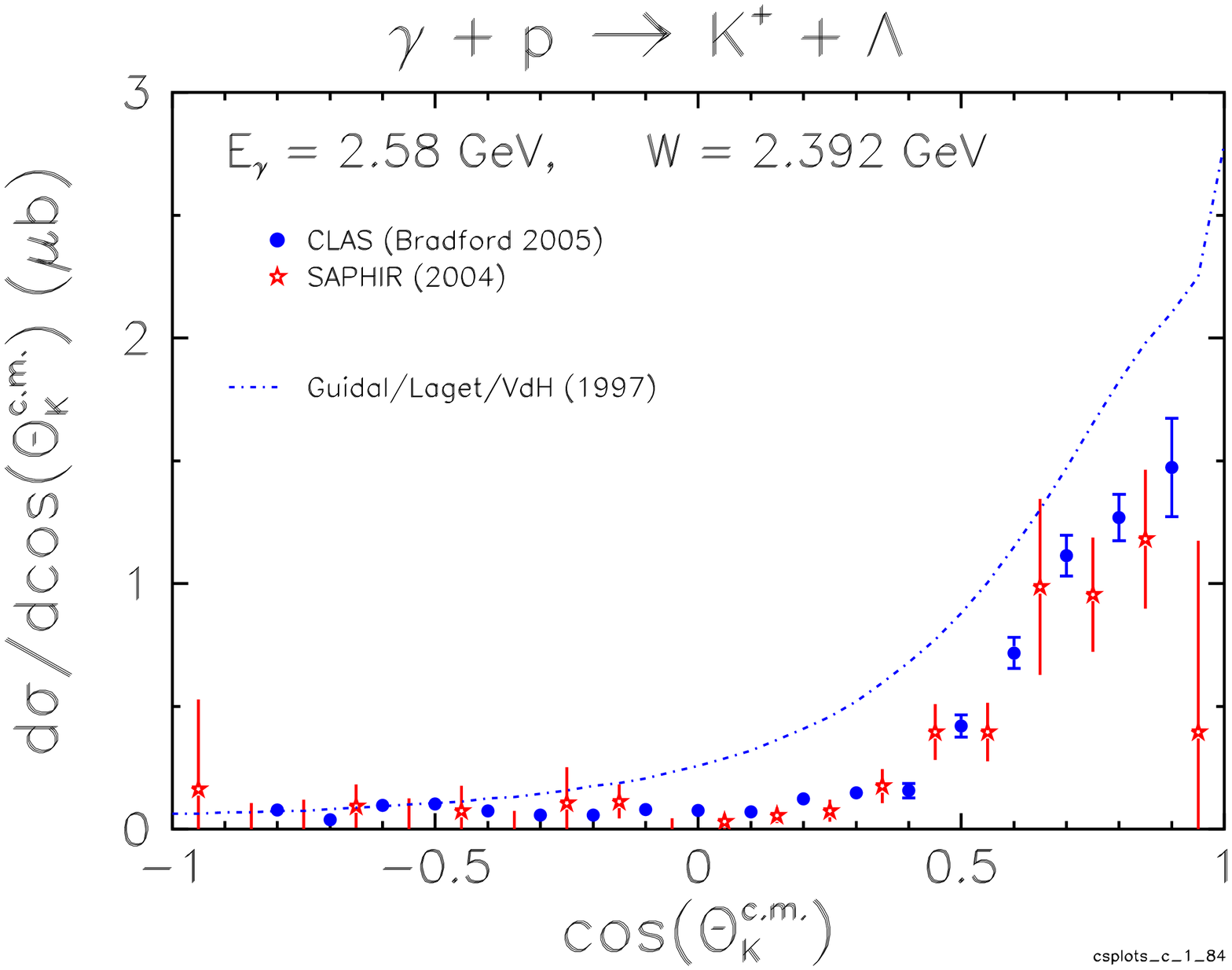}}
\resizebox{0.46\textwidth}{!}{\includegraphics{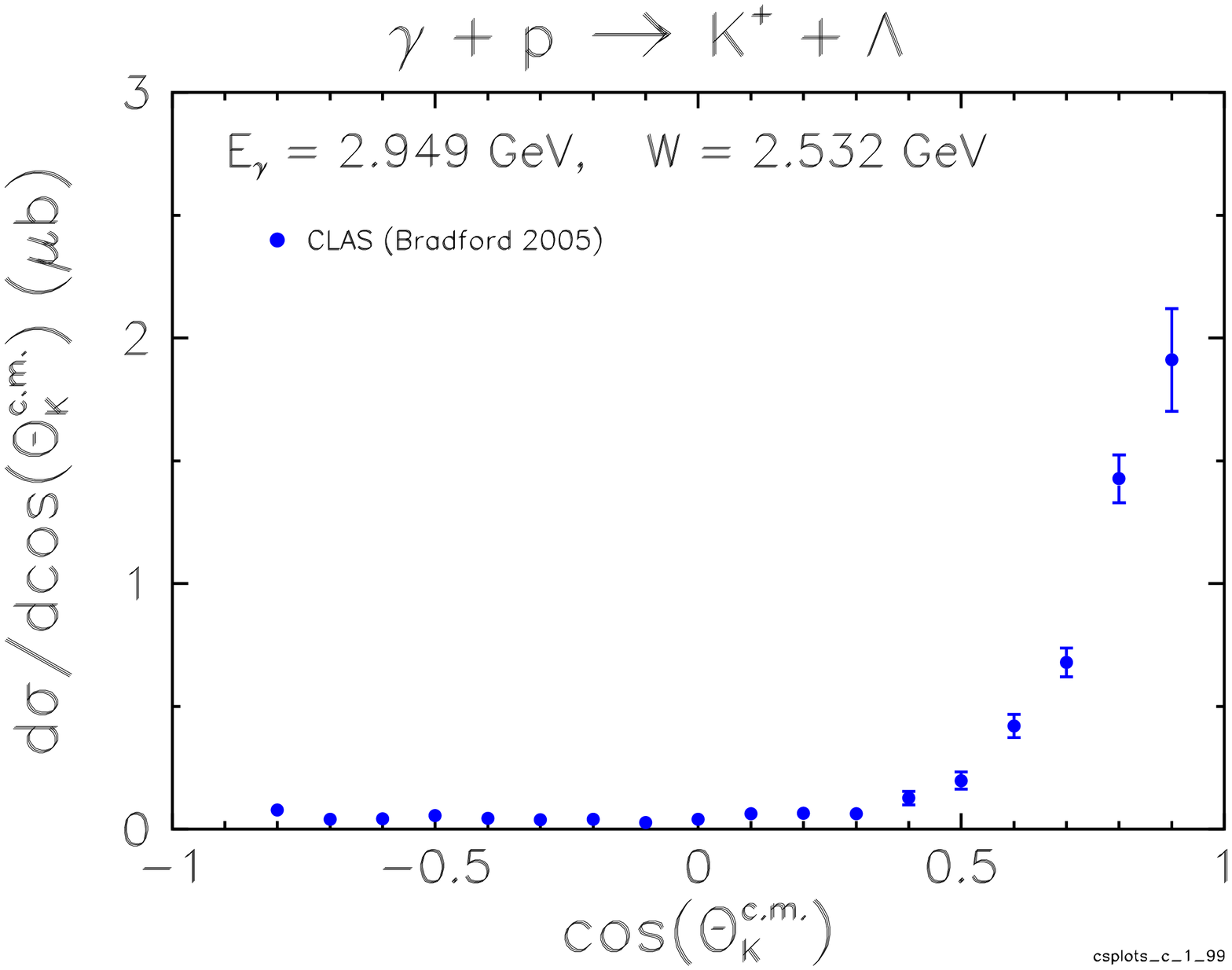}}
\caption{(Color online)
Angular distributions for $\gamma + p \rightarrow K^+ + \Lambda$ for
selected bins of total energy $W$.  The present CLAS results (blue circles)
are shown with statistical and yield-fit uncertainties.  Data from {\small
SAPHIR}~\cite{bonn2} (open stars) and from older
experiments~\cite{land} (black squares) are also shown.  The curves are for
effective Lagrangian calculations computed by Kaon-MAID~\cite{maid}
(solid red) and Ireland {\it et al.}~\cite{ireland} (dashed black), and a
Regge-model calculation of Guidal {\it et al.}~\cite{lag1,lag2}
(dot-dashed blue).  
}
\label{fig:dsdo_l_c2}       
\end{figure*}

Figures~\ref{fig:dsdo_l_c1} and~\ref{fig:dsdo_l_c2} show selected
differential cross sections from this experiment compared to
previous data and with three published model calculations.  The
selected panels show about 1/6 of our data, in increments of $\Delta
W \approx 80$ MeV to show the trends in the cross sections and the
calculations; the exact $W$ values were chosen to emphasize available
comparison data.

\begin{figure*}
\resizebox{0.46\textwidth}{!}{\includegraphics{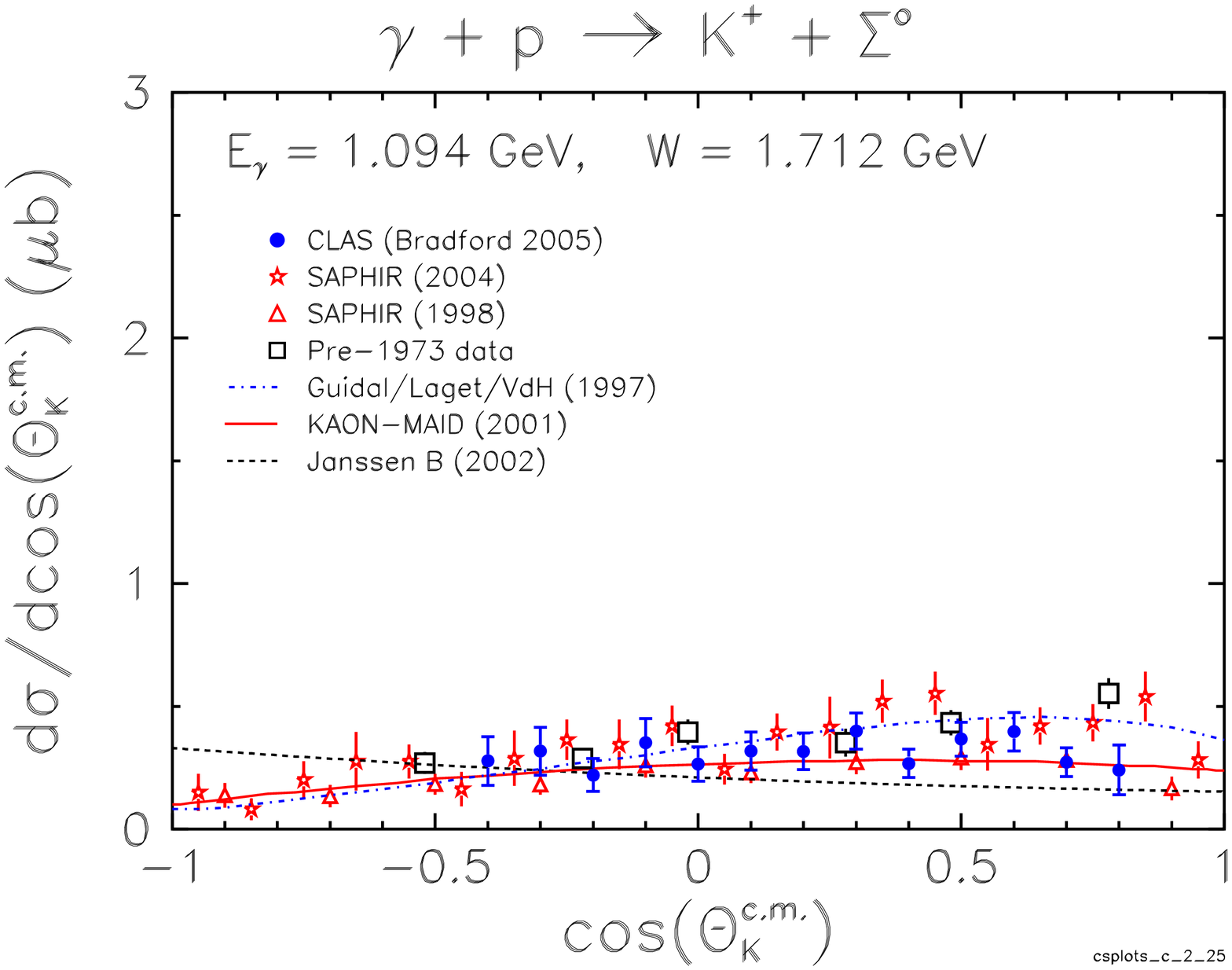}}
\resizebox{0.46\textwidth}{!}{\includegraphics{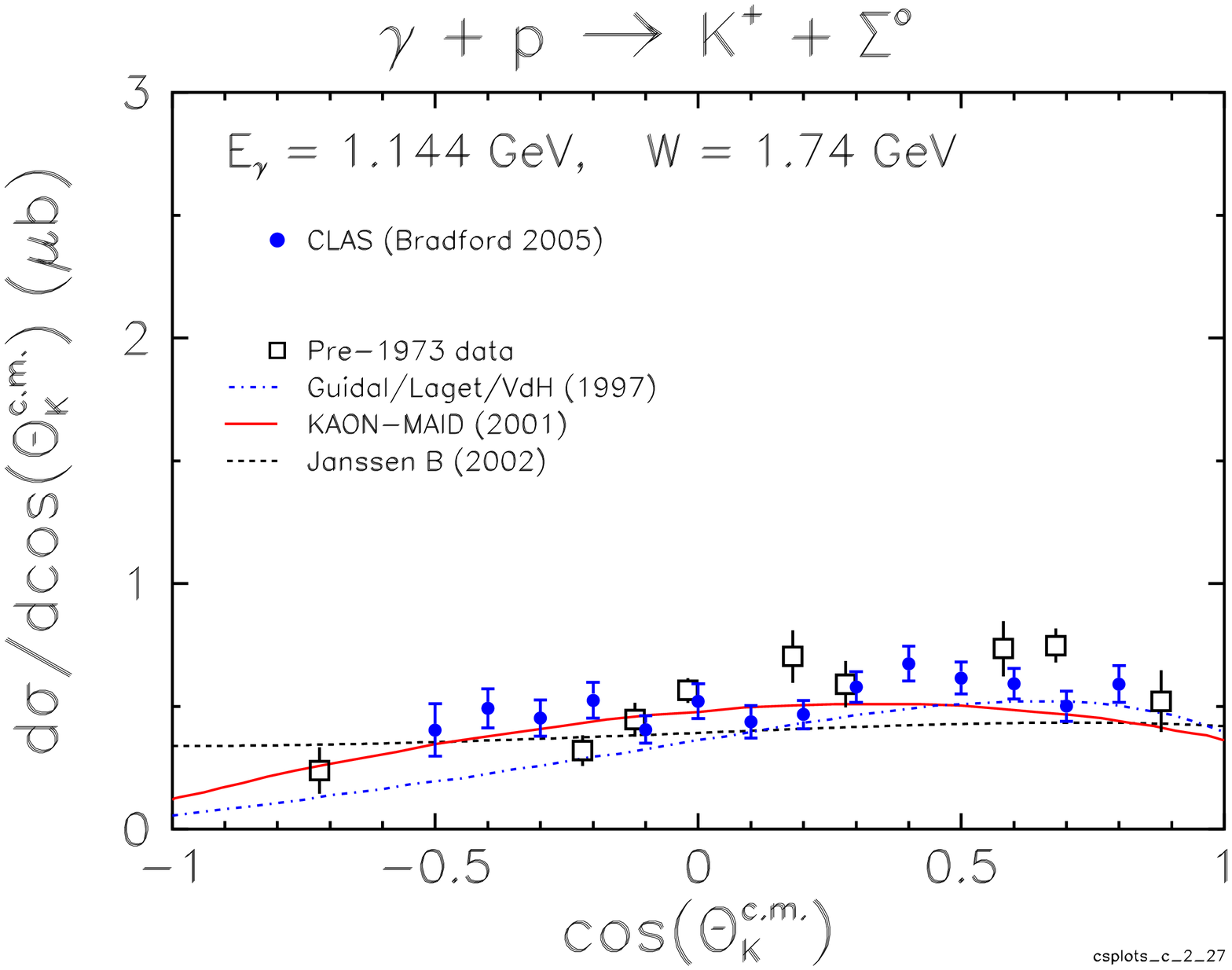}}
\resizebox{0.46\textwidth}{!}{\includegraphics{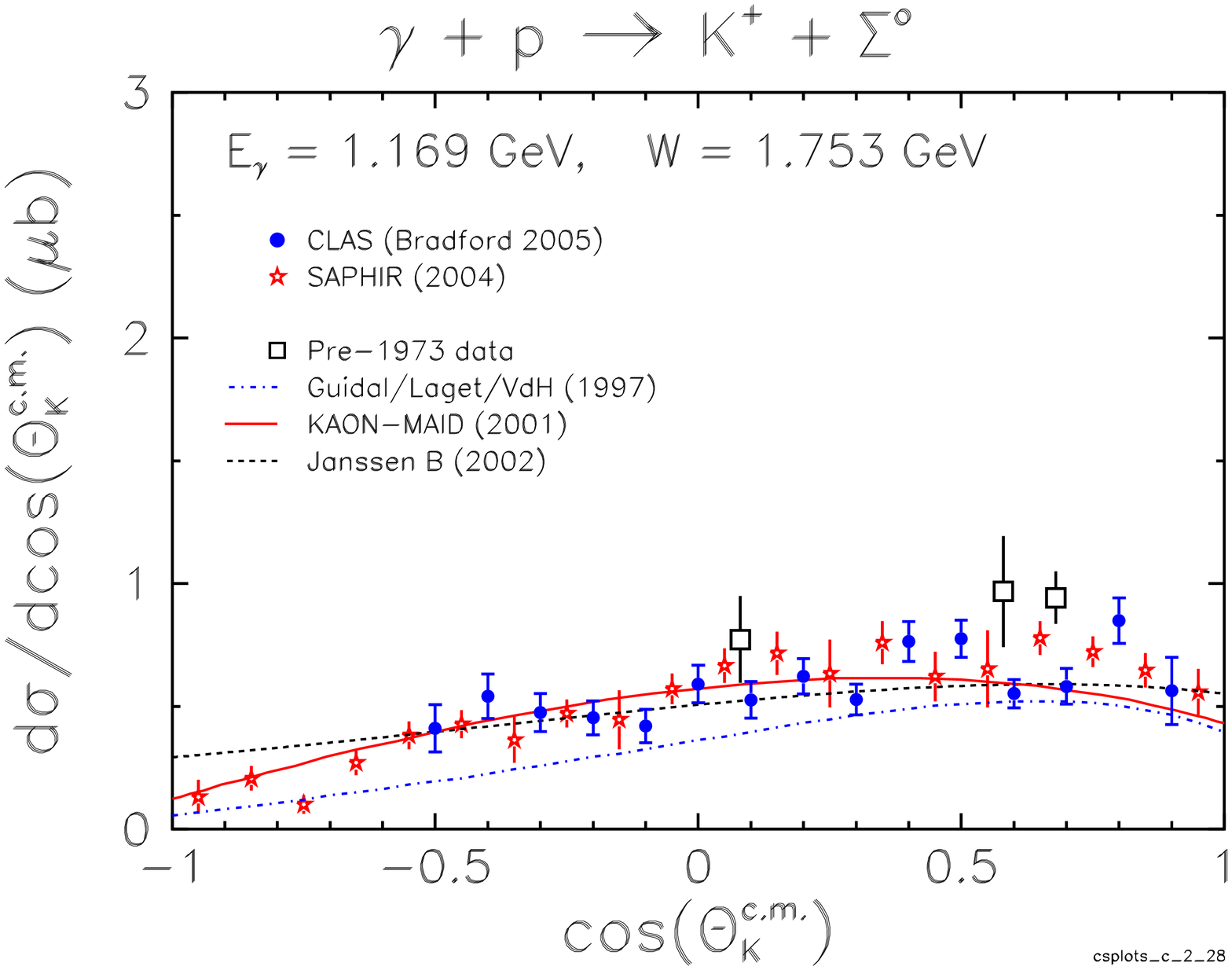}}
\resizebox{0.46\textwidth}{!}{\includegraphics{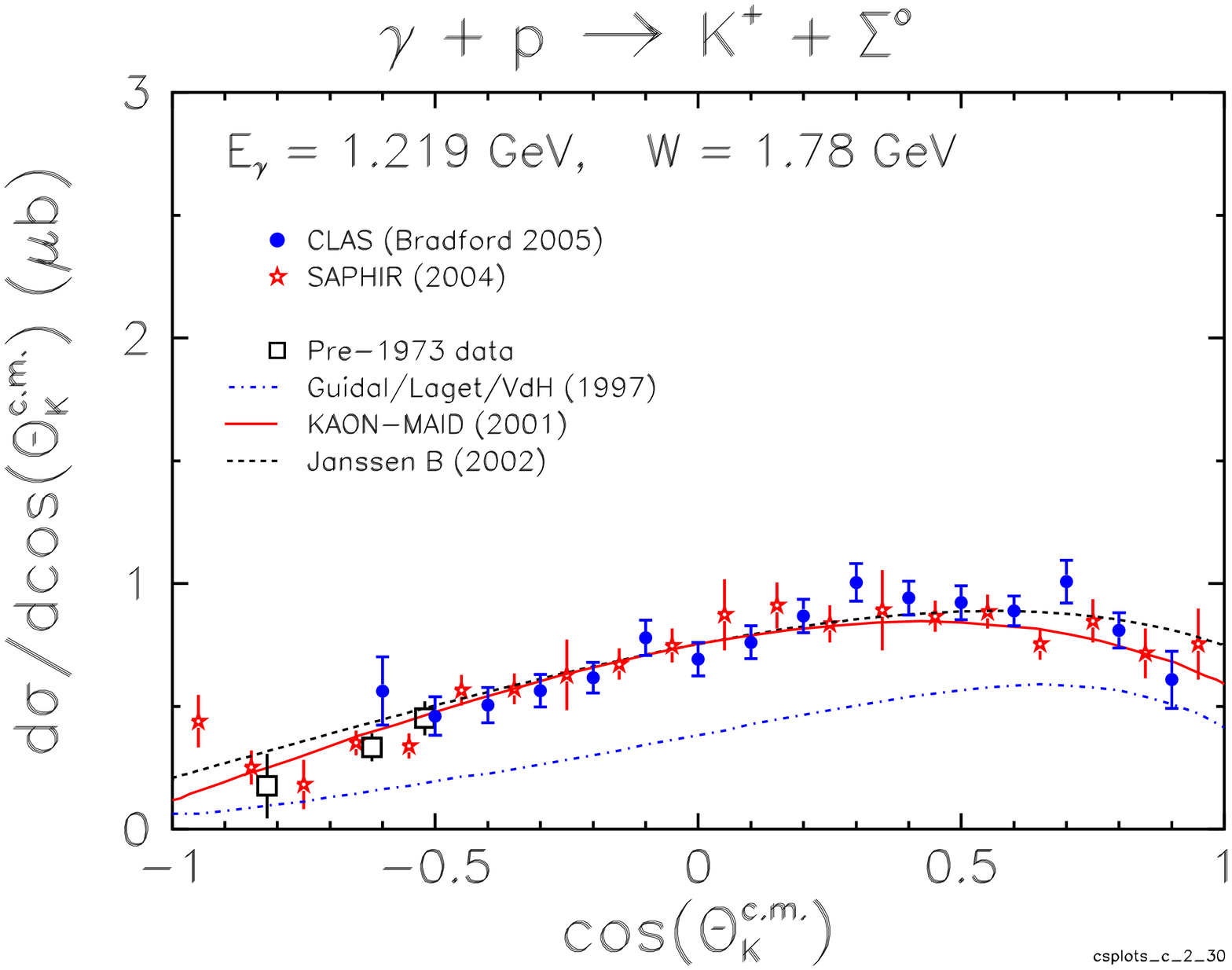}}
\resizebox{0.46\textwidth}{!}{\includegraphics{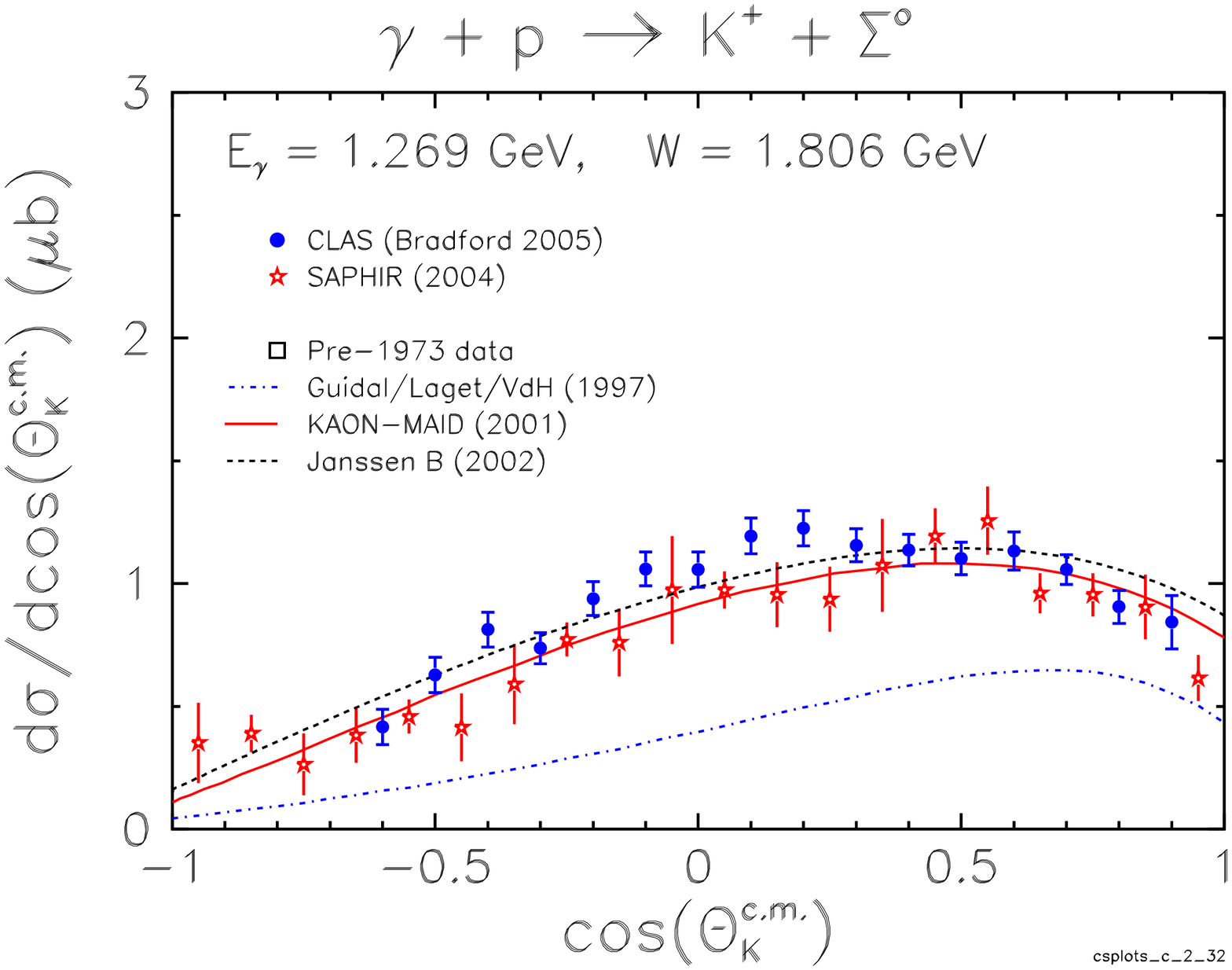}}
\resizebox{0.46\textwidth}{!}{\includegraphics{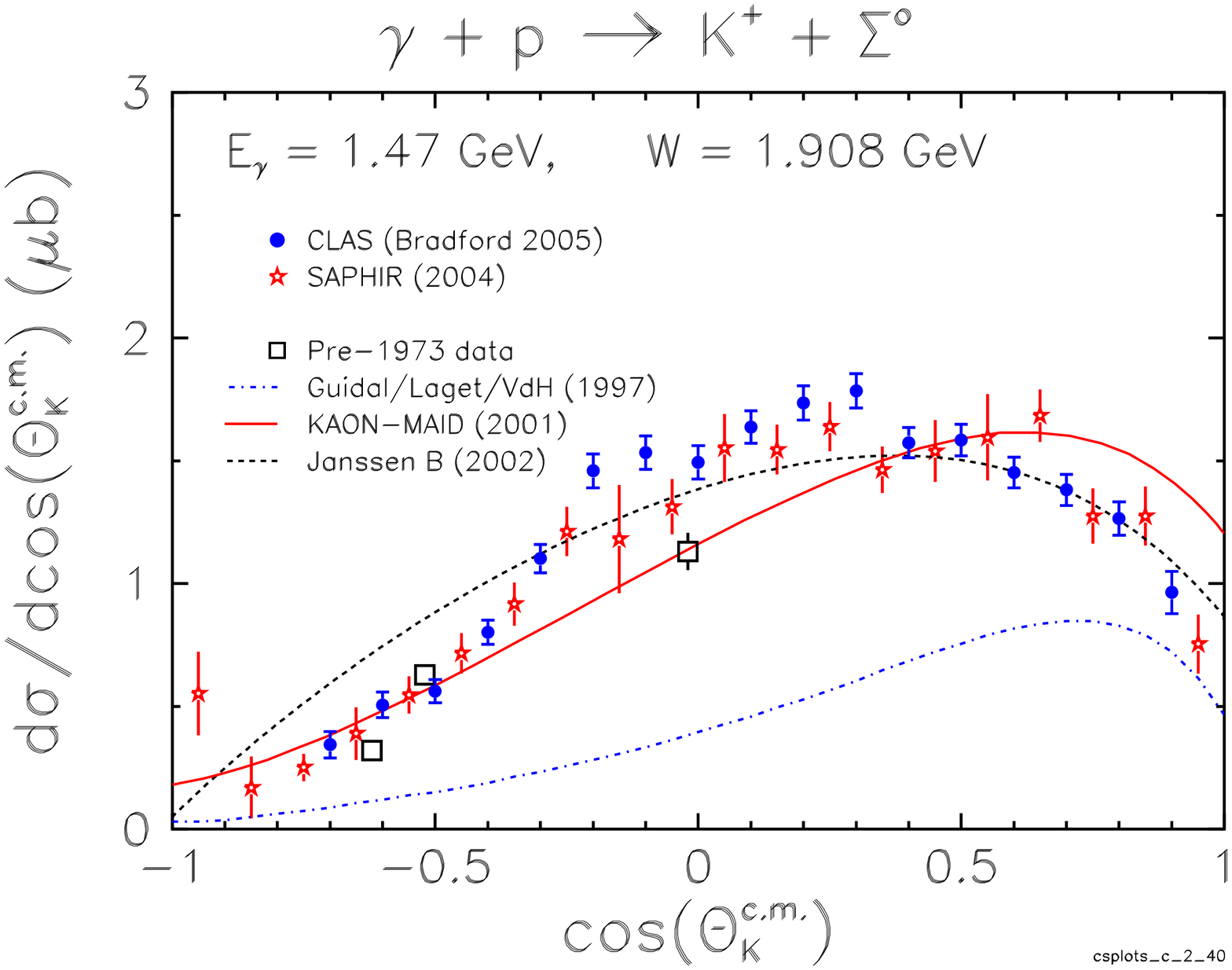}}
\caption{(Color online)
Angular distributions for $\gamma + p \rightarrow K^+ + \Sigma^0$ for
selected bins of total energy $W$.  The present CLAS results (blue circles)
are shown with statistical and yield-fit uncertainties.  
Data from {\small SAPHIR} (open stars~\cite{bonn2}
and triangles~\cite{bonn1}) and from older
experiments~\cite{land} (black squares) are also shown.  The curves are for
effective Lagrangian calculations computed by Kaon-MAID~\cite{maid}
(solid red) and Janssen {\it et al.}~\cite{jan} (dashed black), and a
Regge-model calculation of Guidal {\it et al.}~\cite{lag1,lag2}
(dot-dashed blue).  
}
\label{fig:dsdo_s_c1}       
\end{figure*}

\begin{figure*}
\resizebox{0.46\textwidth}{!}{\includegraphics{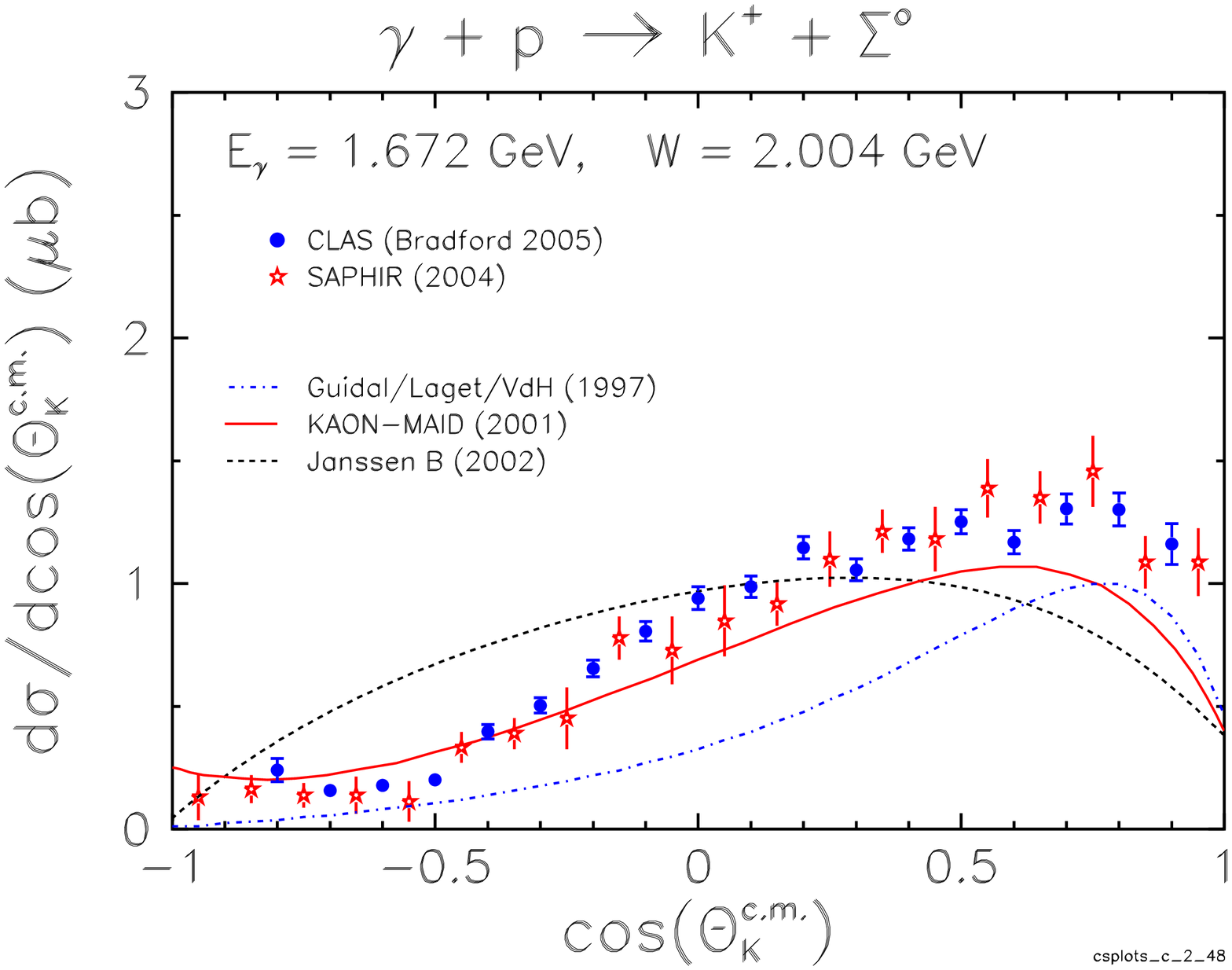}}
\resizebox{0.46\textwidth}{!}{\includegraphics{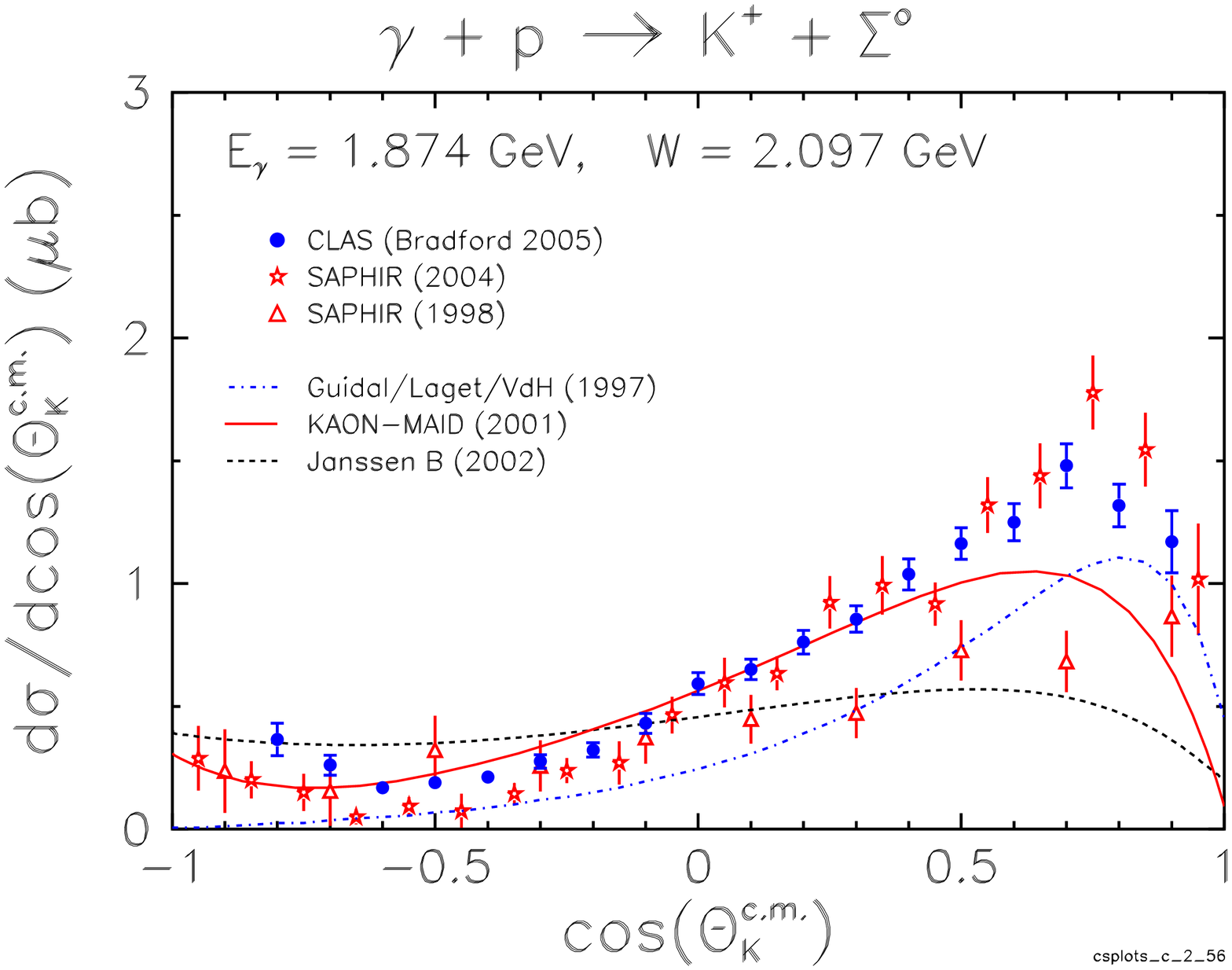}}
\resizebox{0.46\textwidth}{!}{\includegraphics{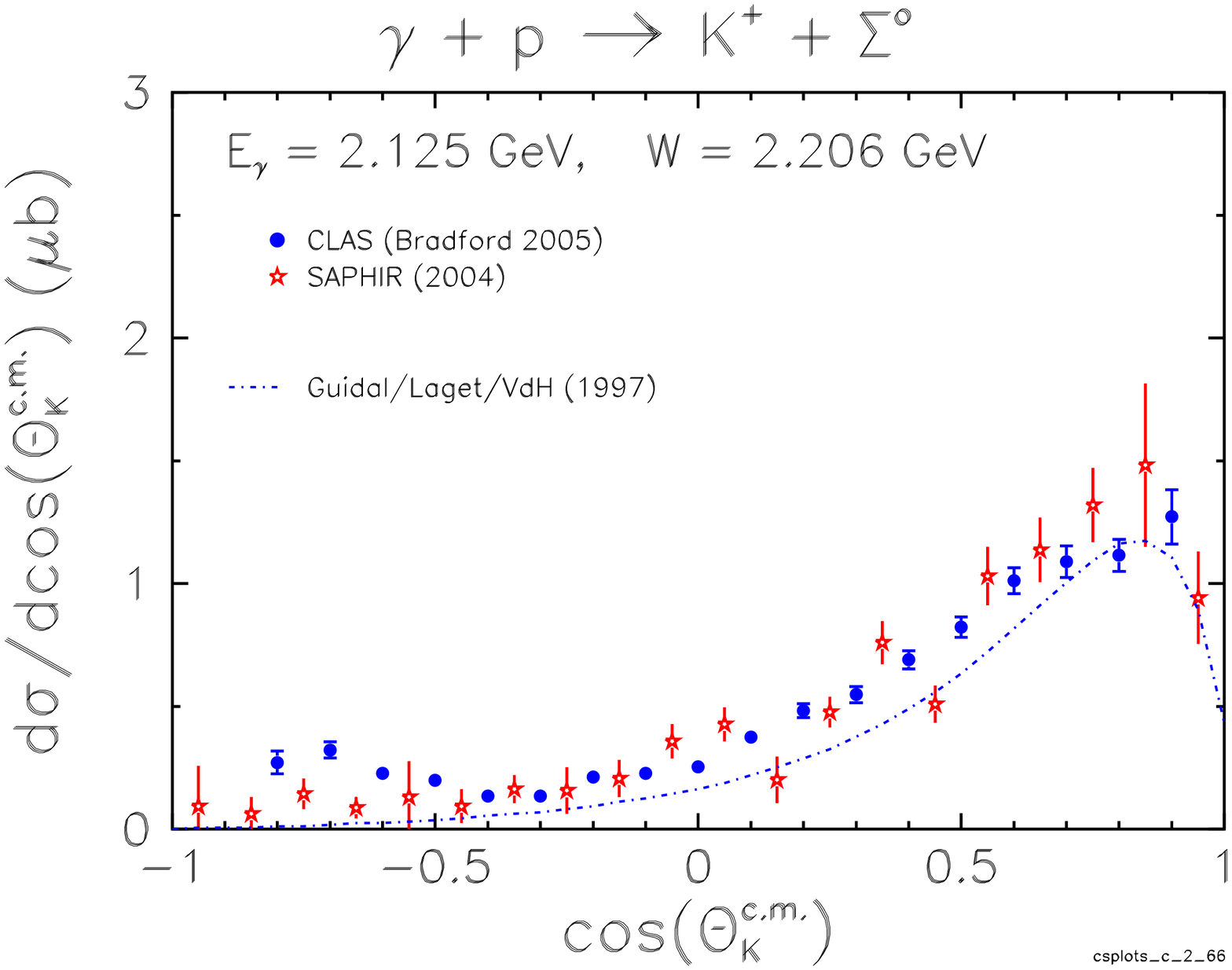}}
\resizebox{0.46\textwidth}{!}{\includegraphics{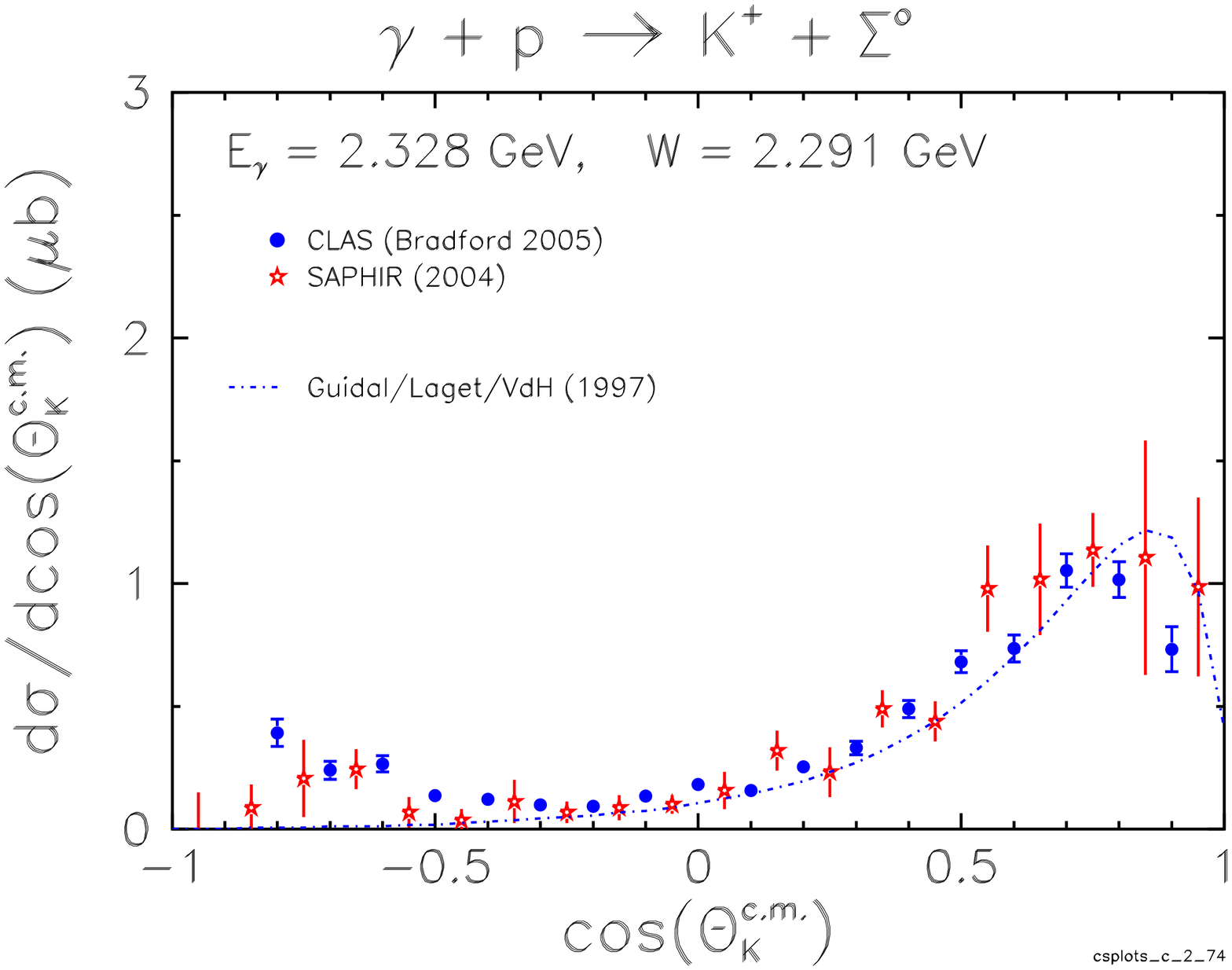}}
\resizebox{0.46\textwidth}{!}{\includegraphics{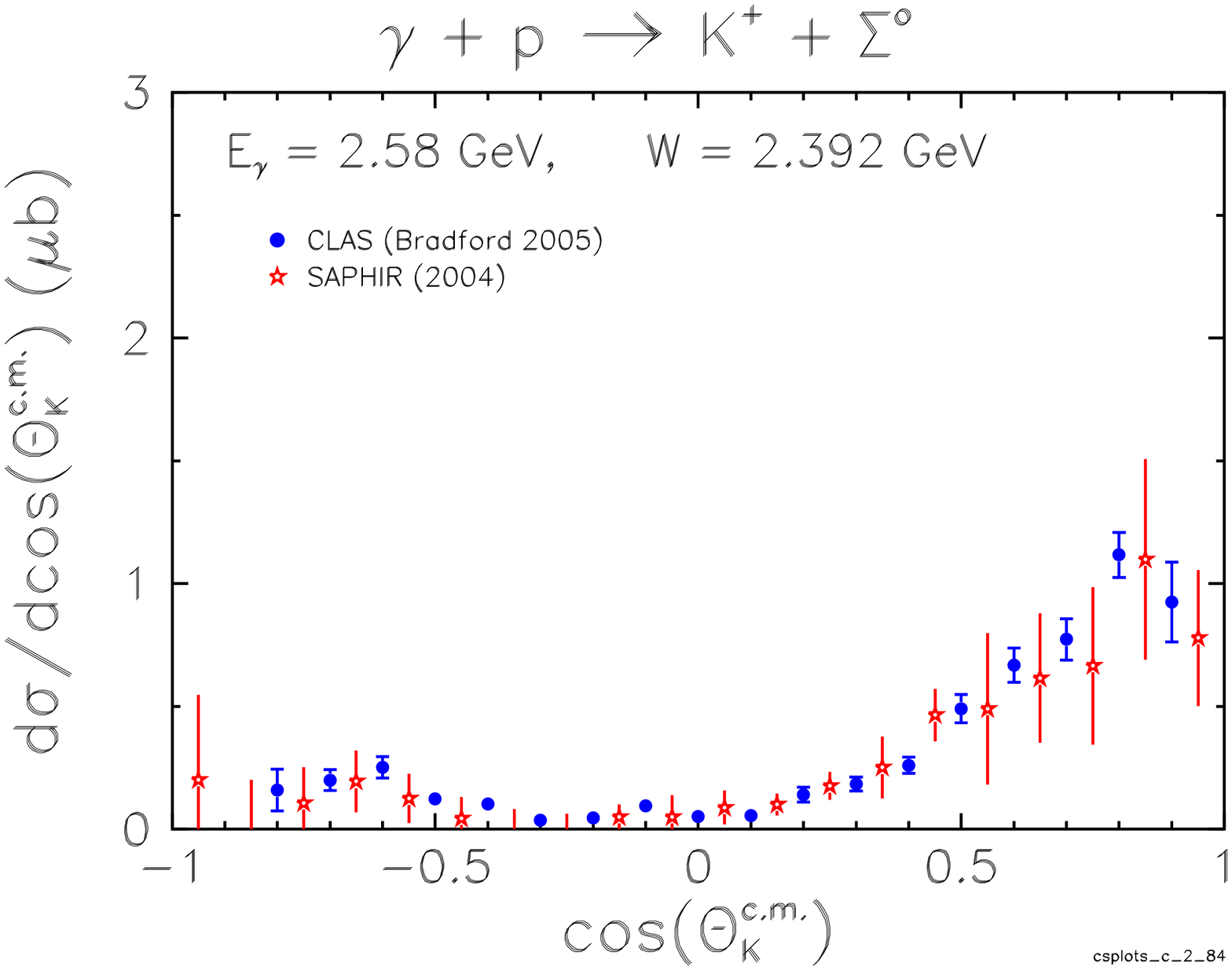}}
\resizebox{0.46\textwidth}{!}{\includegraphics{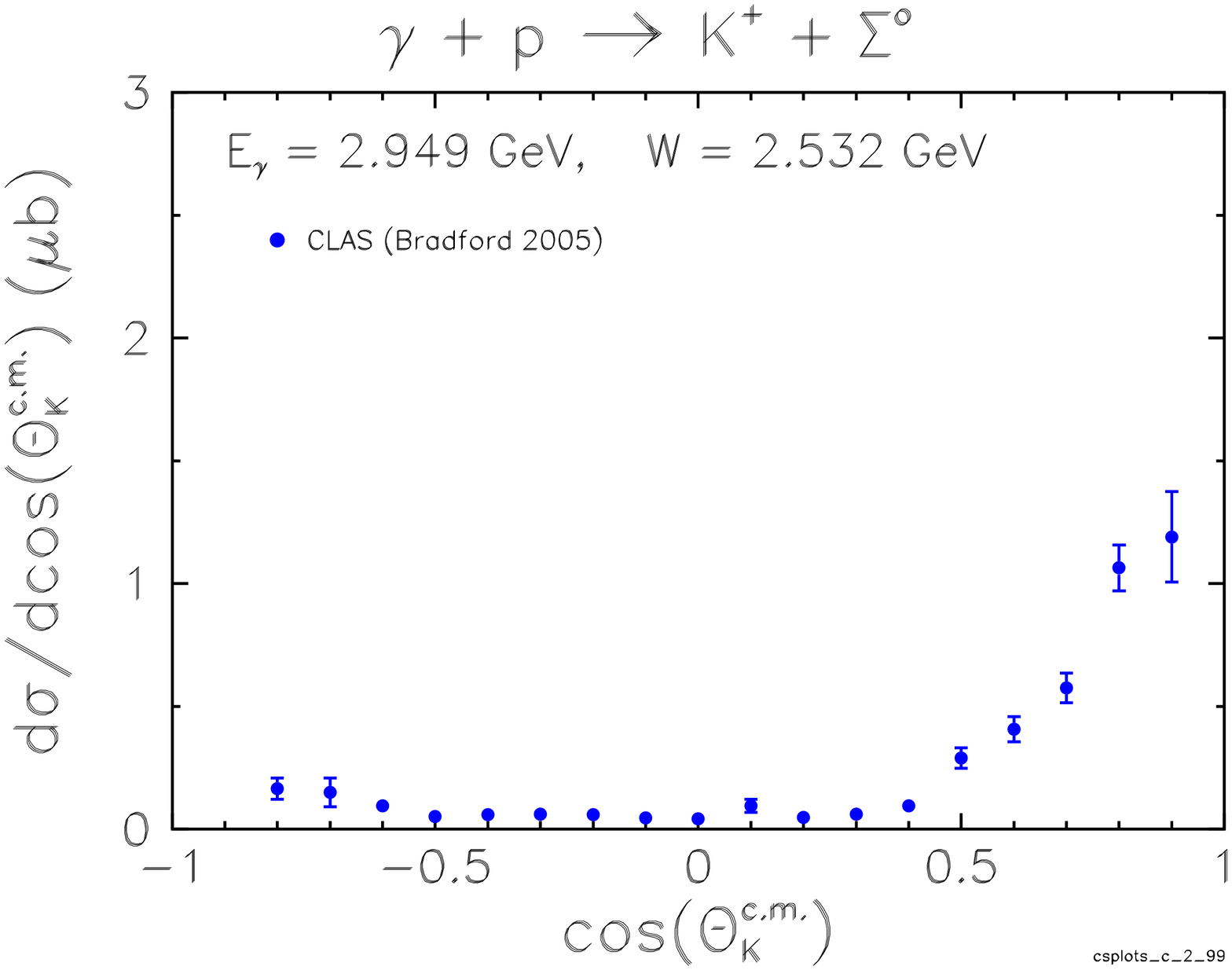}}
\caption{(Color online)
Angular distributions for $\gamma + p \rightarrow K^+ + \Sigma^0$ for
selected bins of total energy $W$.  The present CLAS results (blue
circles) are shown with statistical and yield-fit uncertainties.  Data
from {\small SAPHIR} (open stars~\cite{bonn2} and
triangles~\cite{bonn1}) and from older experiments~\cite{land} (black
squares) are also shown.  The curves are for effective Lagrangian
calculations computed by Kaon-MAID~\cite{maid} (solid red) and Janssen
{\it et al.}~\cite{jan} (dashed black), and a Regge-model calculation
of Guidal {\it et al.}~\cite{lag1,lag2} (dot-dashed blue).  }
\label{fig:dsdo_s_c2}       
\end{figure*}

The results for the angular distributions of photoproduction of
$\Sigma^0$ are shown in Figs.~\ref{fig:dsdo_s_c1} and
~\ref{fig:dsdo_s_c2}.  Again, the panels are selected to increase in
steps of about 80 MeV in $W$, also to allow comparison to previous
data.

\subsection{\label{sec:energy}$W$ Dependence}

\begin{figure*}
\vspace{-0.05in}
\resizebox{0.40\textwidth}{!}{\includegraphics{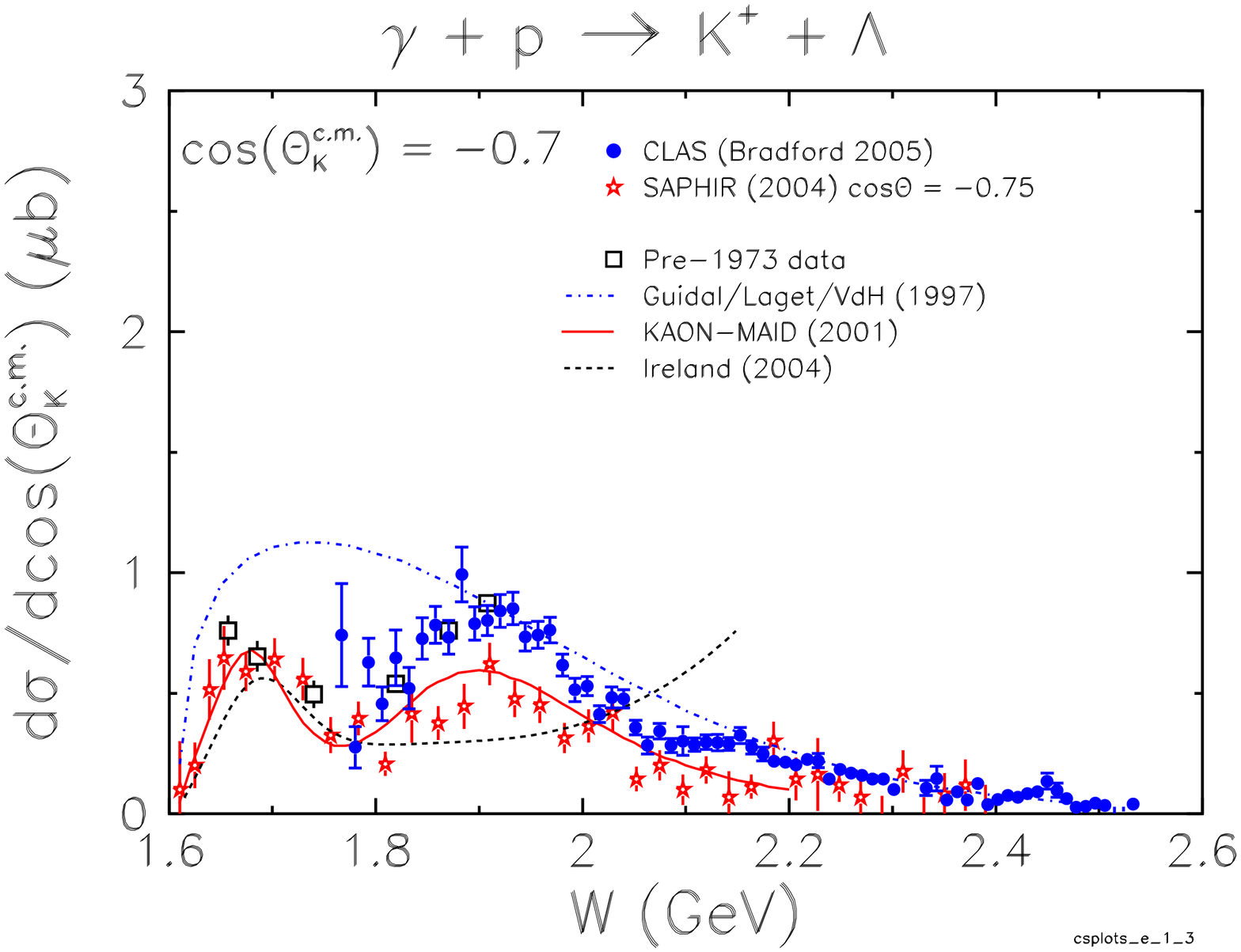}}
\vspace{-0.05in}
\resizebox{0.40\textwidth}{!}{\includegraphics{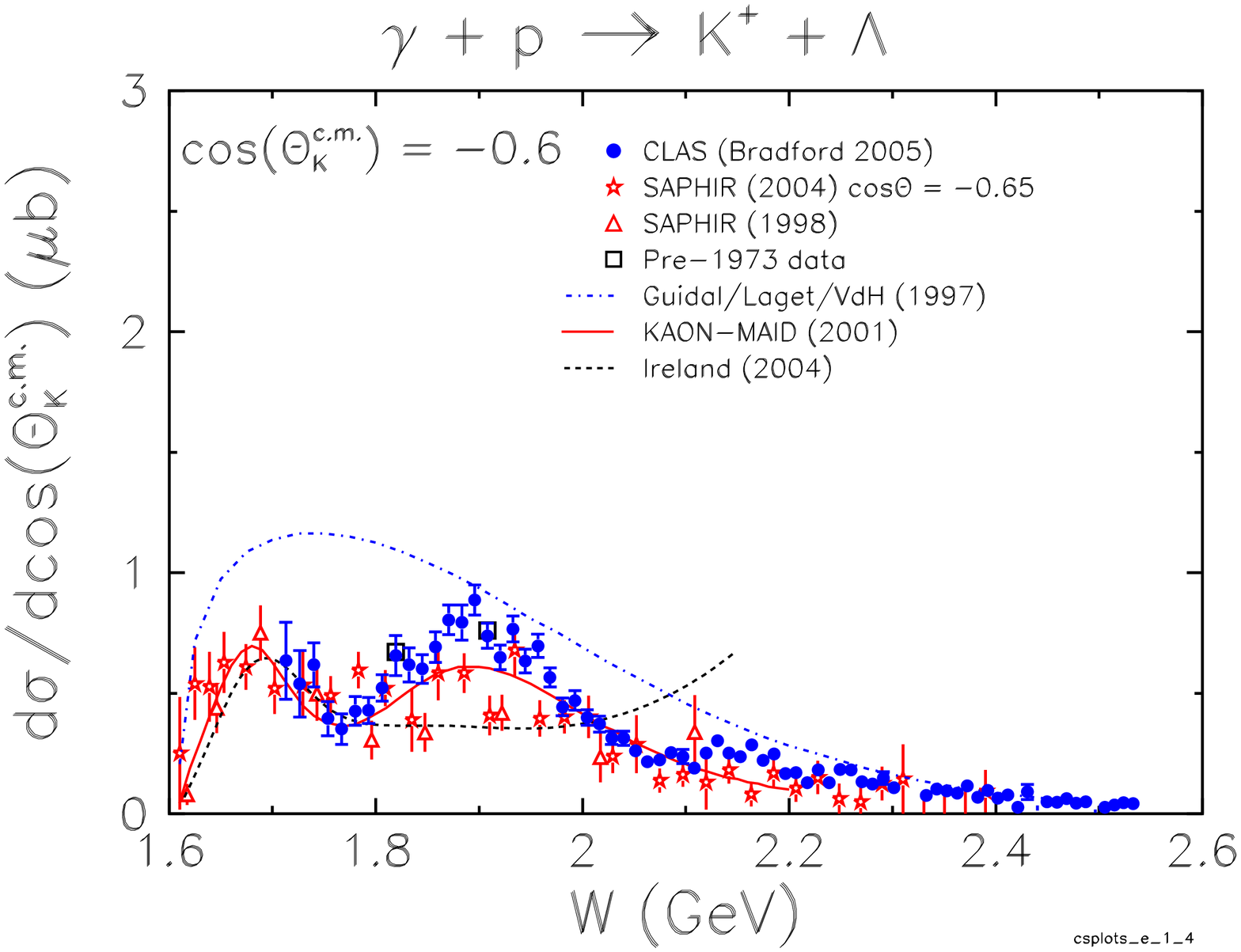}}
\vspace{-0.05in}
\resizebox{0.40\textwidth}{!}{\includegraphics{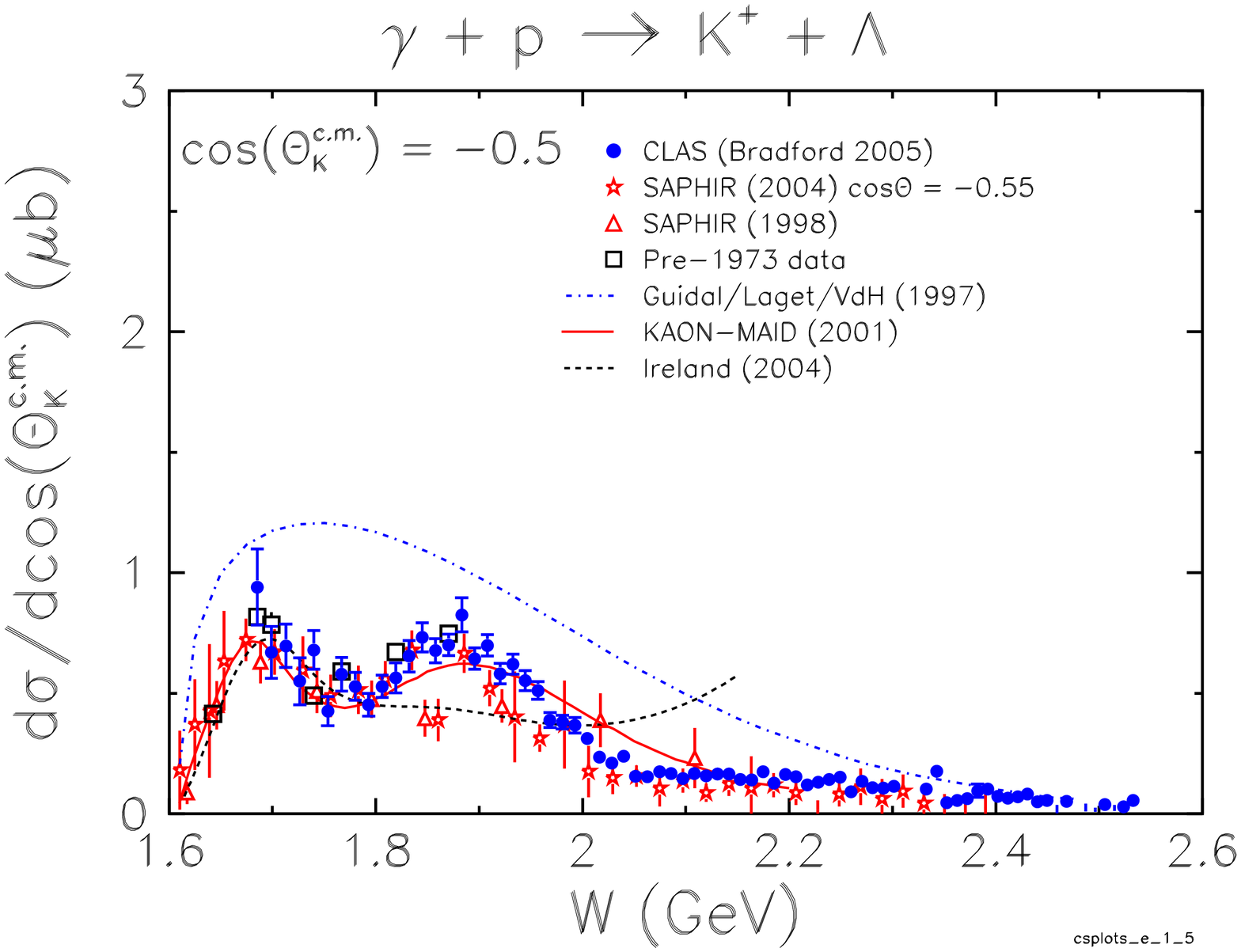}}
\vspace{-0.05in}
\resizebox{0.40\textwidth}{!}{\includegraphics{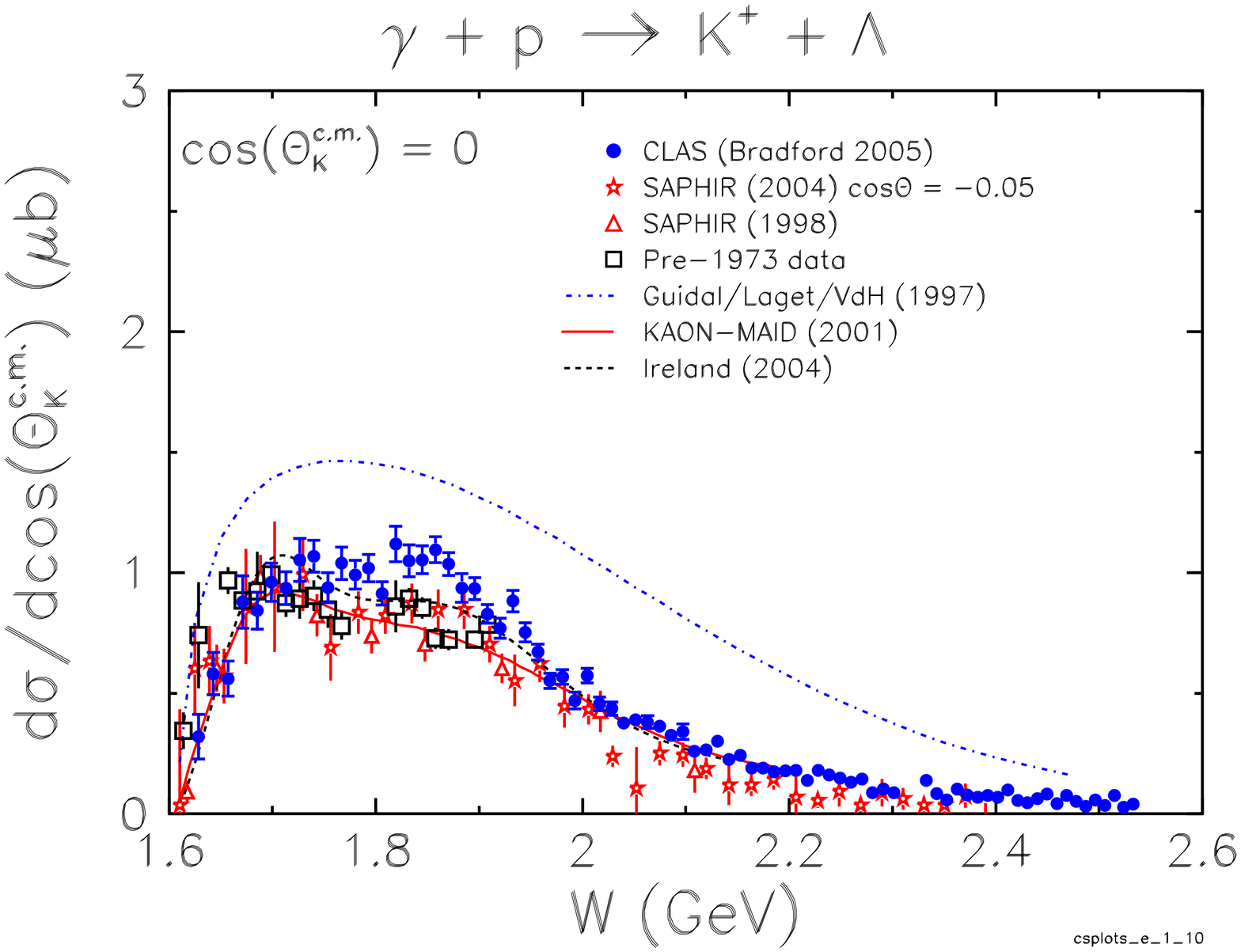}}
\vspace{-0.05in}
\resizebox{0.40\textwidth}{!}{\includegraphics{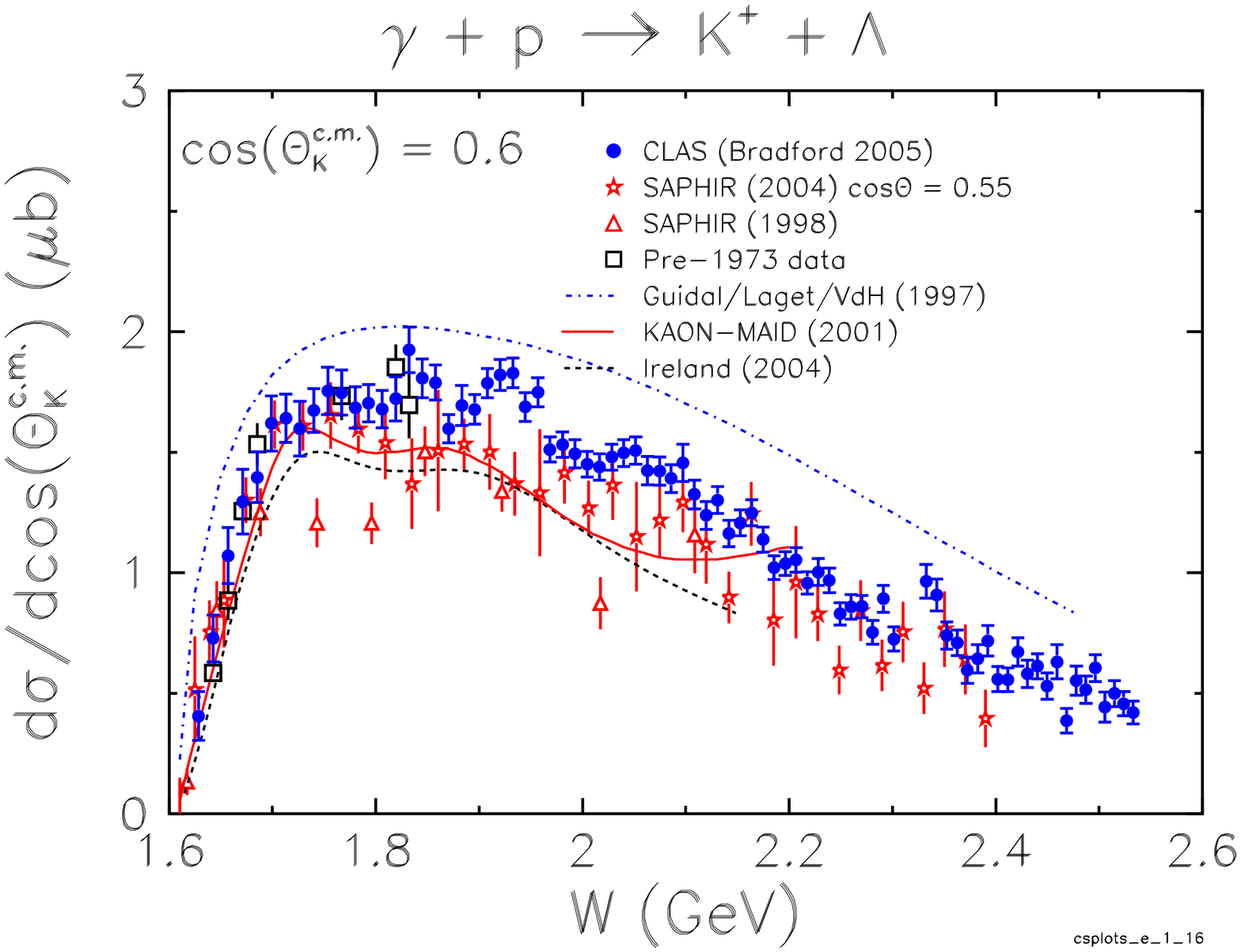}}
\vspace{-0.05in}
\resizebox{0.40\textwidth}{!}{\includegraphics{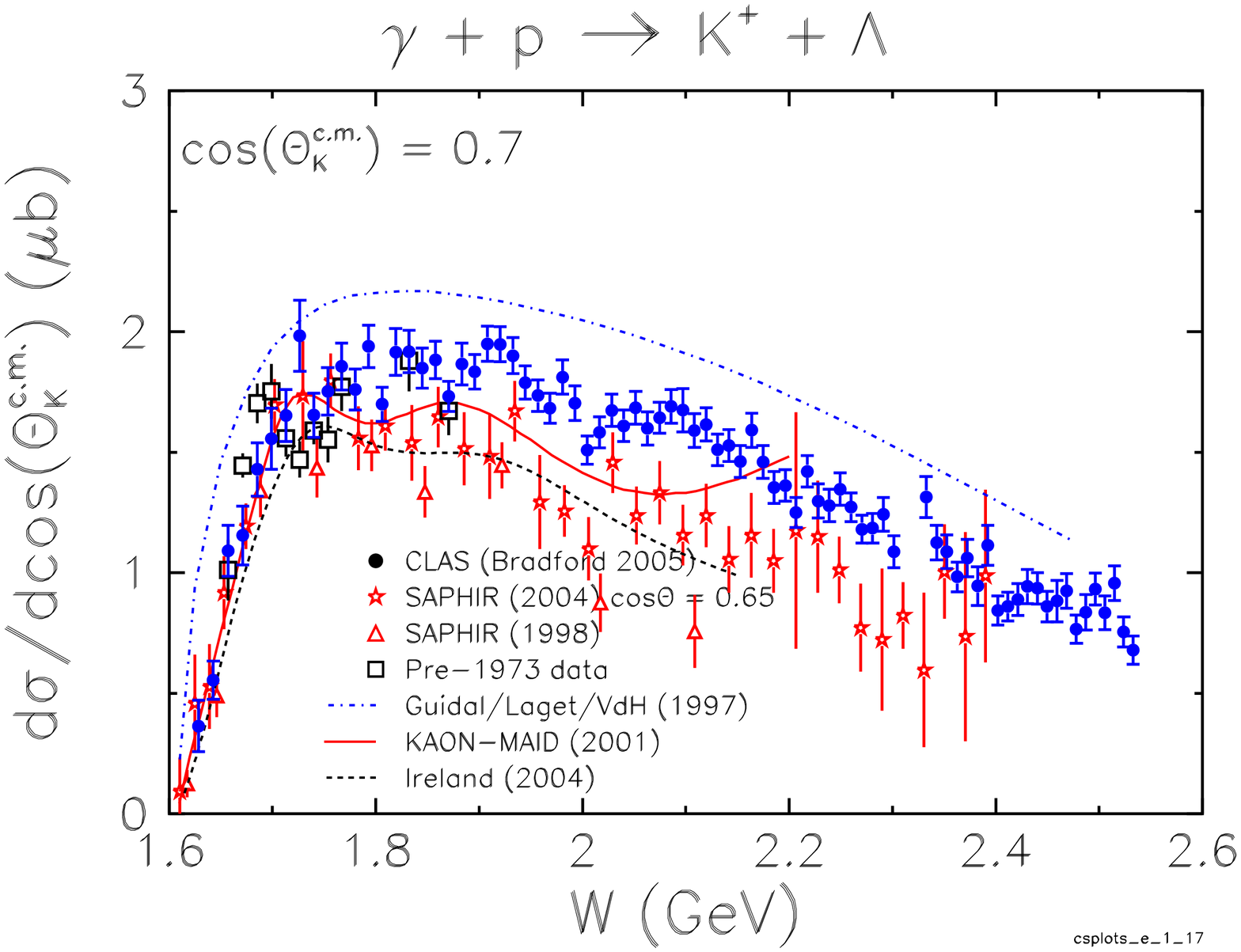}}
\vspace{-0.05in}
\resizebox{0.40\textwidth}{!}{\includegraphics{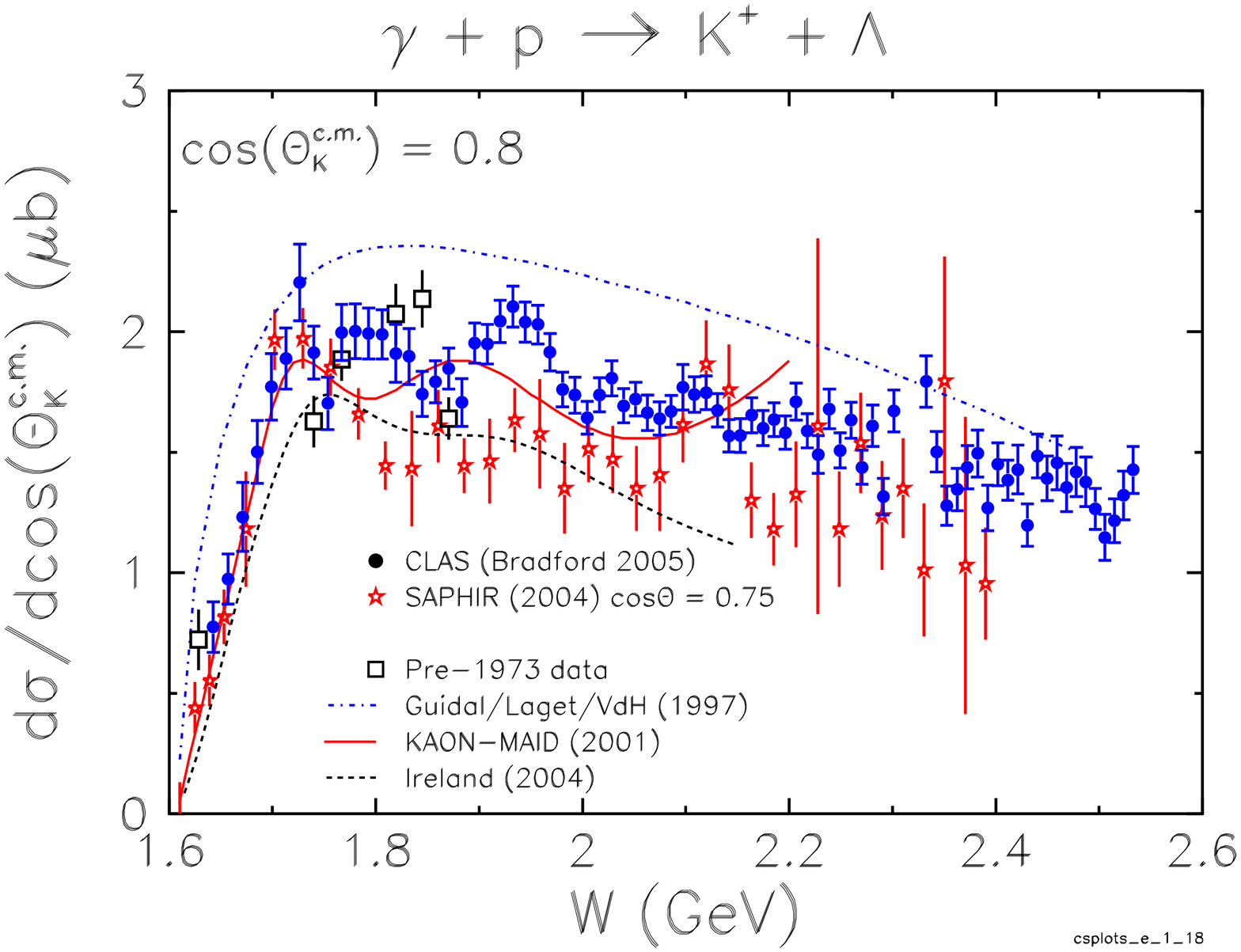}}
\vspace{-0.05in}
\resizebox{0.40\textwidth}{!}{\includegraphics{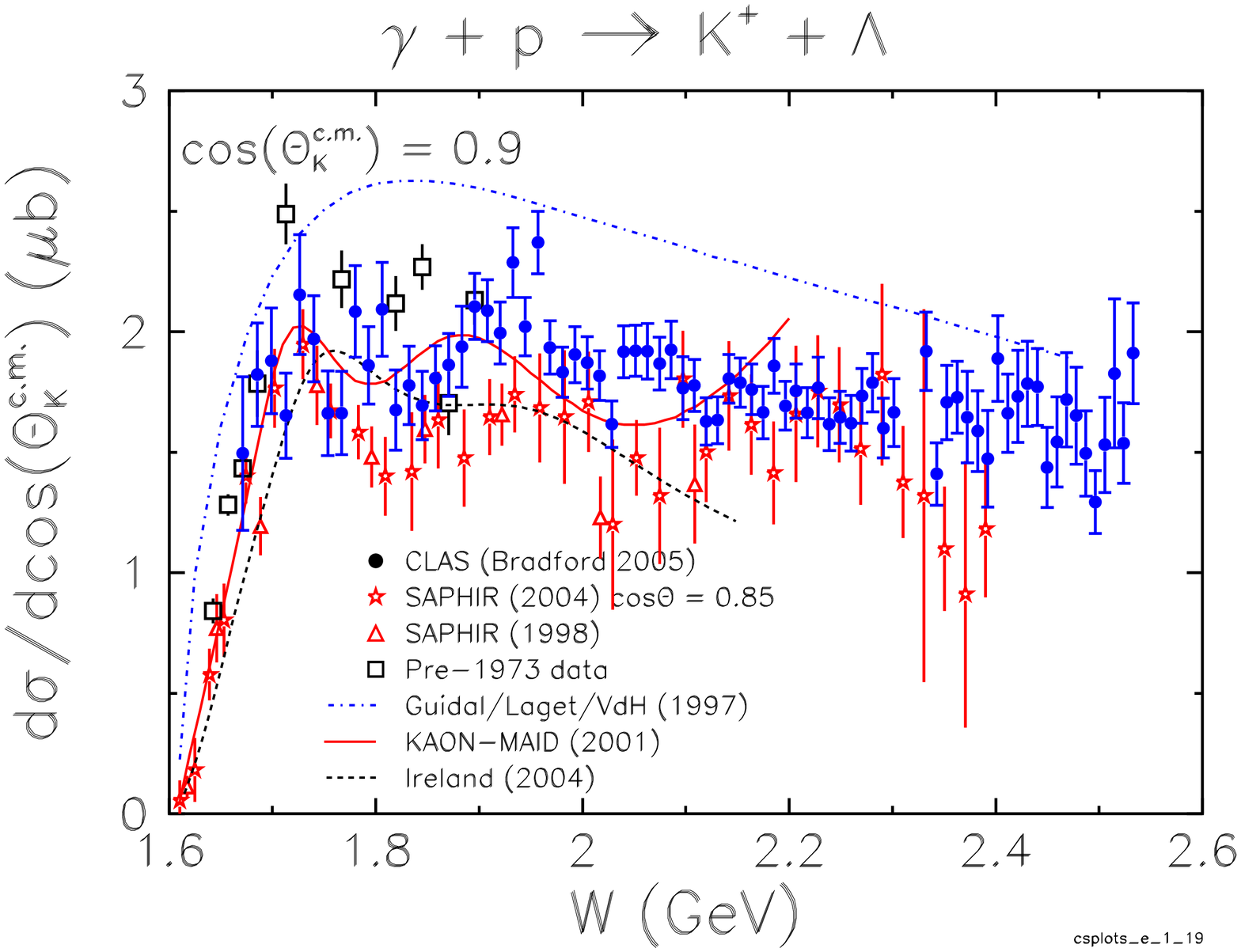}}
\caption{(Color online) 
Energy distributions for $\gamma + p
\rightarrow K^+ + \Lambda$ for selected c.m. kaon angles.  CLAS
results (blue circles) are shown with statistical and yield-fit
uncertainties.  Data from {\small SAPHIR} (open stars~\cite{bonn2},
triangles~\cite{bonn1}) and older experiments~\cite{land} (black
squares) are also shown.  The curves are for effective Lagrangian
calculations computed by Kaon-MAID~\cite{maid} (solid red) and Ireland
{\it et al.}~\cite{ireland} (dashed black), and a Regge-model
calculation of Guidal {\it et al.}~\cite{lag1,lag2} (dot-dashed blue).
}
\label{fig:dsdo_l_w}       
\end{figure*}

\begin{figure*}
\vspace{-0.05in}
\resizebox{0.40\textwidth}{!}{\includegraphics{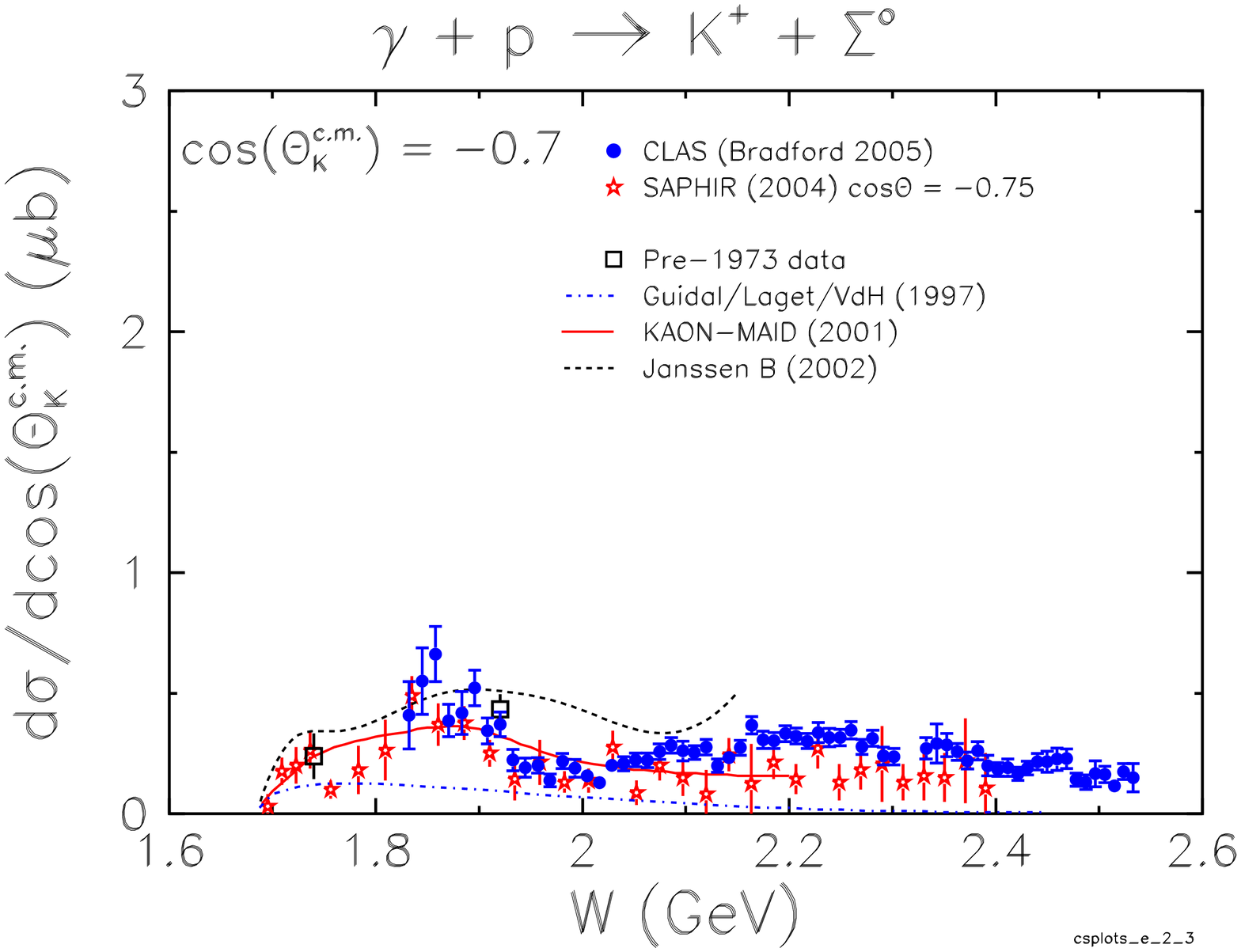}}
\vspace{-0.05in}
\resizebox{0.40\textwidth}{!}{\includegraphics{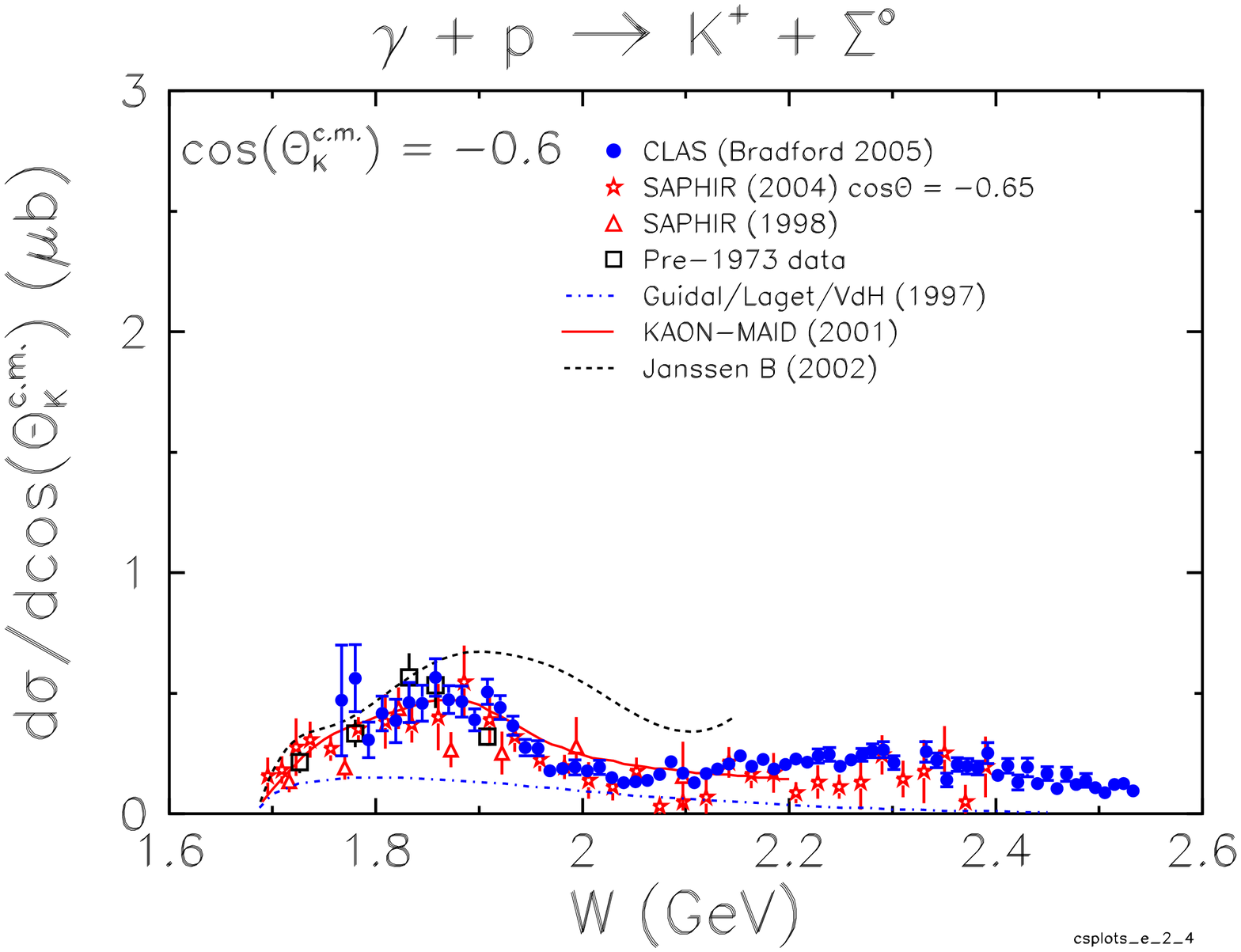}}
\vspace{-0.05in}
\resizebox{0.40\textwidth}{!}{\includegraphics{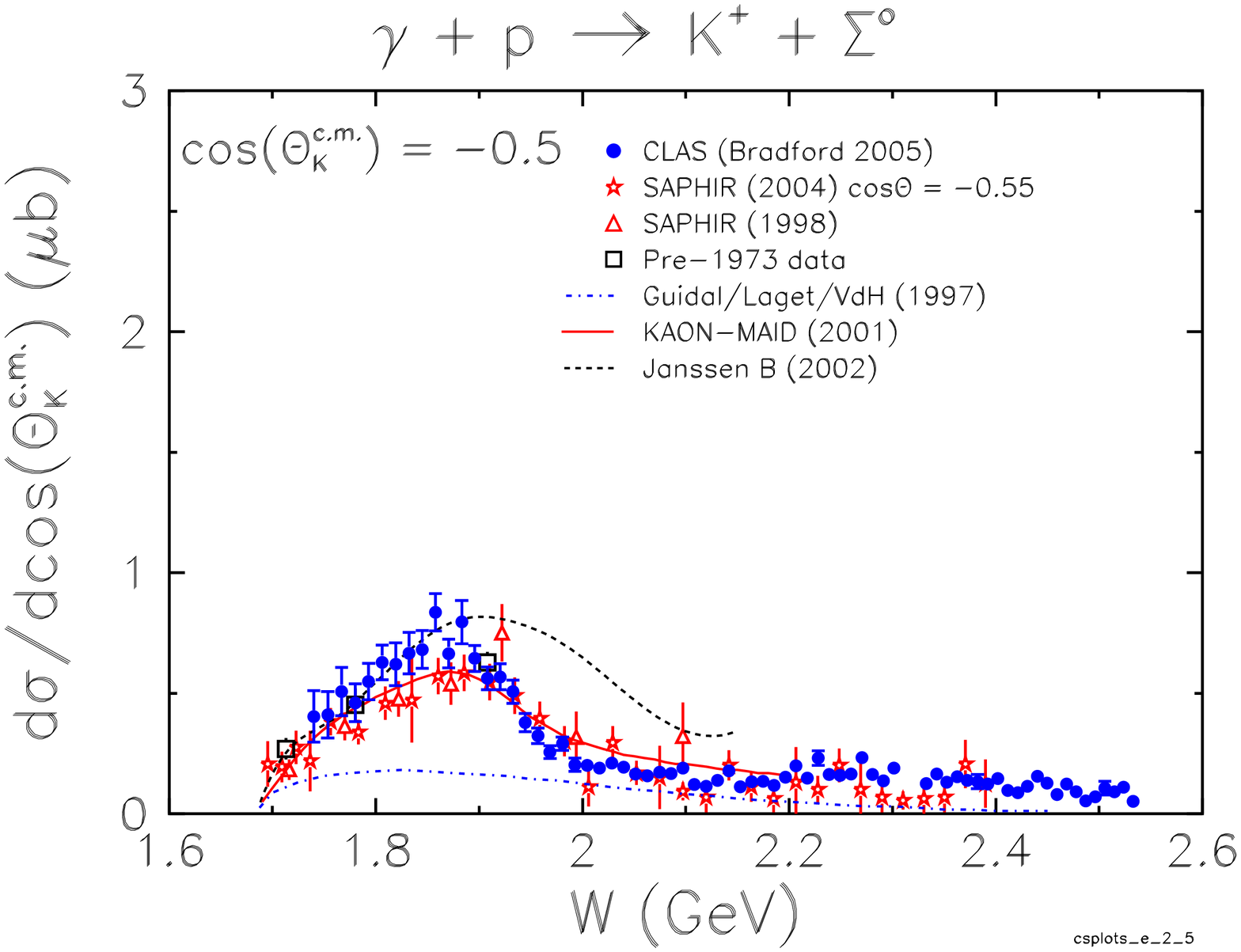}}
\vspace{-0.05in}
\resizebox{0.40\textwidth}{!}{\includegraphics{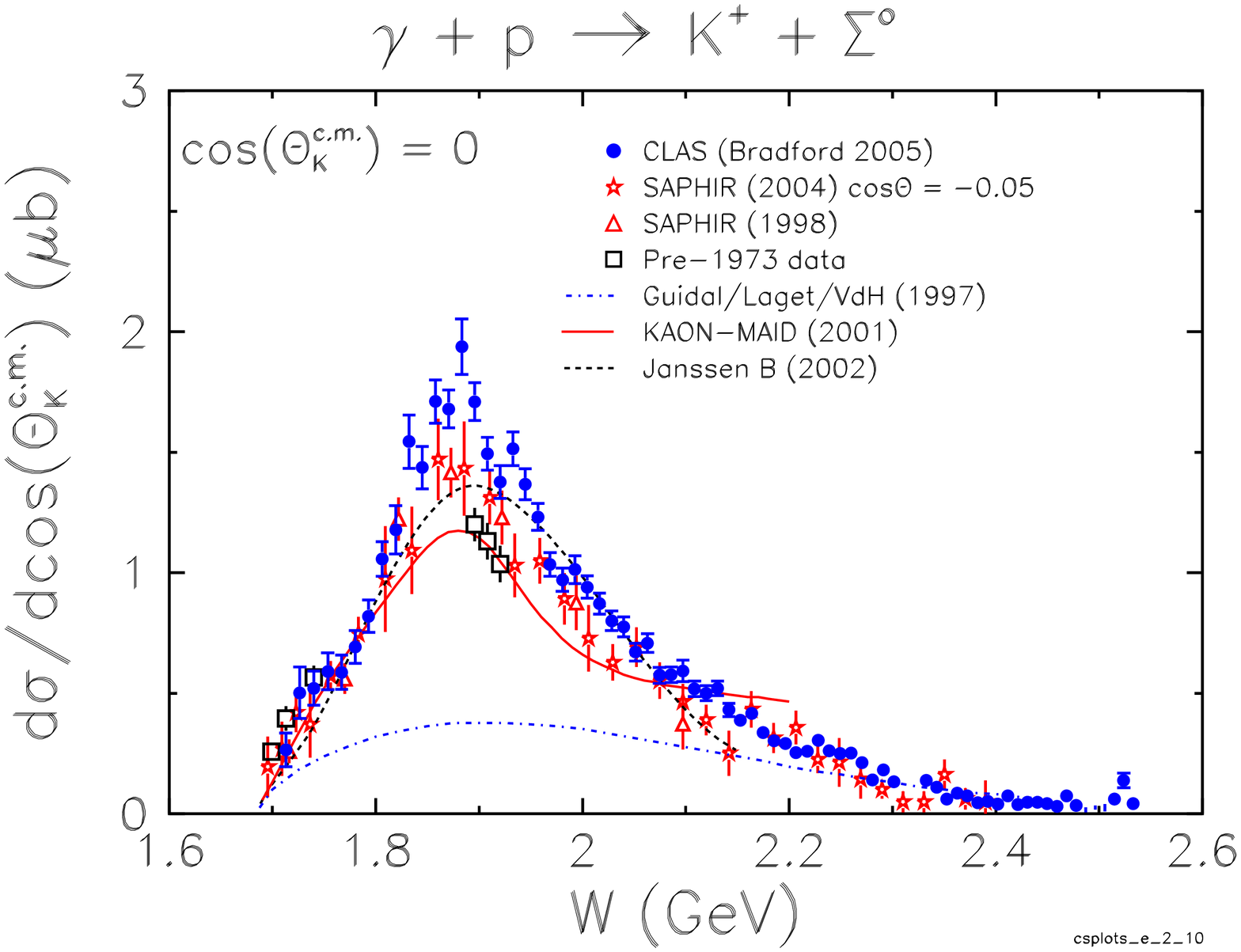}}
\vspace{-0.05in}
\resizebox{0.40\textwidth}{!}{\includegraphics{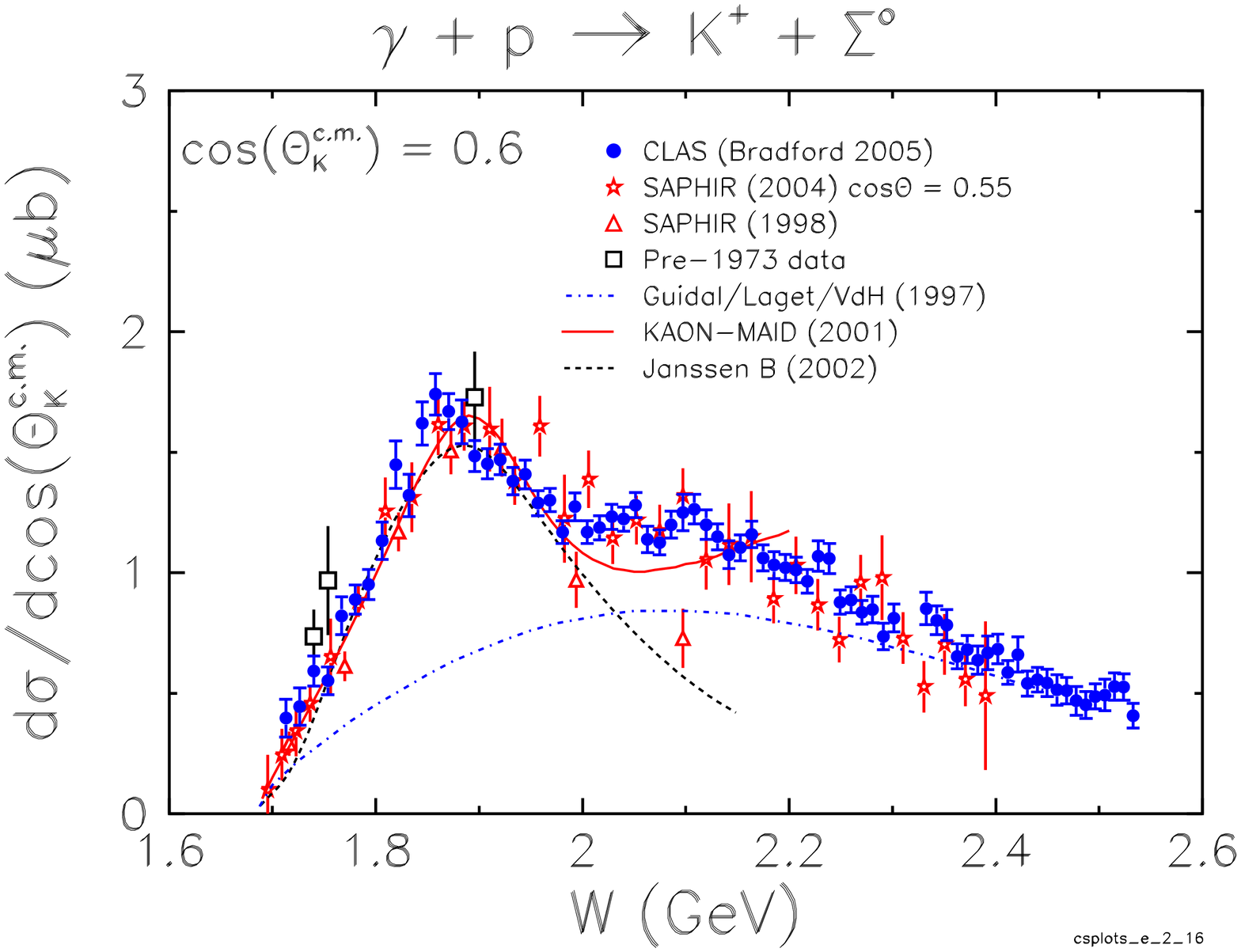}}
\vspace{-0.05in}
\resizebox{0.40\textwidth}{!}{\includegraphics{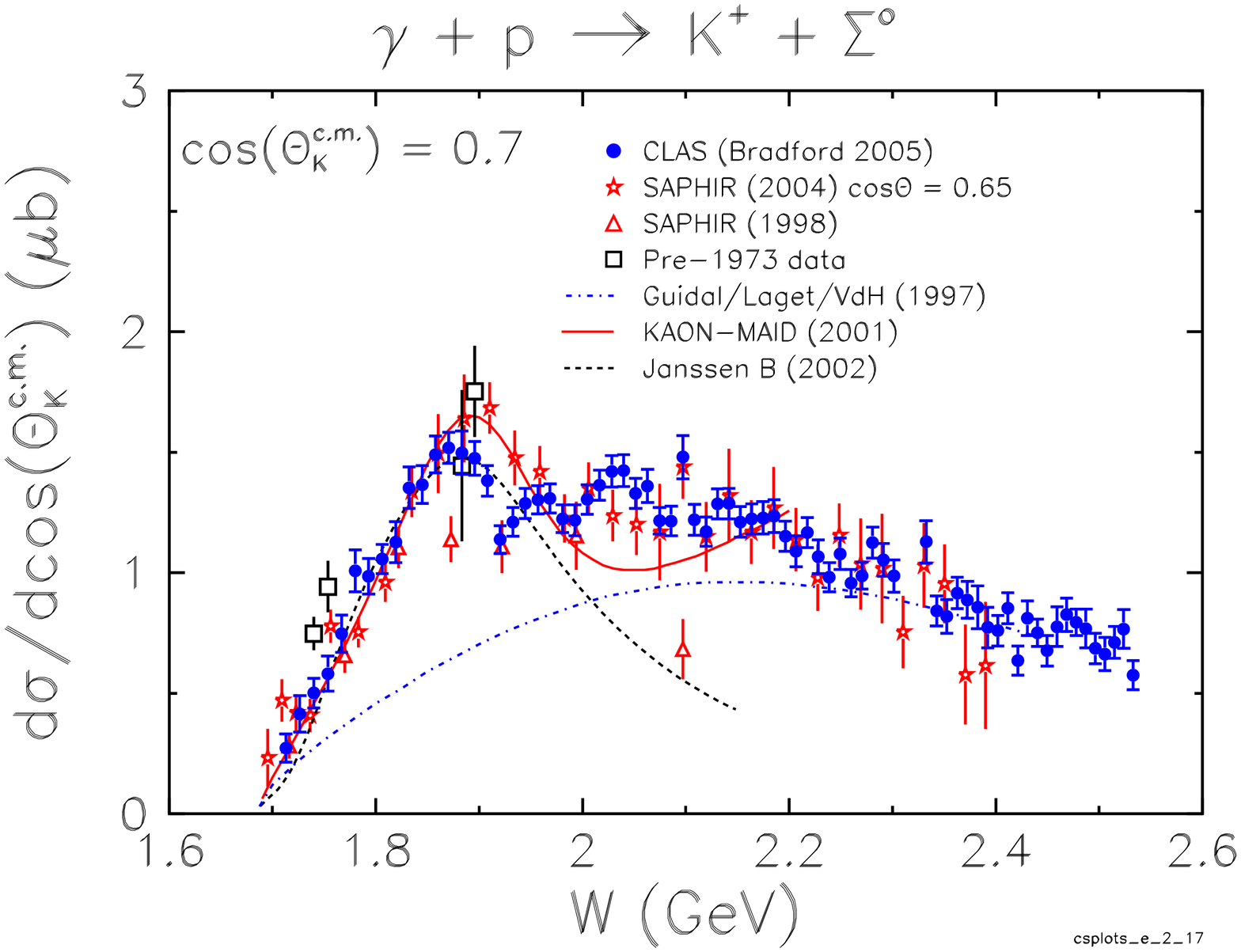}}
\vspace{-0.05in}
\resizebox{0.40\textwidth}{!}{\includegraphics{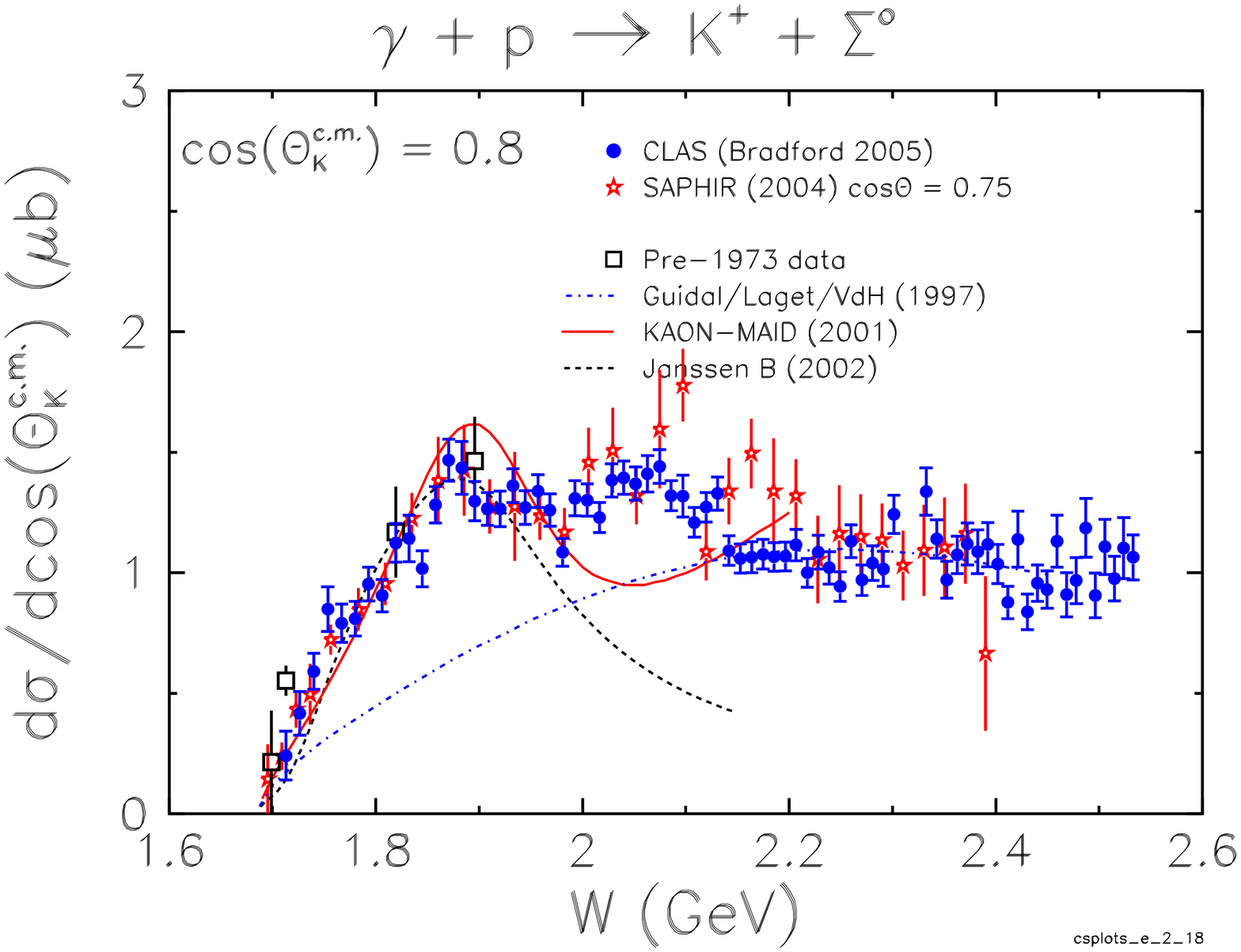}}
\vspace{-0.05in}
\resizebox{0.40\textwidth}{!}{\includegraphics{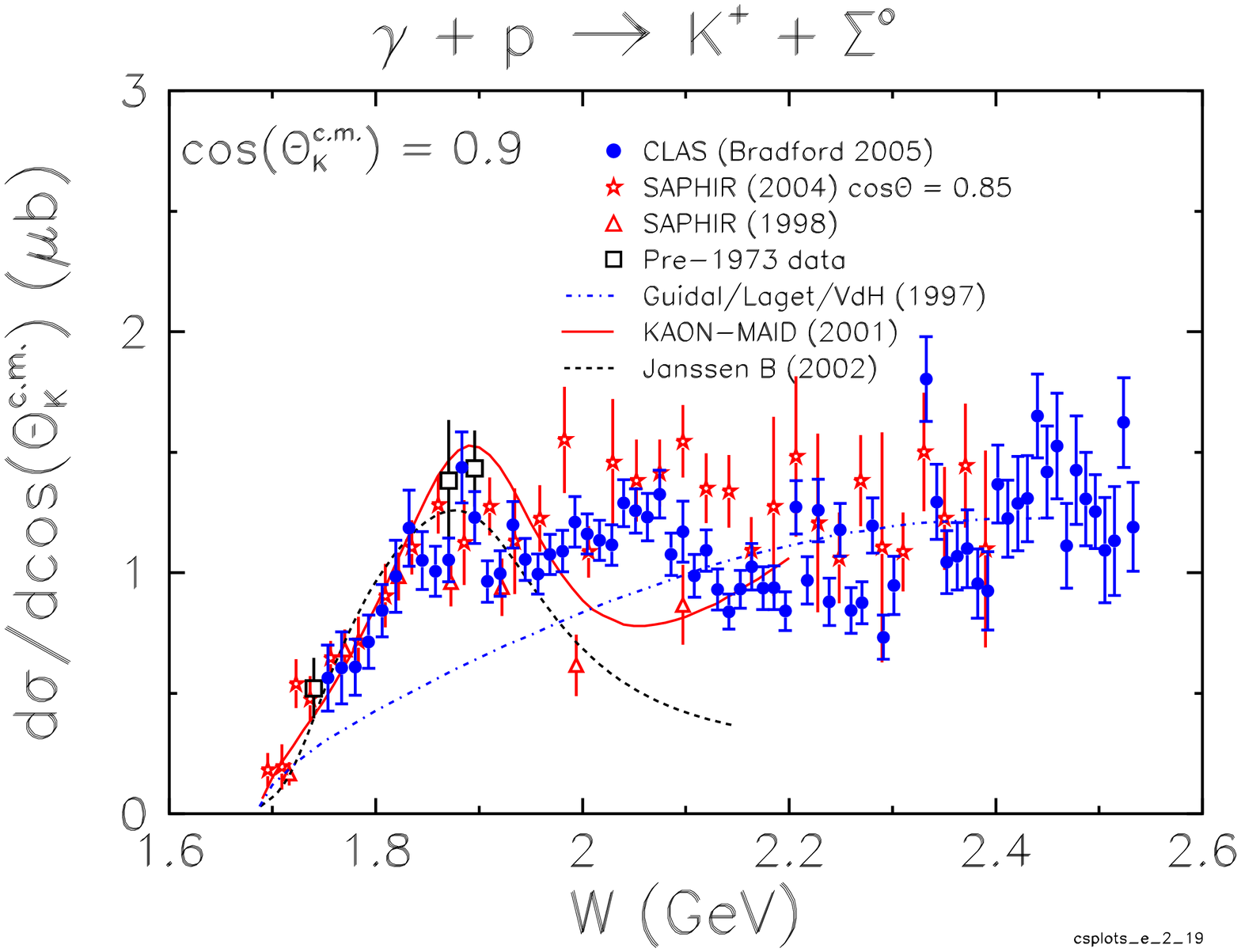}}

\caption{(Color online) 
Energy distributions for $\gamma + p
\rightarrow K^+ + \Sigma^0$ for selected c.m. kaon angles.  CLAS
results (blue circles) are shown with statistical and yield-fit
uncertainties. Data from {\small SAPHIR} (open stars~\cite{bonn2},
triangles~\cite{bonn1}) and older experiments~\cite{land} (black
squares) are also shown.  The curves are for effective Lagrangian
calculations computed by Kaon-MAID~\cite{maid} (solid red) and Janssen
{\it et al.}~\cite{jan} (dashed black), and a Regge-model calculation
of Guidal {\it et al.}~\cite{lag1,lag2} (dot-dashed blue).  }
\label{fig:dsdo_s_w}       
\end{figure*}

Resonance structure in the $s$-channel should appear most clearly in
the $W$ dependence of the cross sections.  In
Fig.~\ref{fig:dsdo_l_w} we show the $K^+ \Lambda$ cross section at
selected angles.  The corresponding information for the $K^+ \Sigma^0$
channel is shown in Fig.~\ref{fig:dsdo_s_w}.  We discuss these results in
the next section.

The full set of numerical results from this experiment are available
from various archival sources, including a Ph.D. thesis~\cite{bradford},
the CLAS online database~\cite{clasdb}, or private
communication~\cite{contact}.

\section{\label{sec:discussion}DISCUSSION}

\subsection{\label{sec:olddata}Comparison to Previous Data}

Figures \ref{fig:dsdo_l_c1},~\ref{fig:dsdo_l_c2},~\ref{fig:dsdo_s_c1},
and \ref{fig:dsdo_s_c2} show a sample of the differential cross
section for $\Lambda$ and $\Sigma^0$ hyperon photoproduction as a
function of angle for a set of $W$ values.  For comparison, we can
examine the previous large-acceptance experiment from {\small SAPHIR}
at Bonn~\cite{bonn1,bonn2}.  There is also a set of measurements that
was accumulated from the late 1950's to the early 1970's using
small-aperture magnetic spectrometers at
CalTech~\cite{brody,peck,groom,donoho}, Cornell~\cite{anderson,thom},
 Bonn~\cite{bleckman,feller}, Orsay~\cite{decamp}, DESY~\cite{going},
and Tokyo~\cite{fujii}.  These results are compiled, for example, in
Ref~\cite{land}.

The agreement with data from {\small SAPHIR} is fair or good, but
there are some discrepancies.  The CLAS results are generally more
precise, having statistical uncertainties that are about 1/4 as large,
with about twice as many energy bins.  The {\small SAPHIR} experiment
had better backward-angular coverage at low energies as well as
coverage at extreme forward angles where CLAS has an acceptance hole.
The measurements agree within the estimated uncertainties at some
angles and generally near threshold energies, but CLAS measures
consistently larger $K^+\Lambda$ cross sections at most kaon angles
and for $W > 1.75$ GeV.  This is discussed in more detail below in the
context of the total cross sections, where it appears that there is an
energy-independent scale factor of about 3/4 in going from the CLAS to
the {\small SAPHIR} $K^+\Lambda$ results.  The data for the
$K^+\Sigma^0$ channel are generally in better agreement overall: the
two experiments agree within their stated systematic uncertainties.

We collected the historic (pre-1973) results from different measurements
and plotted them together.  The error bars are taken as the quoted
random uncertainties, with no consideration of the quoted systematic
uncertainties.  While these early experiments did not span the large $W$ and
angular range of the recent experiments, they did make high-precision
measurements at selected kinematics.  There are 144 $K^+\Lambda$
points and 57 $K^+\Sigma^0$ points that, overall, are in fair
agreement with the CLAS results.  At backward angles the historic data
are in very good agreement with the present results from CLAS; at
forward angles the agreement is fair or good.  In the mid-range of
angles, the historic results are lower than our results, and more similar
to the {\small SAPHIR} data.

The fit coefficients presented in Figs.~\ref{fig:lamp} and
\ref{fig:samp} are in good qualitative agreement with results
published by {\small SAPHIR}, apart from an arbitrary overall change
in sign.  The CLAS results generally have finer binning and smaller
estimated uncertainties away from threshold.  However, our vertical
scales do not agree with {\small SAPHIR}, though it is clear their
units are incorrect as given, since they should be $\sqrt{\mu b}$.

Total cross sections, $\sigma_{tot}$, for $\gamma + p \rightarrow K^+
+ \Lambda$ and $\gamma + p \rightarrow K^+ + \Sigma^0$ can be
calculated from the integrated angular distributions.  There is some
danger in the integration procedure since (i) it requires some model
of the reactions which may bias the resulting fit, and (ii) in the
absence of complete angular coverage there is also the problem of
extrapolating the fit into the unmeasured section of phase space.  Our
procedure for extracting and calculating the total cross sections was
based on fitting $d\sigma/d\cos(\theta_{K^+}^{c.m.})$ in two ways:
using Eq.~\ref{eq:mag} to fit the magnitude directly, and Eq.~\ref{eq:amp}
to fit the partial wave amplitudes. In the magnitude fit, one of the
coefficients directly gives $\sigma_{tot}$ and its associated error.
In the amplitude fit $\sigma_{tot}$ is easily computed from the set of
fit parameters, but the error is difficult to extract since the fit
parameters and their errors are correlated.  We estimated the
systematic bias in our integrations by taking the standard deviation
of the two resultant values as an additional uncertainty, and this was
added in quadrature to the other estimated uncertainties.

\begin{figure*}
\resizebox{0.7\textwidth}{!}{\includegraphics{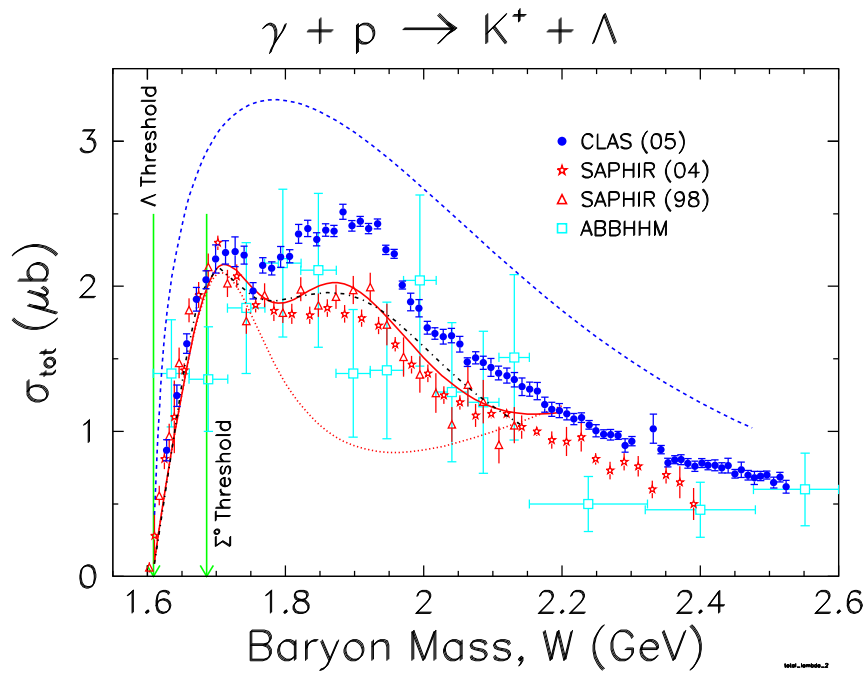}}
\caption{(Color online) 
Total cross section for $\gamma + p \rightarrow K^+ + \Lambda$.  The
data from CLAS (blue circles) are shown with combined statistical and
fitting uncertainties.  Also shown are results from two publications
from {\small SAPHIR} (red stars (2004)~\cite{bonn2}, red triangles
(1998)~\cite{bonn1}), and the ABBHHM Collaboration (light blue
squares)~\cite{erbe}.  The curves are from a Regge model (dashed
blue)~\cite{lag1,lag2}, Kaon-MAID (solid red)~\cite{maid}, Kaon-MAID
with the $D_{13}(1895)$ turned off (dotted red), and Saghai {\it et al.}
(dot-dashed black)~\cite{sag}.
}
\label{fig:ltot}       
\end{figure*}

\begin{figure*}
\resizebox{0.7\textwidth}{!}{\includegraphics{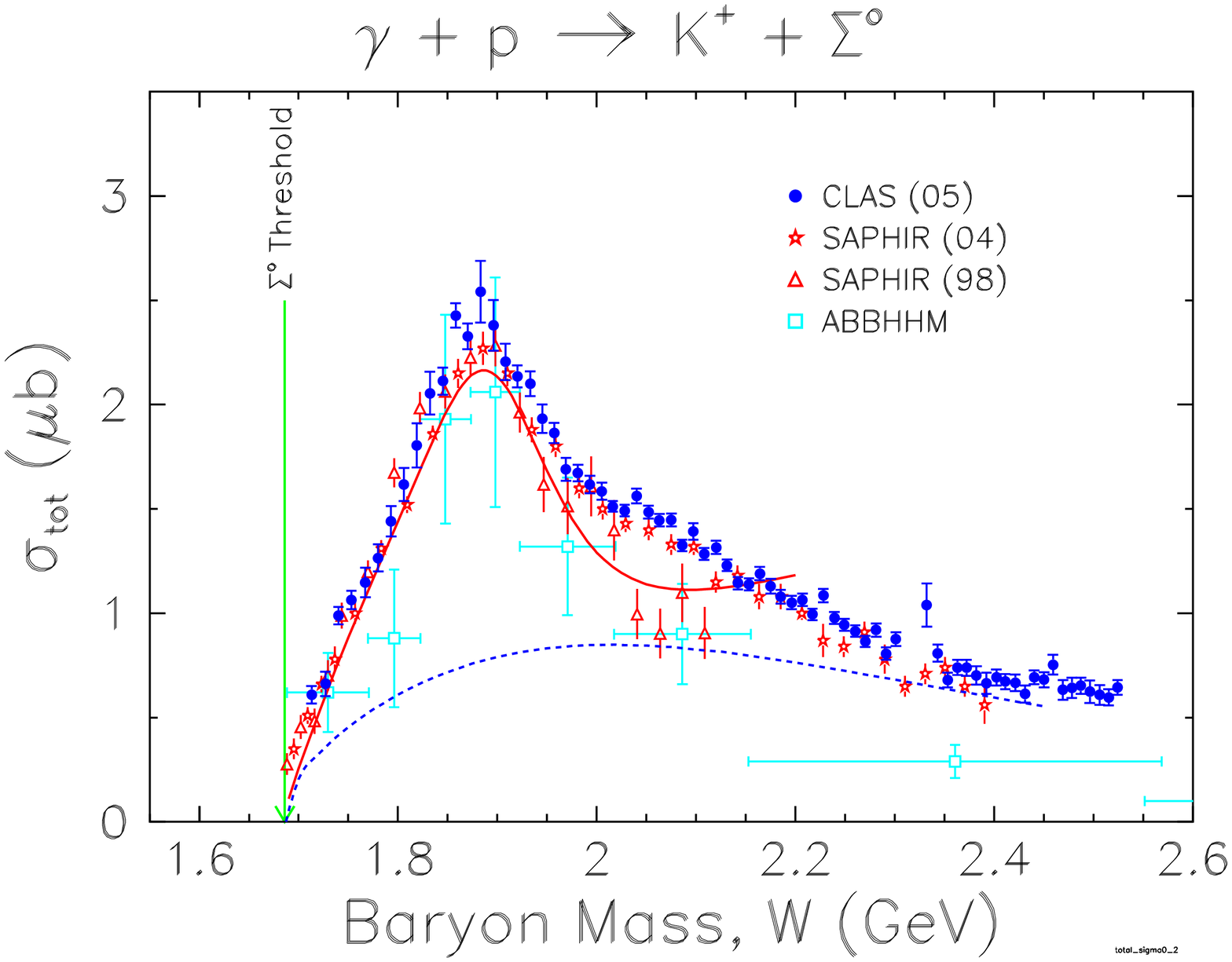}}
\caption{(Color online) 
Total cross section for $\gamma + p \rightarrow K^+ + \Sigma^0$.  The
data are from CLAS (blue circles) are shown with combined statistical
and fitting uncertainties, Also shown are results from two publications
from {\small SAPHIR} (red stars (2004)~\cite{bonn2}, red triangles
(1998)~\cite{bonn1}), and the ABBHHM Collaboration (light blue
squares)~\cite{erbe}.  The curves are from a Regge model (dashed
blue)~\cite{lag1,lag2} and Kaon-MAID (solid red)~\cite{maid}.
}
\label{fig:stot}       
\end{figure*}

The total cross section results are shown in Figs.~\ref{fig:ltot} and
~\ref{fig:stot}.  The error bars combine statistical and estimated
systematic uncertainty due to the fitting procedure.  The gaps in the
spectra at $W=2.375$ and $2.400$ GeV stem from photon tagger failures
at those energies.  For comparison we show two previously published
data sets from Bonn~\cite{bonn1,bonn2}~\cite{bonnnote}.  Also, bubble chamber
data for the total cross sections came from Erbe {\it et al.}
(ABBHHM)~\cite{erbe}.  Also shown are model curves for two
calculations, the effective Lagrangian model embodied in
Kaon-MAID~\cite{maid} and the Regge model of Guidal, Laget, and
Vanderhaeghen~\cite{lag1,lag2}.  The CLAS results for $\sigma_{tot}$
differ from the Bonn results in an unexpected way, namely that the
Bonn $K^+\Lambda$ cross section is smaller than the CLAS result by a
factor of close to $3/4$. This is in contrast to the $K^+\Sigma^0$
results, where the CLAS and the Bonn results are in good agreement:
the values of $\sigma_{tot}$ agree well within their quoted systematic
uncertainties.  We note that the CLAS results for the two hyperons
used exactly the same photon normalizations, and that the hyperon
yield extractions for both cases were made together, as discussed
above.  The acceptance calculations for the CLAS results used the same
software as well, differing only in the input events used for the
calculations.  In short, we have not found any reason within the CLAS
analysis for one channel agreeing well with previous work and the
other not.  Both results are consistent with the ABBHHM data~\cite{erbe}.


The CLAS $\sigma_{tot}$ results for $K^+\Lambda$ show a prominent
peak centered near 1.9 GeV.  It does not resemble a simple single
Lorentzian, reflective of the expectation that several resonant
structures are present in this mass range.  The peak near 1.7 GeV is
consistent with contributions from the $P_{11}(1710)$ and
$P_{13}(1720)$.  In the case of $K^+\Sigma^0$, the $\sigma_{tot}$
curve shows the previously seen strong peak centered at 1.88 GeV, and
in addition there is a slight shoulder at about 2.05 GeV.  The
location of the strong peak is consistent with the mass of several
well-established $\Delta$ resonances which may contribute to an
isospin 3/2 final state.

\subsection{\label{sec:compare}Comparison to Reaction Models}

The model calculations shown in this paper were not fitted to the
present results.  The effective Lagrangian calculations, in
particular, were fitted to the previous data shown in this paper, and
have, therefore, at least fair agreement with those earlier results.
However, since in the case of $K^+\Lambda$ production we have some
disagreement with the {\small SAPHIR} data in the mid-range of angles,
we cannot expect these calculations to be in quantitative agreement
with us.  It is nevertheless interesting to see what the more copious
CLAS results seem to indicate in comparison to a few of these previous
models.

The Regge-model calculation~\cite{lag1,lag2} shown in the preceding
figures uses only $K$ and $K^*$ exchanges, with no $s$-channel
resonances.  The model was constructed to fit high-energy kaon
photoproduction data~\cite{slac}, for $W$ between 5 and 16 GeV, and
may be expected to reproduce the average behavior of the cross section
in the nucleon resonance region.  However, extrapolated down to the
resonance region, the model overpredicts the size of the $\Lambda$
cross section and underpredicts that of the $\Sigma^0$.  This is
evident in all the graphs, but is especially easily seen in the total
cross sections, Figs.~\ref{fig:ltot} and \ref{fig:stot}.  Since it is
a pure $t$-channel reaction model, it cannot produce a rise at back
angles as seen for the $\Lambda$, and illustrates the need for $s$-
and $u$-channel contributions to understand that feature.

Two hadrodynamic models \cite{mart, jan} based on similar effective
Lagrangian approaches are also shown. Both emphasize the addition of a
small set of $s$-channel resonances to the non-resonant Born terms,
and differ in their treatment of hadronic form factors and gauge
invariance restoration. As both were fitted to the previous data from
{\small SAPHIR}~\cite{bonn2}, they are expected to be in somewhat
poorer agreement with our $K^+ \Lambda$ than our with $K^+ \Sigma^{0}$ data.

Both models contain a set of known $s$-channel $N^*$ resonances:
$S_{11}(1650)$, $P_{11}(1710)$, and $P_{13}(1720)$. The model of Mart
\emph{et al.}~\cite{mart} which is used in the Kaon-MAID calculations
contains an additional $D_{13}$(1895) resonance in its $K^+ \Lambda$
description.  In the $K^+ \Lambda$ case, the calculations of Ireland
\emph{et al.}~\cite{ireland} are shown since they represent an update
of the earlier work of Janssen \emph{et al.}~\cite{jan}. These
calculations included photon beam asymmetry~\cite{zegers} and
electroproduction~\cite{mohring} data points in the dataset used for
fitting. The curves displayed on Figs.~\ref{fig:dsdo_l_c1},
~\ref{fig:dsdo_l_c2}, and~\ref{fig:dsdo_l_w} contain the set of known
resonances plus an additional $P_{11}$(1895) resonance. This
combination was found to give the best quantitative agreement with the
dataset used for fitting. The analysis of Ref.~\cite{ireland} was
restricted to a study of the $K^+ \Lambda$ channel, so for comparison
with the present $K^+ \Sigma^{0}$ data, we use slightly older
calculations~\cite{jan} which contain an additional $D_{13}$(1895).

The CLAS $K^+\Lambda$ results, which show a structure that varies in
width and position with kaon angle, suggests an interference
phenomenon between several resonant states in this mass range, rather
than a single, well-separated resonance. This should be expected,
since several $N^*$ resonances with one- and two-star PDG ratings
occupy this mass range. From Fig.~\ref{fig:dsdo_l_w}, the best
qualitative modeling of the structure near 1.9 GeV at backward angles
is given by Kaon-MAID~\cite{maid}, but the model seems to diverge from
the trends of the data at forward angles. The calculation of
Ref.~\cite{ireland} gives a poor description of the data in the 1.9
GeV region at backward angles, but at forward angles it is similar to
the Kaon-MAID calculation.  Using the model curves as a guide, we see
that a fixed position for a single isolated resonance near 1.9 GeV is
not consistent with the small ($\sim 50$ MeV) variation with angle of
the feature seen in the cross sections.

In the $\Sigma^0$ case there is some indication of a structure above the
large peak at 1.9 GeV between 2.0 and 2.1 GeV.  This shoulder or small
bump in the cross section, seen in Fig.~\ref{fig:dsdo_s_w} and in the total
cross section Fig.~\ref{fig:stot}, is not reproduced by either of the
hadrodynamic reaction models.

\subsection{\label{sec:scaling}Phenomenological $t$-Scaling}

The forward peaking of the $K^+\Lambda$ cross section suggests that
there is substantial contribution to the reaction mechanism by
$t$-channel exchange, even in the nucleon resonance region.  To test
this idea, the data can be cast into the form
of $d\sigma/dt$ vs. $-t$, where $t$ is the Mandelstam invariant
that gives the 4-momentum squared of the kaonic exchange
particle(s). The conversion of the cross section was done using 

\begin{equation}
\frac{d\sigma}{dt} = \frac{d\sigma}{d\cos \theta_{K^+}^{c.m.}} \times 
\frac{1}{2 k q}
\end{equation}

\noindent
where $k$ is the center of mass momentum of the incoming photon and
$q$ is the center of mass momentum of the produced kaon.  In the
simplest Regge picture involving the exchange of a single trajectory,
the cross section can be written as~\cite{perkins}

\begin{equation}
\frac{d\sigma}{dt} = D(t)\left(\frac{s}{s_0}\right)^{2\alpha(t) -2} 
\end{equation}

\noindent
where $D(t)$ is a function of $t$ only, $s_0$ is a baryonic scale
factor taken to be 1 GeV$^2$, and $\alpha(t)$ is the Regge trajectory
itself that describes how the angular momentum of the exchange varies
with $t$. At our kinematics for small $|t|$ we find $\alpha(t) \approx
0$, so the leading behavior of the cross section is that it
approximately scales with $s^2$.

\begin{figure*}
\resizebox{0.7\textwidth}{!}{\includegraphics{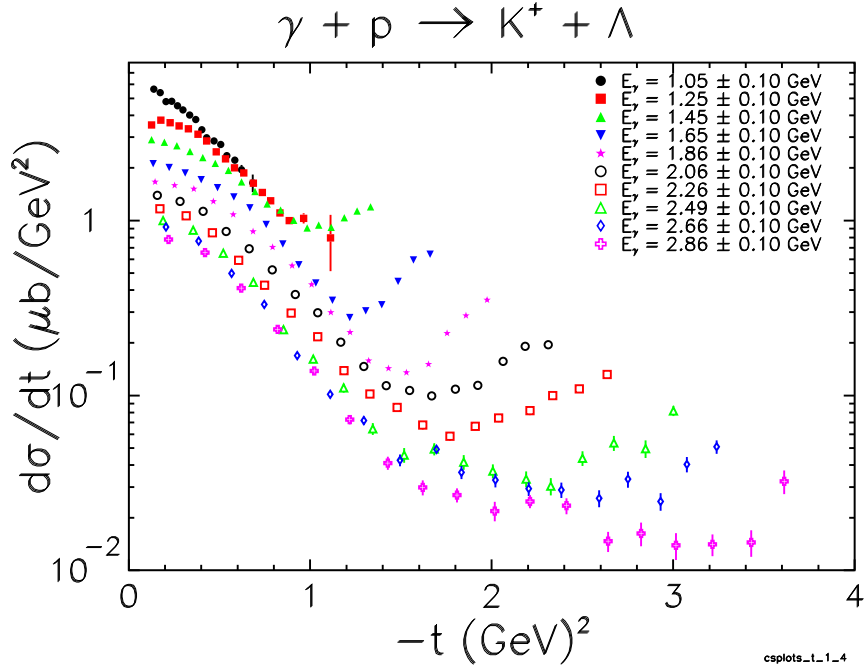}}
\caption{(Color online)
The entire $\gamma + p \rightarrow K^+ + \Lambda$ data set shown as
$d\sigma/dt$ vs. $-t$, for ten bands of photon energy with $\Delta
E_\gamma = 0.20$ GeV.  No scaling was applied.
}
\label{fig:dsdt_l1}       
\end{figure*}

\begin{figure*}
\resizebox{0.7\textwidth}{!}{\includegraphics{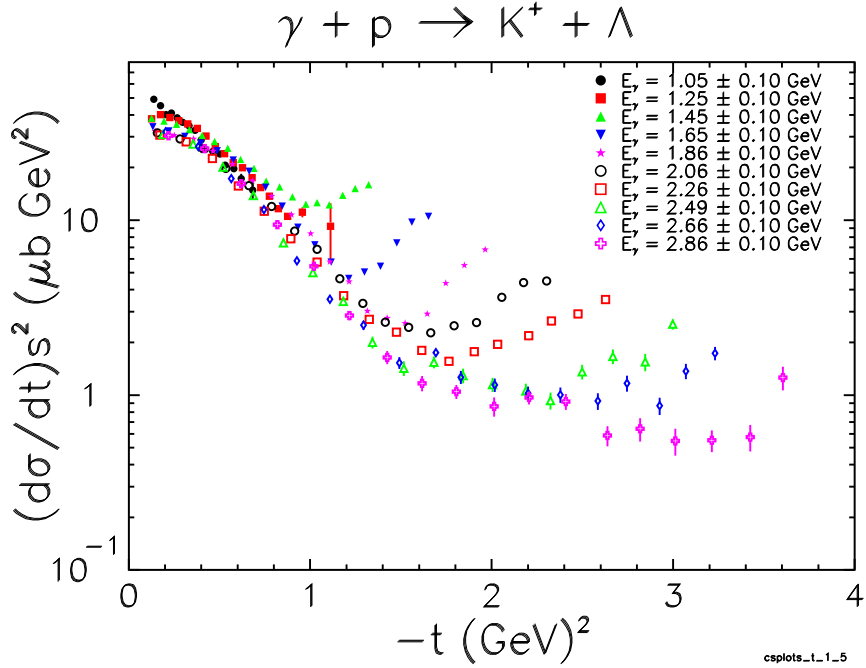}}
\caption{(Color online)
The entire $\gamma + p \rightarrow K^+ + \Lambda$ data set shown as
$d\sigma/dt$ vs. $-t$, for ten bands of photon energy with $\Delta
E_\gamma = 0.20$ GeV.  The cross sections were scaled by $s^2 = W^4$,
resulting in a well-defined band of data for low $-t$ values.
}
\label{fig:dsdt_l2}       
\end{figure*}

The cross section $d\sigma/dt$ for $K^+\Lambda$ production is plotted
in Fig.~\ref{fig:dsdt_l1}.  To obtain sufficient statistical precision,
bands of width 200 MeV were combined as weighted averages (amounting
to groups of 8 of our actual bins).  The lowest band, for $E_\gamma =
1.05\pm0.10$ GeV, starts 40 MeV above the reaction threshold.  We
observe in the figure how the cross section values fall on
smoothly-varying contours as a function of $-t$.  There is an
inflected fall-off from the minimum $-t$ that is similar for all
photon energy bands, but as $|-t|$ increases the fall-off flattens and
then becomes a rise.  Fig.~\ref{fig:dsdt_l2} shows the cross sections scaled
by $s^2$, and it is seen that there is a clear indication of a locus
$D(t)$ describing the data over a range of $-t$.  We interpret the
departures from this locus as the onset of the $s$- and $u$-channel
contributions to the reaction mechanism. At a given value of $-t$ the
residual spread of the points can be used to determine $\alpha(t)$ for
this reaction; this work is in progress and will be published
separately.

Examination of Figs.~\ref{fig:dsdt_l1} and \ref{fig:dsdt_l2} shows
a progressive flattening of the slope in the cross section as $|-t|
\rightarrow 0$.  This same ``plateau'' phenomenon was seen in data from
SLAC~\cite{slac} taken at $E_\gamma = 5, 8, 11, 16$ GeV, that is, well
above the energies of the present results.  In the model of Guidal,
Laget, and Vanderhaeghen~\cite{lag1}, this effect arose from the
interplay of degenerate $K$ and $K^*$ Regge trajectories and the
requirements imposed by gauge invariance in the model.  The intercepts
of these trajectories are at $\alpha(0) = -0.20$ and $+0.25$,
respectively, so their average is indeed at about 0, leading to the
observed $s^2$ scaling.  We note that this plateau effect persists well into
the nucleon resonance region, which suggests the importance of $K$ and
$K^*$ exchange throughout this kinematic region.

\begin{figure*}
\resizebox{0.7\textwidth}{!}{\includegraphics{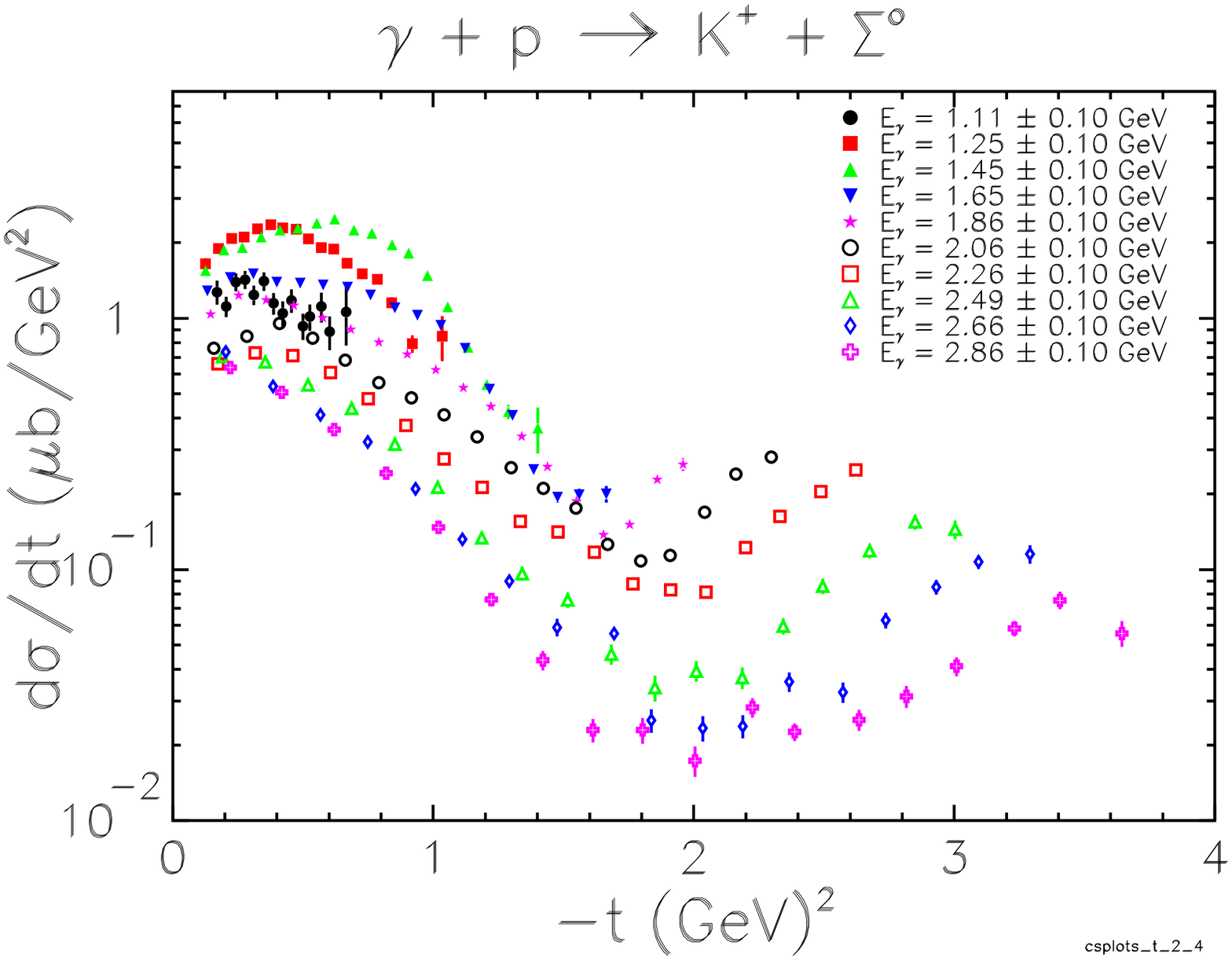}}
\caption{(Color online)
The entire $\gamma + p \rightarrow K^+ + \Sigma^0$ data set shown as
$d\sigma/dt$ vs. $-t$, for ten bands of photon energy with $\Delta
E_\gamma = 0.20$ GeV.  No scaling was applied.
}
\label{fig:dsdt_s1}       
\end{figure*}

\begin{figure*}
\resizebox{0.7\textwidth}{!}{\includegraphics{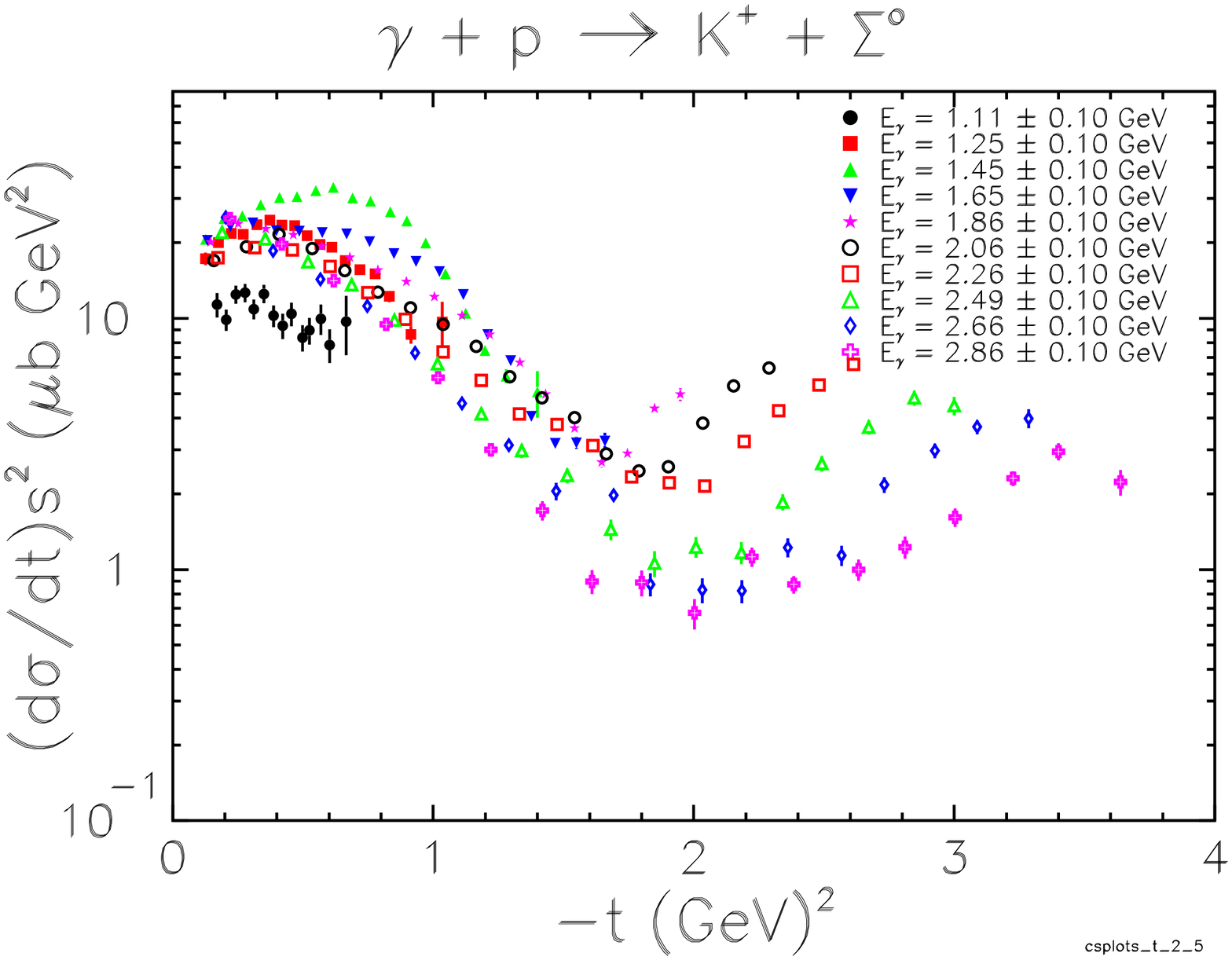}}
\caption{(Color online)
The entire $\gamma + p \rightarrow K^+ + \Sigma^0$ data set shown as
$d\sigma/dt$ vs. $-t$, for ten bands of photon energy with $\Delta
E_\gamma = 0.20$ GeV.  The cross sections were scaled by $s^2 = W^4$,
showing a less well-defined band of points than in the $K^+ \Lambda$ case.
}
\label{fig:dsdt_s2}       
\end{figure*}

The cross section $d\sigma/dt$ for the $\Sigma^0$ channel is shown in
Fig.~\ref{fig:dsdt_s1}.  In this case, the data do not fall in
monotonically shifting contours as $E_\gamma$ increases, as was the
case for the $\Lambda$ in Fig.~\ref{fig:dsdt_l1}.  Instead, a more
nucleon-resonance dominated picture is suggested by the crossing of
the bands of data points.  This is emphasized again in
Fig.~\ref{fig:dsdt_s2} that shows the $s^2$ scaled cross sections,
which in this case do not form a tight band of points.  There is no
consistent trend toward a flattening of the slope, as was the case in
$K^+ \Lambda$ production; in the previously cited theory~\cite{lag1}
this is because in $K^+ \Sigma^0$ production the $K$ plays little role
compared to $K^*$ since $g_{K \Sigma N} < g_{K\Lambda
N}$. Furthermore, the large ``resonant'' rise in the $\Sigma^0$ cross
section near $W=1.90$ GeV is serving to cover up any simple
$t$-channel behavior for this hyperon.

\begin{figure}
\resizebox{0.45\textwidth}{!}{\includegraphics{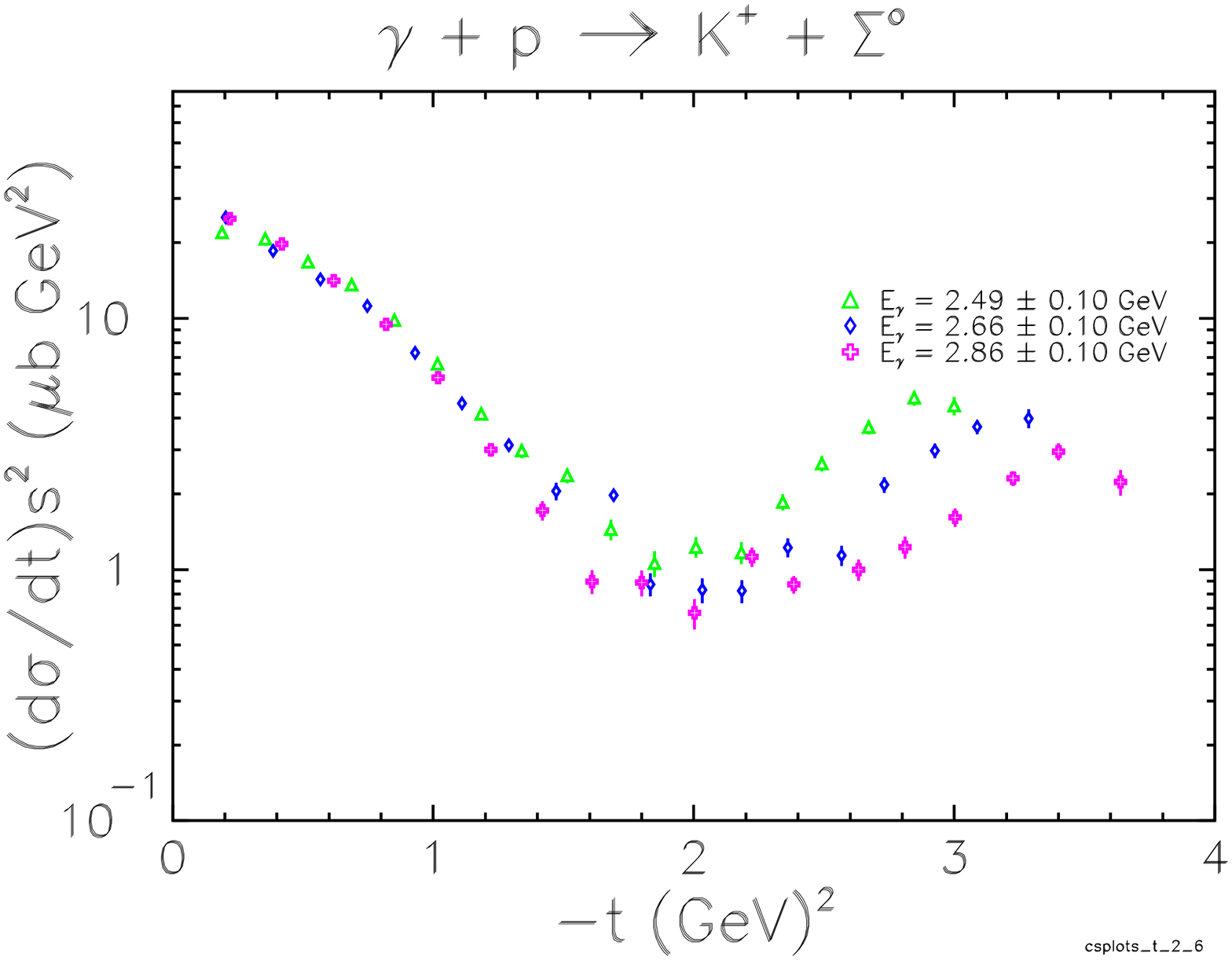}}
\caption{(Color online)
The $\gamma + p \rightarrow K^+ + \Sigma^0$ data shown as
$d\sigma/dt$ vs. $-t$, for the top three bands of photon energy with $\Delta
E_\gamma = 0.20$ GeV.  The cross sections were scaled by $s^2 = W^4$,
showing that for the highest energies the same scaling
phenomenon is apparent.
}
\label{fig:dsdt_s3}       
\end{figure}

At high enough energies, it is expected, however, that the $\Sigma^0$
cross section should also behave as expected by $t$-channel dominance.
In Fig.~\ref{fig:dsdt_s3} we show the subset of the data from the previous
figure for $E_\gamma > 2.39$ GeV, where the scaling by $s^2$ does seem
to work.  We note that this is well above the large ``$\Delta$'' peak
in the total cross section, Fig.~\ref{fig:stot}, and spans the range
where the Regge calculation~\cite{lag1,lag2} is successful in
explaining these data.

\section{\label{sec:conclusions}CONCLUSIONS}

In summary, we present results from an experimental investigation of
$\Lambda$ and $\Sigma^0$ hyperon photoproduction from the proton in
the energy range where nucleon resonance physics should dominate.  We
provide the to-date largest body of data for these reactions in
coverage over energy and meson angle.  Our $K^+\Lambda$ cross section
results reveal an interesting $W$-dependence: double-peaked at forward
and backward angles, but not at central angles.  We see that the
structure near $W=1.9$ GeV shifts in position and shape from forward
to backward angles.  This finding cannot be explained by a $t$-channel
Regge-based model or by the addition of a single new resonance in the
$s$ or $u$ channel.  The $\Sigma^0$ results confirm a single large
maximum in the cross section near 1.9 GeV, with weak indications of
more structure between 2.0 and 2.1 GeV.  The results are in fair or good
agreement with several older experiments.  For the $K^+\Lambda$ case
we see that a phenomenological scaling of the $t$-dependence of the
cross section by $s^2$ is quite successful in describing the full
range of forward-angle data, and that this scaling does not work as
well for the $K^+\Sigma^0$ data.  Our results show that hyperon
photoproduction can reveal resonance structure previously ``hidden''
from view, thereby improving our understanding of nucleonic
excitations in the higher mass region where data are sparse.
Comprehensive partial wave analysis and amplitude modeling for these
results can therefore be hoped to firmly establish the mass and possibly
the quantum numbers of these states.


\begin{acknowledgments}
We thank the staff of the Accelerator and the Physics Divisions at
Thomas Jefferson National Accelerator Facility who made this
experiment possible.  We thank D. Seymour for help with the angular
distribution fits.  Major support came from the U.S. Department of
Energy and National Science Foundation, the Italian Instituto
Nazionale di Fisica Nucleare, the French Centre National de la
Recherche Scientifique, the French Commissariat \`a l'Energie
Atomique, the German Deutsche Forschungs Gemeinschaft, and the Korean
Science and Engineering Foundation.  The Southeastern Universities
Research Association (SURA) operates Jefferson Lab under United States
DOE contract DE-AC05-84ER40150.
\end{acknowledgments}


\begin{thebibliography}{9}

\bibitem{mcnabb} J. W. C. McNabb, R. A. Schumacher, L . Todor (CLAS
Collaboration), {\it et al.},
Phys. Rev. C {\bf 69} 042201(R) (2004).

\bibitem{cap}  S. Capstick and W. Roberts, Phys. Rev. {\bf D58}, 074011
  (1998), and references therein.

\bibitem{pdg}  S. Eidelman \emph{et al.}, Phys. Lett. {\bf B592}, 1 (2004).

\bibitem{klempt} See for example: E. Klempt, {\it nucl-ex/0203002},
  and references therein. 

\bibitem{maid} T. Mart, C. Bennhold, H. Haberzettl
and L. Tiator, ``KaonMAID 2000'' at 
www.kph.uni-mainz.de/MAID/kaon/kaonmaid.html.

\bibitem{mart} T. Mart and C. Bennhold, Phys. Rev. {\bf C61}, 012201 (2000);
C. Bennhold, H. Haberzettl, and T. Mart, 
{\it nucl-th/9909022} and 
Proceedings of the 2nd Int'l Conf. on Perspectives in Hadronic
Physics, Trieste, S. Boffi, ed.,  World Scientific, (1999).

\bibitem{loring}  U. L\"oring, B. Ch. Metsch, and H. R. Petry,
  Eur. Phys. J. A {\bf 10}, 395 (2001).

\bibitem{bonn1} M. Q. Tran {\it et al.}, Phys. Lett. {\bf B445}, 20 (1998);
M. Bockhorst {\it et al.}, Z. Phys. {\bf C63}, 37 (1994).

\bibitem{sag} B. Saghai, {\it nucl-th/0105001}, AIP Conf. Proc. {\bf
59}, 57 (2001).  
See also: J. C. David, C. Fayard, G. H. Lamot, and
B. Saghai, Phys. Rev. {\bf C53}, 2613 (1996); $\sigma_{tot}$ curve by
private communication.

\bibitem{jan} S. Janssen, J. Ryckebusch, D. Debruyne, and T. Van
  Cauteren, Phys. Rev {\bf C65}, 015201 (2001); 
  S. Janssen {\it et al.}, 
  Eur. Phys. J. {\bf A 11}, 105 (2001); curves via private communication. 

\bibitem{jan_a}
  S.~Janssen, D.~G.~Ireland, and J.~Ryckebusch,
  Phys.\ Lett.\ B {\bf 562}, 51 (2003)
  [arXiv:nucl-th/0302047].

\bibitem{ireland}
  D.~G.~Ireland, S.~Janssen, and J.~Ryckebusch,
  Nucl.\ Phys.\ A {\bf 740} (2004) 147.


\bibitem{zegers}
  R.~G.~T.~Zegers {\it et al.}  [LEPS Collaboration],
  Phys.\ Rev.\ Lett.\  {\bf 91}, 092001 (2003)
  [arXiv:nucl-ex/0302005].

\bibitem{mohring}
  R.~M.~Mohring {\it et al.}  [E93018 Collaboration],
  Phys.\ Rev.\ C {\bf 67}, 055205 (2003)
  [arXiv:nucl-ex/0211005].

\bibitem{penner} G. Penner and U. Mosel, Phys. Rev. C {\bf 66}, 055212 (2002).

\bibitem{chiang} W. T. Chiang, Ph.D. Thesis, University of Pittsburgh
(2000) (unpublished); Wen-Tai Chiang, B. Saghai, F. Tabakin,
T. S. H. Lee, Phys. Rev. C {\bf 69}, 065208 (2004).

\bibitem{shklyar} 
  V.~Shklyar, H.~Lenske, and U.~Mosel,
  Phys.\ Rev.\ C {\bf 72}, 015210 (2005)
  [arXiv:nucl-th/0505010].

\bibitem{bonn2} K.-H. Glander {\it et al.}, Eur. Phys. J. {\bf 19}, 251 (2004).

\bibitem{sarantsev}
  A.~V.~Sarantsev, V.~A.~Nikonov, A.~V.~Anisovich, E.~Klempt, and U.~Thoma,
  arXiv:hep-ex/0506011.

\bibitem{lag1} M. Guidal, J.-M. Laget, and 
M. Vanderhaeghen, Nucl. Phys. {\bf A627}, 645 (1997).

\bibitem{lag2} M. Guidal, J.-M. Laget, and M. Vanderhaeghen,
 Phys Rev. {\bf C61}, 025204 (2000).

\bibitem{tagger}D. Sober {\it et al.}, Nucl. Instrum. Methods {\bf A440}, 263 (2000).

\bibitem{clas0} B. Mecking {\it et al.}, Nucl. Instrum. Methods 
                {\bf A503}, 513 (2003), and references therein. 

\bibitem{said} R.A. Arndt, W.J. Briscoe, I.I. Strakovsky, and
R.L. Workman, Phys. Rev. {\bf C66}, 055213 (2002).

\bibitem{john} J. W. C. McNabb, Ph.D. Thesis, Carnegie Mellon 
University (2002) (unpublished).  Available at
www.jlab.org/Hall-B/general/clas\_thesis.html.  

\bibitem{bradford} R. K. Bradford, Ph.D. Thesis, Carnegie Mellon 
University (2005) (unpublished).  Available at
www.jlab.org/Hall-B/general/clas\_thesis.html.  

\bibitem{clasdb} The CLAS Database collects all data from CLAS.  It is
  reachable via {\tt http://clasweb.jlab.org/physicsdb}.

\bibitem{contact}Text file available by sending email request to 
{\tt schumacher@cmu.edu}.

\bibitem{brody} H. M. Brody, A. M. Wetherell, and R. L. Walker, 
                Phys. Rev. {\bf 119}, 1710 (1960).

\bibitem{peck} C. W. Peck, Phys. Rev.  {\bf 135}, B830 (1964).

\bibitem{groom} D. E. Groom and J. H. Marshall, Phys. Rev. {\bf 159}, 1213 (1967).

\bibitem{donoho} P. L. Donoho and R. L Walker, Phys. Rev. {\bf 112},
                 981 (1958); Phys. Rev. {\bf 107}, 1198 (1957);

\bibitem{anderson} R. L. Anderson {\it et al.}, Phys. Rev. Lett. 
                  {\bf 9}, 131 (1962); also:  Phys. Rev. {\bf 123}, 1003 (1961).

\bibitem{thom} H. Thom, Phys Rev. {\bf 151}, 1322 (1966); 
               H. Thom, Phys. Rev. Lett. {\bf 11}, 432 (1963).

\bibitem{bleckman} A. Bleckmann {\it et al.}, Z. Phys. {\bf 239}, 1 (1970).

\bibitem{feller} P Feller, D. Menze, U. Opara, W. Schulz, and
                 W. J. Schwille, Nucl Phys. {\bf B39}, 413 (1972).

\bibitem{decamp} D. Decamp {\it et al.}, Preprint LAL 1236, Orsay (1970); 
                 Th. Fourneron, LAL 1258, Orsay (1971).

\bibitem{going} H. G\"oing, W. Schorsch, J. Tietge, and W. Weilnb\"ock,
                   Nucl. Phys. {\bf B26}, 121 (1971).

\bibitem{fujii} T. Fujii {\it et al.} Phys. Rev. D {\bf 2}, 439 (1970).

\bibitem{land} ``Photoproduction of Elementary Particles,''
  Landolt-B\"ornstein, New Series I/8, editted by H. Genzel, P. Joos,
  and W. Pfeil, Landolt-B\"ornstein, New Series I/8 (Springer Verlag,
  NY) (1973).

\bibitem{bonnnote} The $\sigma_{tot}$ values were read from the published graphs.

\bibitem{erbe} R. Erbe {\it et al.} (ABBHHM Collaboration) Phys. Rev. 188,
  2060 (1969);  data tabulated in  ``Numerical data and
  Functional Relationships in Science and Technology'', editted by
  H. Schopper, Landolt-B\"ornstein, New Series  I/12b (Springer Verlag, NY), 388. 

\bibitem{slac} A. M. Boyarski {\it et al.}, Phys. Rev. Lett. {\bf 22}, 1131 (1969).

\bibitem{perkins} See for example: D. H. Perkins, \underline {Introduction to
  High Energy Physics}, 2nd Ed., Addison Wesley, pp 166 (1982).





\end{thebibliography}
\end{document}